\documentclass[11pt]{article}

\usepackage{eucal}
\usepackage{color}

\usepackage{tikz}
\usepackage{tikz}
\usetikzlibrary{arrows}
\usetikzlibrary{decorations.pathmorphing}
\usetikzlibrary{decorations.markings}
\usetikzlibrary{patterns}
\usetikzlibrary{automata}
\usetikzlibrary{positioning}
\usepackage{tikz-cd}

\usepackage{pgfplots}

\usetikzlibrary{arrows}
\usetikzlibrary{calc}
\usepackage{amsmath,bbm}
\usepackage{amssymb}
\usepackage{epic}

\usepackage{graphicx}
\usepackage{pict2e}

\usepackage{pdfsync}

\usepackage{float}
\floatstyle{boxed} 
\restylefloat{figure}

\usepackage{amsmath}
\usepackage{upgreek}

\title{Lozenge tilings of hexagons with cuts and asymptotic fluctuations: a new universality class}

\author{Mark Adler\thanks{2000
{\em Mathematics Subject Classification}. Primary:
60G60, 60G65, 35Q53; secondary: 60G10, 35Q58. {\em Key
words and Phrases}:Lozenge tilings, non-convex polygons, kernels. \newline
 Department of Mathematics, Brandeis University,
Waltham, Mass 02453, USA. E-mail: adler@brandeis.edu.
The support of a Simons Foundation Grant 
 \# 278931 is
 gratefully acknowledged. M.A. thanks the Simons Center for Geometry and Physics for its hospitality.}~~~~~~Kurt Johansson\thanks{Department of Mathematics,
KTH Royal Institute of Technology, Stockholm, Sweden. E-mail: kurtj@kth.se. The support of the Swedish Research Council (VR) and grant KAW 2010.0063 of the Knut and Alice Wallenberg Foundation are gratefully acknowledged.} ~~~~~ Pierre
van Moerbeke\thanks{ Department of Mathematics,
Universit\'e de Louvain, 1348 Louvain-la-Neuve, Belgium
and Brandeis University, Waltham, Mass 02453, USA. E-mail: pierre.vanmoerbeke@uclouvain.be . The support of a Simons Foundation Grant 
 \# 280945 is
gratefully acknowledged. PvM thanks the Simons Center for Geometry and Physics, Stony Brook, and the Kavli Institute of Physics, Santa Barbara, for their hospitality.\newline
 }
}

\date{}

\newcommand{\MAT}[1]{\left(\begin{array}{*#1c}}
\newcommand{\mat}{\end{array}\right)}
\newcommand{\qed}{\leavevmode\unskip\nobreak\penalty200\hskip2pt\null
\nobreak\hfill\rule{1.1ex}{1.1ex}
\medbreak }

\newcommand{\x}{ \textcolor[rgb]{0.00,0.00,1.00}{\Large \! /\!\!  /\!\! /\!\!   /\!\! /\!\! /\! }
}

\newcommand{\y}{ \textcolor[rgb]{0.00,0.00,1.00}{\Large \backslash \! \backslash\!\!  \backslash\!\! \backslash\!\!   \backslash\!\! \backslash\!\! \backslash\! }
}

\newcommand{\I}{{\rm i}}

\newcommand{\CR}{{\cal C}}

\newcommand{\LR}{{\cal L}}
\newcommand{\RR}{{\cal R}}
\newcommand{\MR}{{\cal M}}

\newcommand{\PR}{{\cal P}}

\newcommand{\BC}{{\mathbb C}}

\newcommand{\BH}{{\mathbb H}}

\newcommand{\BP}{{\mathbb P}}

\newcommand{\BZ}{{\mathbb Z}}
\newcommand{\Sg}{\Sigma}

\newcommand{\pl}{\partial}
\newcommand{\al}{\alpha}

 \newcommand{\Om}{\Omega}

\newcommand{\tom}{\widetilde \omega}
\newcommand{\haom}{\widehat \omega}
\newcommand{\tze}{\widetilde \zeta}
\newcommand{\haze}{\widehat \zeta}
\newcommand{\tU}{\widetilde U}
\newcommand{\haU}{\widehat U}

\newcommand{\Id}{\mathbbm{1}}

 \newcommand{\bl}{\begin{aligned}}
  \newcommand{\el}{\end{aligned}}

\newenvironment{proof}{\medskip\noindent{\it Proof:\/} }{\qed}
\newenvironment
        {remark}{\medskip\noindent\underline{\it Remark:\/} }{\medbreak}

\newcommand{\om}{\omega}

\newcommand{\ga}{\gamma}
\newcommand{\Ga}{\Gamma}
\newcommand{\dt}{\delta}
\newcommand{\Dt}{\Delta}
 \newcommand{\vr}{\varepsilon}
\newcommand{\sg}{\sigma}
\newcommand{\ze}{\zeta}
\newcommand{\BR}{{\mathbb R}}
\newcommand{\lb}{\lambda}

\newcommand{\dis}{\displaystyle}

\newcommand{\BK}{{\mathbb K}}

\def\be#1\ee{\begin{equation}#1\end{equation}}
\def\bea#1\eea{\begin{eqnarray}#1\end{eqnarray}}
\def\bean#1\eean{\begin{eqnarray*}#1\end{eqnarray*}}

 \newtheorem{definition}{Definition}[section]
 \newtheorem{theorem}[definition]{Theorem}
 \newtheorem{lemma}[definition]{Lemma}
 \newtheorem{corollary}[definition]{Corollary}
 \newtheorem{proposition}[definition]{Proposition}

\catcode `!=11

\newdimen\squaresize
\newdimen\thickness
\newdimen\Thickness
\newdimen\ll! \newdimen \uu! \newdimen\dd! \newdimen \rr! \newdimen
\temp!

\def\sq!#1#2#3#4#5{%
\ll!=#1 \uu!=#2 \dd!=#3 \rr!=#4
\setbox0=\hbox{%
 \temp!=\squaresize\advance\temp! by .5\uu!
 \rlap{\kern -.5\ll!
 \vbox{\hrule height \temp! width#1 depth .5\dd!}}%
 \temp!=\squaresize\advance\temp! by -.5\uu!
 \rlap{\raise\temp!
 \vbox{\hrule height #2 width \squaresize}}%
 \rlap{\raise -.5\dd!
 \vbox{\hrule height #3 width \squaresize}}%
 \temp!=\squaresize\advance\temp! by .5\uu!
 \rlap{\kern \squaresize \kern-.5\rr!
 \vbox{\hrule height \temp! width#4 depth .5\dd!}}%
 \rlap{\kern .5\squaresize\raise .5\squaresize
 \vbox to 0pt{\vss\hbox to 0pt{\hss $#5$\hss}\vss}}%
}
 \ht0=0pt \dp0=0pt \box0
}

\def\vsq!#1#2#3#4#5\endvsq!{\vbox to \squaresize{\hrule
width\squaresize height 0pt%
\vss\sq!{#1}{#2}{#3}{#4}{#5}}}

\newdimen \LL! \newdimen \UU! \newdimen \DD! \newdimen \RR!

\def\vvsq!{\futurelet\next\vvvsq!}
\def\vvvsq!{\relax
  \ifx     \next l\LL!=\Thickness \let\continue=\skipnexttoken!
  \else\ifx\next u\UU!=\Thickness \let\continue=\skipnexttoken!
  \else\ifx\next d\DD!=\Thickness \let\continue=\skipnexttoken!
  \else\ifx\next r\RR!=\Thickness \let\continue=\skipnexttoken!
  \else\def\continue{\vsq!\LL!\UU!\DD!\RR!}%
  \fi\fi\fi\fi
  \continue}

\def\skipnexttoken!#1{\vvsq!}

\def\place#1#2#3{\vbox to 0pt{\vss
\rlap{\kern#1\squaresize
  \raise#2\squaresize\hbox{$#3$}}
\vss}}

\def\Young#1{\LL!=\thickness \UU!=\thickness \DD! = \thickness \RR! =
\thickness \vbox{\smallskip\offinterlineskip
\halign{&\vvsq! ##
\endvsq!\cr #1}}}

\catcode `!=12

\begin{document}

 \sloppy

\maketitle

\vspace*{-.4cm}

 \tableofcontents
 
 \newpage
 
  \begin{abstract}This paper investigates lozenge tilings of non-convex hexagonal regions and more specifically  the asymptotic fluctuations of the tilings within and near the strip formed by opposite cuts in the regions, when the size of the regions tend to infinity, together with the cuts. It leads to a new kernel, which is expected to have universality properties. 
 
  \end{abstract}


\section{Introduction and main results}


The work of MacMahon on the number of tilings of hexagons has gained a considerable interest in the physics community in the 50-60's, thinking of the work of Kaufman and Onsager\cite{Onsager} on the spontaneous magnetization of the square-lattice Ising model, thinking of the Kac-Ward formula \cite{Kac} for the
partition function of the Ising model on planar graphs and thinking of Kasteleyn's work \cite{Kasteleyn, Kast} on a full covering of a two-dimensional planar lattice with dimers, for which he computed the entropy. 

 This last quarter of a century Random Matrix activity has given us new tools and techniques to get insights in tiling problems, their phase transitions, their critical behaviors and this from both, the macroscopic and the microscopic point of view.
Many of these models show two phases (liquid and solid) and some of them three phases (gaseous, liquid and solid)\cite{CJ,BCJ}.  Indeed, experience has shown that the statistical fluctuations of the tiles near the singularities obey new probability laws, which tend to have a universal character. Tiling models are a rich source of new phenomena: they have sufficient complexity to have interesting features, and yet are simple enough to be tractable! For an overview, see \cite{Jo16}.

Tiling of non-convex domains were investigated by Okounkov-Reshetikhin \cite{OR} and Kenyon-Okounkov \cite{KO} from a macroscopic point of view. Further important phenomena for nonconvex domains appear in the work of Borodin, Gorin and Rains \cite{BGR}, Defosseux \cite{Defos}, Metcalfe \cite{Metc}, Petrov \cite{Petrov1,Petrov}, Gorin \cite{Gorin1}, Novak \cite{Nov}, Bufetov and Knizel \cite{BuK}, Duse and Metcalfe \cite{Duse,Duse1}, and Duse, Johansson and Metcalfe \cite{DJM}; see also the recent paper by Betea, Bouttier, Nejjar and Vuletic \cite{BBNV}.

This paper aims at lozenge tilings of non-convex polygonal regions and more specifically at the asymptotic fluctuations of the tilings near the non-convexities (cuts in the regions), when the size of the regions tend to infinity, together with one or several cuts. Do new statistical fluctuations appear near these cuts and are they universal? 

Our  work on random tilings of Aztec diamonds and double Aztec diamonds gave rise to such new fluctuations, leading to the tacnode-Airy \cite{AJvM} and the tacnode-GUE statistics\cite{ACJvM,AvM}; it should undoubtedly also lead to Pearcey statistics. Random lozenge tilings of polygons with cuts have led to the Cusp-Airy statistics \cite{OR, DJM}. All of these phenomena appear as result of non-convexities or cuts in the regions. In \cite{AJvM3} we obtained a kernel for the lozenge tilings of hexagons with several cuts along opposite sides. In this paper, we address the question of their asymptotics when the size of the polygon tends to infinity together with the cuts along opposite sides, while keeping certain geometric data fixed in order to guarantee interaction beyond the limit. It is our belief that this is universal statistics, more general than the tacnode-GUE statistics for overlapping Aztec diamonds and that a number of other statistics can be obtained from this one. There will be some indication of this later in this section.

\vspace*{2cm}
\setlength{\unitlength}{0.012in}\begin{picture}(100, 0) 
\put(  150,  35){\makebox(0,0) { $\fbox{$  \begin{array}{ccccc}
& \mbox{discrete-tacnode kernel for} 
\\& \mbox{non-convex hexagons} 
\end{array}
 $ }$}}
 \put(100,   10){\vector(-0.5, -1){20}}
 \put(160,   10){\vector( 0.7, -1){40}}
  \put(220,   10){\vector( 1.2, -1){60}}
   \put(140,   10){\vector( .2, -1){20}}
 
 \put(  60,  -65){\makebox(0,0) {
  $\fbox{$  \begin{array}{ccccc}
& \mbox{GUE-tacnode for} 
\\& \mbox{double Aztec-diamonds} 
\end{array}
 $ }$
 }}

\put(  230,  -65){\makebox(0,0) {$\fbox{Cusp-Airy kernel}$
}}

\put(  330,  -65){\makebox(0,0) {$\fbox{tacnode kernel}$
}}

\put(  180,  -115){\makebox(0,0) {$\fbox{Pearcey kernel}$
}}

  \end{picture}
 
 \vspace*{4cm}
  Fig.1. Is the statistics associated with the discrete-tacnode kernel for non-convex hexagons universal ?  Does it imply in some appropriate limit all these known statistics?
  
  \vspace*{.7cm}
  
 In this paper, we consider a hexagon with  cuts as in Fig.3, and a tiling with lozenges of the shape as in Fig.2, colored blue, red and green. Notice there is an affine transformation from our tiles to the usual ones in the literature; see e.g. the simulation of Fig.5. The usual right-leaning blue tiles turn into our blue ones, the usual up-right red ones into our red ones and the usual left-leaning green tiles (30${}^o$) to our green tiles (45${}^o$), all as in Fig.2.

Two different determinantal discrete-time processes, a ${\mathbb K}$-process and an ${\mathbb L}$-process, will be considered, depending on the angle at which one looks at the polygons; south to north for the ${\mathbb K}$-process or south-west to north-east

  \newpage
 
 \vspace*{-2.4cm}

\setlength{\unitlength}{0.017in} \begin{picture}(0,60)
\put(145,-70){\makebox(0,0) {\rotatebox{0}{\includegraphics[width=160mm,height=225mm]
 {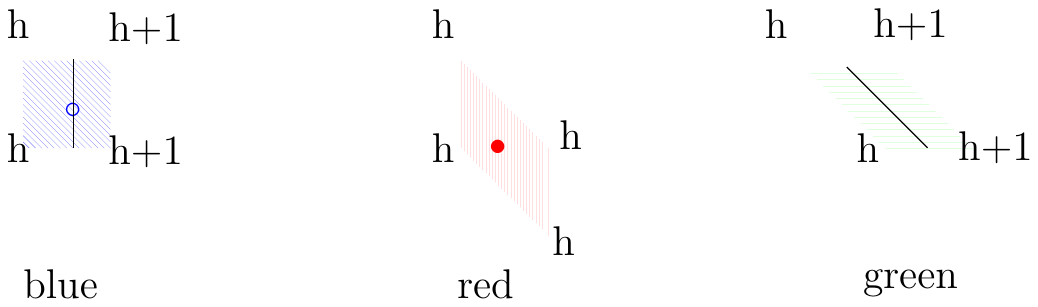} }}}

 \end{picture}
 
 \vspace*{.5cm}
Fig. 2. Three types of tiles, with a height function and with a level line.  

\vspace*{-4.4cm}


\setlength{\unitlength}{0.017in}\begin{picture}(0,170)
\put(150,-70){\makebox(0,0) {\rotatebox{0}{\includegraphics[width=135mm,height=225mm]
 {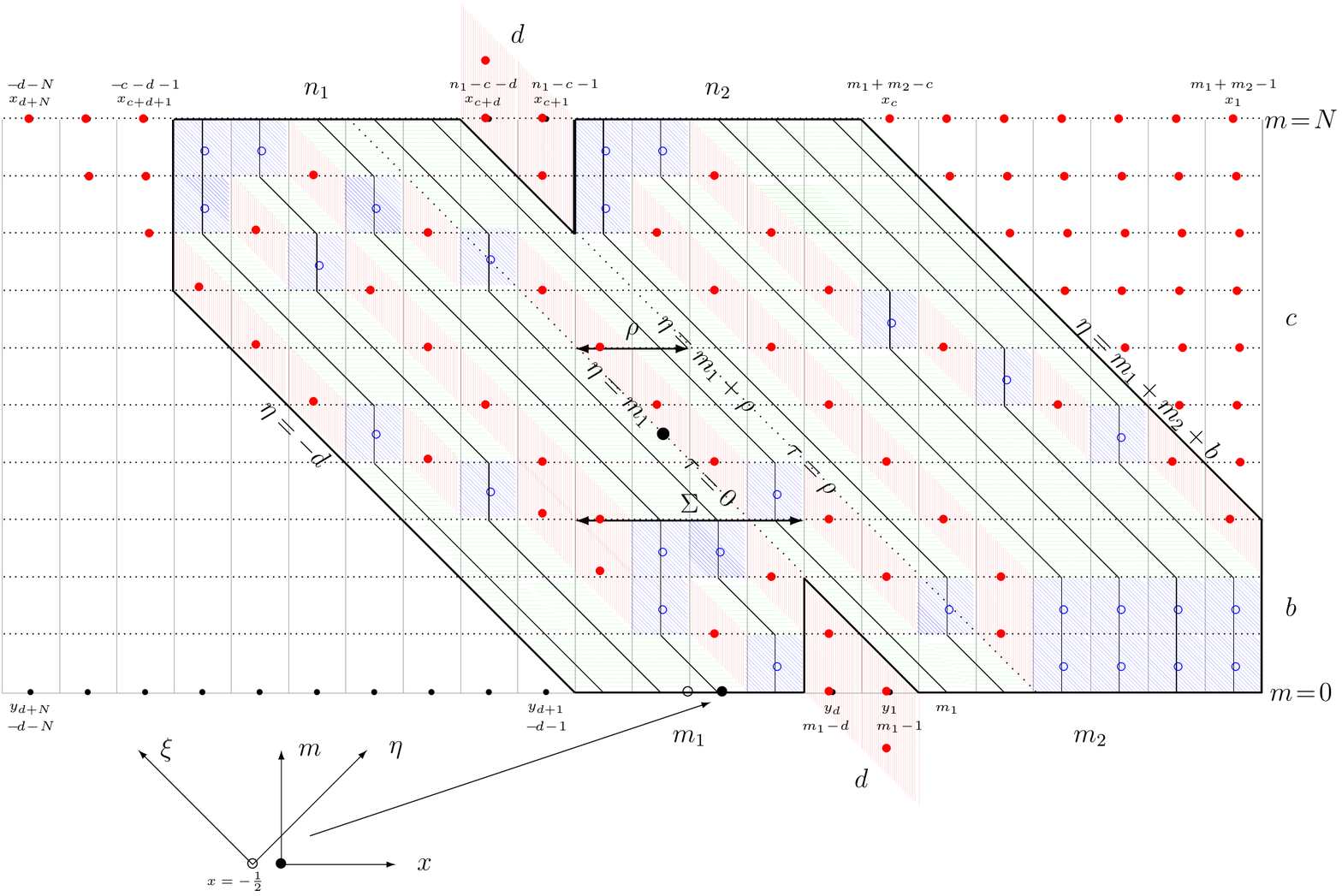} }}}
 

\end{picture} 


\vspace*{8.8cm}


Fig. 3. Tiling of a hexagon with two opposite cuts of equal size ({\em Two-cut case}), with red, blue and green tiles. Here $d=2$, $n_1=n_2=5,~m_1=4,~m_2=6,~ b=3,~c=7$, and thus $~ r=1,~ \rho=2,~\Sigma=4.$ The $(m,x )$-coordinates have their origin at the black dot and the $(\eta,\xi )$-coordinates at the circle given by $(m,x )=(0,-\tfrac 12 )$. 
Red tiles carry red dots on horizontal lines $m=k$  for $0\leq k\leq N$ (${\mathbb K}$-process) and blue tiles blue dots on oblique lines $\eta=k$ for $-d+1\leq k\leq m_1+m_2+b-1$ (${\mathbb L}$-process). The left and right boundaries of the strip $\{\rho\}$ are given by the dotted oblique lines $\eta=m_1$ and $\eta=m_1+\rho$. Asymptotics will be performed about the black dot in the middle of the hexagon.

\newpage

 \vspace*{-4.5cm}

  \setlength{\unitlength}{0.017in}\begin{picture}(0,60)
\put(145,-70){\makebox(0,0) {\rotatebox{0}{\includegraphics[width=115mm,height=160mm]
 {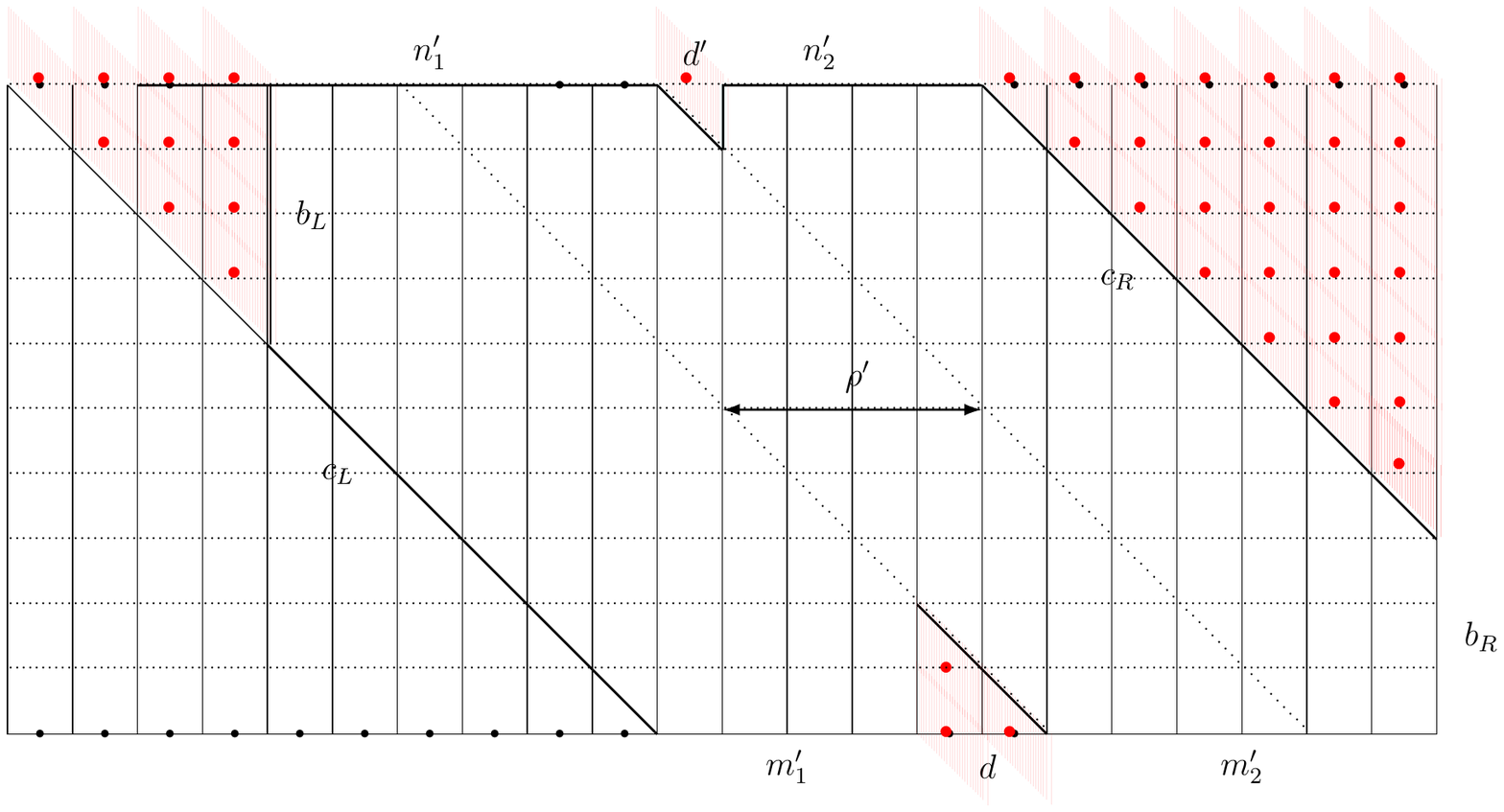}}}}

 \end{picture}

   \vspace{4.8cm}
  
  \noindent Fig. 4. The polygon with two cuts of different sizes $\bf P$, satisfying $c_L+d=c_R+d'$ and the quadrilateral $\widetilde {\bf P}={\bf P}\cup \{4\mbox{ red triangles}\}$.

 
 
  \vspace*{-4.5cm}
  
 

   \setlength{\unitlength}{0.017in}\begin{picture}(0,60)
 \put(125,-70){\makebox(0,0) {\rotatebox{-90}{\includegraphics[width=63mm,height=90mm]
  {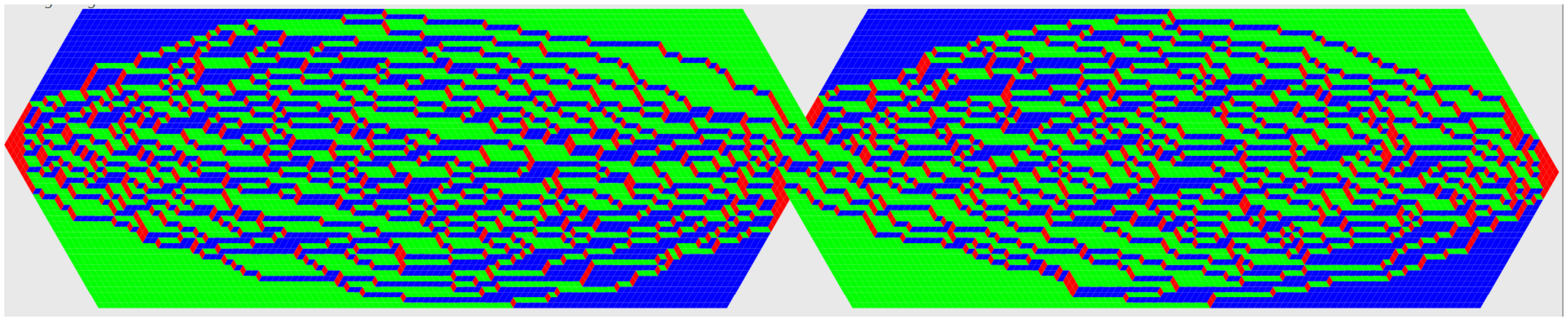
  }}}}

  \put(125,-180){\makebox(0,0) {\rotatebox{0}{\includegraphics[width=80 mm,height=70 mm]
 {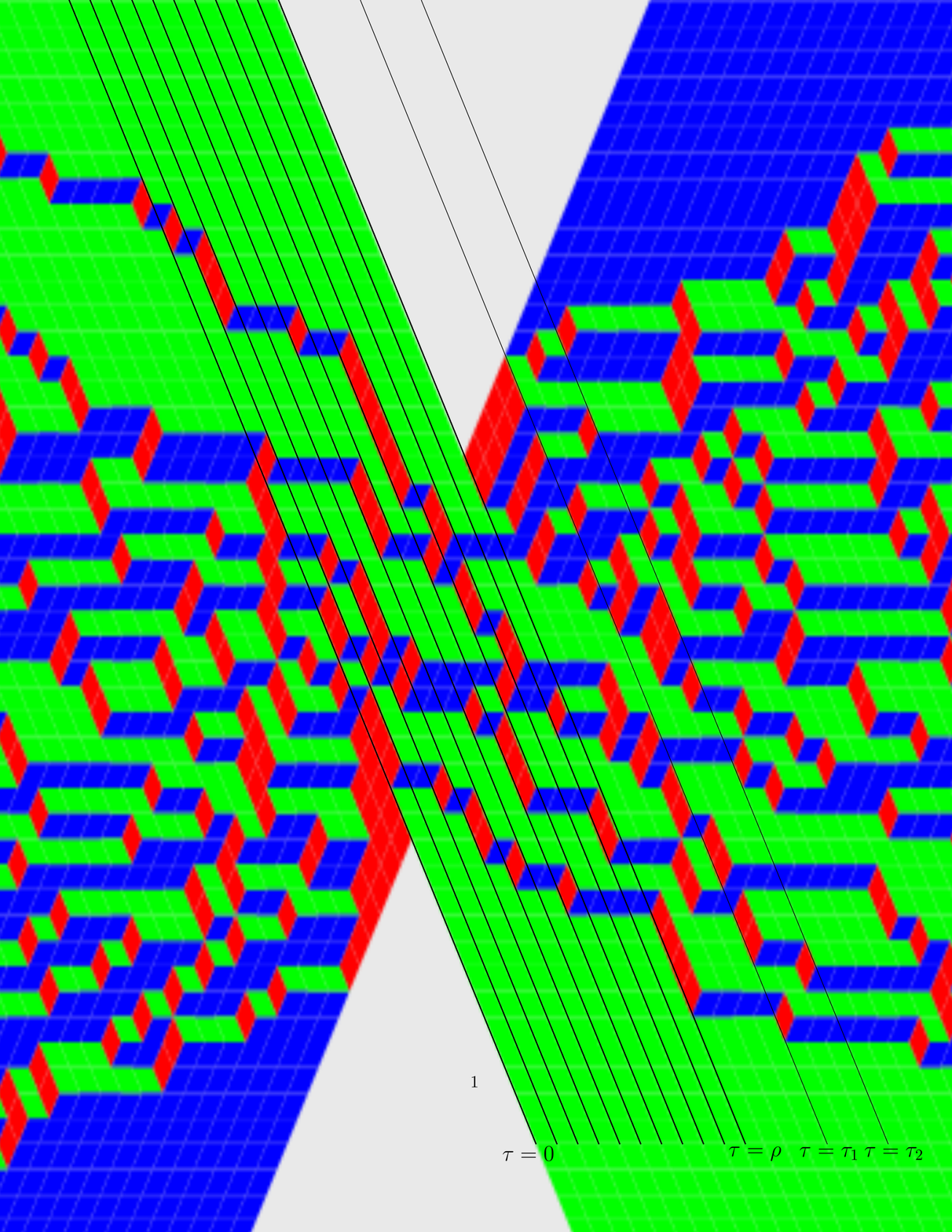} }}}
 

  

 \end{picture}
 
 \vspace{12cm}
 
\noindent Fig. 5. Computer simulation with $n_1 =  105, ~n_2=95,~m_1=m_2=100,~b=25,~c=30,~d=20 ~$, with $\rho=10 $ and $r=5$ {(upper-picture)}. Zooming into the intersection of the the two hexagons, with $\rho+1 =11$ oblique lines $0\leq \tau\leq \rho$, carrying $r=5$ blue tiles each  (lower-picture). The lines $\tau=\tau_\al\geq \rho$ for $\al=1,2$ each carry $\tau_\al-\rho+r$ blue tiles. (Courtesy of Antoine Doeraene)
 
 

 
 %
 
  \vspace*{1cm}
 %



\newpage

\vspace*{-2cm}

\noindent
for  the ${\mathbb L}$-process, both as an evolution in time. Paper \cite{AJvM3} focused on the 
${\mathbb K}$-process, whereas this one on the ${\mathbb L}$-process and its asymptotic limit in the neighborhood and in between the non-convexities. In \cite{AJvM3} we show the ${\mathbb K}$-process is determinantal and its kernel is given. We believe the  ${\mathbb L}$-kernel is a {\em master-kernel}, from which many new and old statistics can be deduced after an appropriate asymptotic limits.

{\bf The multi-cut model} is a  general non-convex polygonal region $\bf P$ consisting of taking a hexagon where two opposite edges have $ u-1$ cuts $b_1,~b_2,\dots,b_{ u-1}$ cut out of the upper-part (of sizes $b_1,\dots,b_{ u-1}$ ) and $\ell$ cuts $d_1,~d_2,\dots,d_{\ell}$ cut out of the lower-part (of sizes $d_1, \dots,d_{\ell}$); let $d:=\sum_{1}^\ell d_i$. Introduce the {\bf coordinates} $(m,x)\in \BZ^2$, where $m=0$ and $m=N$ refer to the lower and upper sides of the polygon, with $x$ being the running variable along the lines $m=$integer. Let $b_0$ and $b_{  u}$ be the 
  ``cuts" corresponding to the two triangles added to the left and the right of {\bf P} and let $d_0$ be the size of the lower-oblique side. Then $N:= b_0+d_0$ is the distance between the lower and upper edges. The intervals separating the upper-cuts (resp. lower-cuts) are denoted by $n_i$ (resp. $m_i$) and we require them to satisfy $\sum_1^{\ell+1} m_i=\sum_1^{  u} n_i$, which is equivalent to $\sum_0^{  u }b_i= d +N$.  Define $\widetilde{\bf P}$ to be the quadrilateral (with two parallel sides) obtained by adding triangles to ${\bf P}$, as in Fig.4. The vertices of $\bf P$ and $\widetilde{\bf P}$ and its tiles all belong to the vertical lines $x=\{\mbox{half-integers}\}$ of the grid (in Figs. 3 and 4).
  
  The $d$ integer points in $\{\widetilde{\bf P}\backslash {\bf P}\}\cap \{m=0\}$ are labeled by $y_1>\ldots>y_d$. We complete that set with the integer points to the left of $\widetilde{\bf P}$ along $\{m=0\}$; they are labeled by $y_{d+1} >\dots>y_{d+N} $ and we set $y_{d+1}=-d-1$ and $y_{d+N}=-d-N$. Similarly, the integer points $\{\widetilde{\bf P}\backslash {\bf P}\}\cap \{m=N\}$ are labeled by $x_1>\dots>x_{d+N}=-d-N$. For more details on the multi-cut model\footnote{We assume that $x_i\geq y_i$ for all $1\leq i\leq d+N$,  and that $y_d\notin \{\mbox{$x$-coordinates of an upper-cut}\}$.}, see \cite{AJvM3}.

 The {\bf two-cut model} is a special case where two opposite edges have one cut each ($\ell=1,~u=2$), both of same size $d:=d_1=b_1>0$.  Referring to Fig.3, we define  $b:=b_0,~c:=  b_u=d_0    $ and so $b+c=N$. The upper-cut is at distances $n_1,~n_2$ from the extremities of the upper-edge and the lower-cut at distances $m_1,~m_2$ from those of the lower-edge. The integers $b$ and $c$ determine the sizes of the two triangles which complete the figure into a quadrilateral (with two parallel sides) $\widetilde{\bf P}$, as in Fig. 4. 
 In other terms, the polygon $\bf P$ is now a hexagon with edges of size $m_1+m_2+d,~ b,~c\sqrt{2},~n_1+n_2+d, ~b,~c\sqrt{2}$ with two cuts, one below and one above, both of same size $d$, satisfying $m_1+m_2=n_1+n_2$.

 Assuming
  \be x_{c+1}<y_d< x_c,\label{geom}\ee  
  define polynomials\footnote{\label{ft1} For any integers $k\in \BZ$ and $N\geq  0$ we have $k_{0}=1$ and $k_{ N }=k(k+1)\dots(k+N-1)$.}:
\be \label{P} \mbox{$P(z) :=(z-y_d+1)_{N-d}~~\mbox{and}  ~~
Q(z) :=\prod_1^{d+N}(z-x_i)$}. \ee
 %
The choice of origin $(m,x)=(0,0)$ implies that the left most and right most points of the hexagon for the two-cut case along $m=0$ are given by $x=-d-1/2$ and $x=m_1+m_2-1/2$ respectively. The left-most point of the lower cut and of the upper-cut are located at $(m,x)=(0,m_1-d-\tfrac 12 )$ and $(m,x)=(N,n_1-c-d-\tfrac 12)$ respectively. 
Fig. 3 is such an example, which is covered by tiles of the three shapes.    Note that the right-most  point $y_1=m_1-1$ in the lower-cut will play an important role!

     Besides the $(m,x )$-coordinates, another set of coordinates $(\eta,\xi )$ will be used throughout (see Fig. 3):
\be\eta=m+x+\tfrac 12,~~~\xi=m-x-\tfrac 12 ~~~~~ \Leftrightarrow ~~~~~x=\tfrac 12 (\eta-\xi-1) ,~~~m =\tfrac 12(\eta+\xi).\label{Lcoord}\ee
%
 %


 %
  
  \medbreak

\noindent {\bf The ${\mathbb K}$ and ${\mathbb L}$-processes}. Given a covering of this polygonal shape with tiles of three shapes, colored in red, blue and green tiles, as in Figs. 2 and 3, put a {red dot} in the middle of the red tiles and a {blue dot} in the middle of the blue tiles. The {\bf red dots} belong to the intersections of the vertical lines $x=$ {\em integers} and the horizontal lines $m=0, \ldots , N$; they define a point process $(m,x)$, which was called the {\bf ${\mathbb K}$-process} in \cite{AJvM3}. The initial condition at the bottom $m=0$ is given by the $d$ {\em fixed red dots} at integer locations in the lower-cut, whereas the final condition at the top $m=N$ is given by the  $d+N$ {\em fixed red dots} in the upper-cut, including the red dots to the left and to the right of the figure, all at integer locations. Notice that the process of red dots on $\widetilde {\bf P}$ form an interlacing set of integers starting from 
   $d$ fixed dots (contiguous for the two-cut and non-contiguous for the multi-cut model) and growing linearly to end up with a set of $d+N$ (non-contiguous) fixed dots. This can be viewed as a ``truncated" Gel'fand-Zetlin cone!

 The {\bf blue dots} belong to the intersection $\in {\bf P}$ of the parallel oblique lines $x+m=k-\tfrac 12$ 
  with the horizontal lines $m=\ell-\tfrac 12$ for $k,\ell \in \BZ$
 ; in terms of the coordinates (\ref{Lcoord}), the blue dots are parametrized by $(\eta,\xi  )=(k,2\ell-k-1 )\in \BZ^2$, with $( k,\ell)$   as above. It follows that the $(\eta,\xi )$-coordinates of the blue dots satisfy $\xi+\eta=1,3,\dots,2N-1$.
 This point process defines the {\bf ${\mathbb L}$-process}, as was also discussed in \cite{AJvM3}. The blue dots on the oblique lines also interlace, going from left to right, but their numbers go up, down, up and down again, with a special feature, which will be explained later. 

\medbreak

In {\bf the two-cut model}, two strips within ${\bf P}$ will play a role (see Fig.3):

\noindent(i) an {\bf oblique strip $\{\rho\}$} extending the oblique segments of the upper- and lower-cuts; that is the region between the lines $x+m=-\tfrac 12 +k$ or what is the same $\eta=k$  for $m_1\leq k\leq n_1+b-d$. The strip $\{\rho\}$ has width 
\be \rho:=n_1-m_1+b-d=m_2-n_2+b-d,\label{rho}\ee
and assume $\rho\geq 0$. 
\newline(ii) a {\bf vertical strip $\{\Sg\}$} extending the vertical segments of the upper- and lower-cuts; that is the region between the lines $x = n_1-c-\tfrac 12$ and $x=m_1-d-\tfrac 12. $
The strip $\{\Sg\}$ has width (again same notation for the {\em name and the width} of the strip!)
\be \Sg:=m_1-n_1+c-d=n_2-m_2+c-d\geq 0;\label{sg}\ee
this inequality follows from (\ref{geom}). 

It is natural to {\bf assume} that the strips $\{\rho\}$ (respectively $\{\Sg\}$) have no point in common with the vertical parts (respectively oblique parts) of the boundary $\pl {\bf P}$. %


As shown in \cite{AJvM3}, the integer  \be r:=b-d \geq 0 \label{r} \ee
 equals the number of blue dots on the $\rho+1$ oblique lines $\eta=k$ for $m_1\leq k\leq m_1+\rho$; see Fig. 3. It will play a crucial role in this paper.


In \cite{AJvM3}, we obtained the ${\mathbb K}$-kernel. It is not clear how to obtain the ${\mathbb L}$-kernel from scratch, due to the intricacy of the interlacing pattern, mentioned earlier. Therefore one must first compute the ${\mathbb K}$-kernel and then one hopes to compute the ${\mathbb L}$-kernel by an alternative method. Indeed, we check that the inverse Kasteleyn matrix of the dimers on the associated bipartite graph dual to ${\bf P}$ coincides with the ${\mathbb K}$-kernel. This leads us to the first main statement of the paper. 

\begin{theorem}\label{Theo:L-kernel}
For the {\bf multi-cut case}, the ${\mathbb L}$-process of blue dots and the ${\mathbb K}$-process of red dots have kernels related as follows:

%
\be\begin{aligned}{\mathbb L}(\eta &,\xi ;\eta',\xi') 
 =-{\mathbb K}\left(m -\tfrac 12,x ;
m' +\tfrac 12,x' \right),
\end{aligned}\label{L-kernel'}\ee
where $(m,x)$ and $(m',x')$ are the same geometric points as $(\eta,\xi)$ and $(\eta',\xi')$, expressed in the new coordinates (\ref{Lcoord}); see Fig. 6.
%
\end{theorem}

\newpage


\vspace*{-4cm}

  \setlength{\unitlength}{0.017in}\begin{picture}(0,170)
\put(175,0){\makebox(0,0) {\rotatebox{0}{\includegraphics[width=160mm,height=225mm]
 {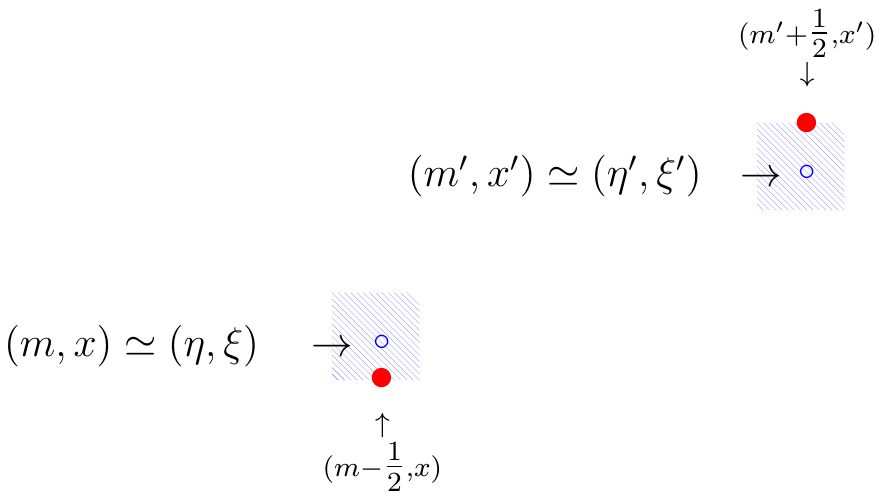} }}}

 \put(140, 50){\makebox(0,0) {$\bl\mbox{Fig. 6.}& ~\mbox{The ${\mathbb L}$-kernel of blue dots expressed in terms of the ${\mathbb K}$-kernel of}\\
 &\mbox{neighbouring red dots.}
 \el $  }}
  \end{picture}
 
 \vspace{-1.5cm}

 Before stating the next main Theorem of the paper (in the two-cut case) and assuming $N$ even, we denote by $\Dt_r$ the Vandermonde determinant, we define the Heaviside function for $m\in \BZ$ ,
\be
\begin{aligned}
   \BH^{m}(z)&:=\frac{z^{m-1}}{(m-1)!}\Id _{z\geq 0}\Id_{m\geq 1},
 \end{aligned} \label{Hm} \ee
 and a new kernel, the discrete Tacnode kernel $ {\mathbb L}^{\mbox{\tiny dTac}}  ( \tau_1, \theta_1 ;\tau_2, \theta_2) $, for $(\tau_i, \theta_i)\in \BZ\times \BR$. 
 It is a sum of a Heaviside function and four double integrals depending themselves on multiple integrals $\Theta_r$, defined below. The integrations are along  upwards oriented vertical lines $\uparrow L _{0+}$ to the right of a (counterclock) contour $\Gamma_0$ about the origin:
  \be \label{Final0}\begin{aligned}
 {\mathbb L}^{\mbox{\tiny dTac}}  (&\tau_1, \theta_1 ;\tau_2, \theta_2)  
  =  - 
{\mathbb H}^{\tau_1-\tau_2}(  \theta_2-  \theta_1) 
\\& +\oint_{\Ga_0}\frac{dV} {(2\pi\I)^2}\oint_{\uparrow L_{0+}}  \frac{ dZ}{Z-V}\frac{V^{\rho-\tau_1}}{Z^{\rho-\tau_2}}
\frac{e^{-V^2  -  \theta_1    V  }}
{e^{-Z^2 - \theta_2   Z  }}
      \frac{ \Theta_r( V, Z )} { \Theta_r(0,0)}  
\\& +\oint_{\Ga_0}\frac{dV} {(2\pi\I)^2}\oint_{\uparrow L_{0+}}\frac{ dZ}{Z-V} \frac{V^{  \tau_ 2}}{Z^{  \tau_1}}
\frac{e^{ -V^2 +  (\theta_2- \beta  )V  }}{e^{- Z^2  + (   \theta_1- \beta    )Z  }}
     \frac{ \Theta_r( V  , Z )} { \Theta_r(0,0)} 
\\& +r\oint_{\uparrow L_{0+}  }    \frac{dV} {(2\pi\I)^2} \oint_{\uparrow L_{0+}}dZ \frac{V^{ -\tau_1}}{Z^{\rho-\tau_2}}
\frac{e^{ V^2 -    (\theta_1- \beta     )V  }}
{e^{-Z^2 - \theta_2    Z  }}
    \frac{ \Theta^+_{r-1}(  V, Z )} {  \Theta_r(0,0)} 
 \\
 & -\tfrac1{r+1}\oint_{\Ga_0}\frac{dV} {(2\pi\I)^2} \oint_{\Ga_0}dZ 
 \frac{V^{ \rho-\tau_1}}{Z^{ -\tau_2}}
\frac{e^{-V^2 - \theta_1 V  }}
{e^{ Z^2 - (\theta_2- \beta    )Z  }}
    \frac{ \Theta^-_{r+1}( V, Z  )} {  \Theta_r(0,0)}   
\\
=&   \sum_{k=0}^4  {\mathbb L}_k^{\mbox{\tiny dTac}}
(\tau_1, \theta_1;\tau_2, \theta_2),
\end{aligned}\ee
where
  \be   \begin{aligned}
 \Theta_r( V, Z)&:=\left[     
\prod_1^r\oint_{\uparrow L_{0+}}\frac{e^{    2W_\al^2+\beta   W_\al}}{    W_\al  ^{\rho }}
 ~\left(\frac{Z\!-\!W_\al}{V\!-\!W_\al}\right) \frac{dW_\al}{2\pi \I}\right]\Dt_r^2(W_1,\dots,W_r)
%
 \\  \Theta^{\pm}_{r\mp1}(  V, Z)     &:=\left[     
\prod_1^{r\mp 1}\oint_{\uparrow L_{0+}}\frac{e^{    2W_\al^2+\beta  W_\al}}{    W_\al  ^{\rho }}
 ~\left( ({Z\!-\!W_\al} )\ ({V\!-\!W_\al})\right)^{\pm 1} \frac{dW_\al}{2\pi \I}\right]
 \\
 &~~ ~~\qquad~ ~\qquad \qquad~~~~~~~~~~~~~~~~~~~~~\Dt_{r\mp1}^2(W_1,\dots,W_{r\mp1}).
 \label{Theta}
\end{aligned}\ee
We now state the second main Theorem:

\begin{theorem}\label{Theorem2} ({\bf Two-cut case}) 
Keeping $r,\rho\geq 0$ fixed and letting the size $d$ of the two cuts go to $ \infty$, consider the following rescaling of the geometric variables $b,c,m_i,n_i>0 $ of the polygon $\bf P$, in terms of $d\to \infty$ and new parameters $1<\gamma<3$,   $a:=2\sqrt{\frac{\ga}{\ga-1}}$, $\bar \beta_1<0$, ~$\bar \beta_2, ~\bar\gamma_1,~\bar\gamma_2\in \BR$, 
\be\begin{array}{lllll}
b=d+r&& c=\ga d
\\
n_1 = m_1+(\rho-r) &&  m_1=\tfrac{\ga+1}{\ga-1}
 ( d+\tfrac a2 \bar\beta_1\sqrt{d}+\bar\ga_1)
\\
n_2 = m_2-(\rho-r) &&  m_2=\tfrac{\ga+1}{\ga-1}
 ( d+\tfrac a2 \bar\beta_2\sqrt{d}+\bar\ga_2).  
\end{array} \label{geomscaling}\ee
%
%
The variables $(\eta,\xi)\in \BZ^2$ with $\xi-\eta\in 2\BZ+1$ get rescaled into new variables $(\tau,\theta)\in \BZ\times\BR$, having their origin at the halfway point $(\eta_0,\xi_0)$ along the left boundary of the strip $\{\rho\} $, shifted by $(-\tfrac 12, \tfrac 12)$  : 
\be\begin{aligned}
 (\eta_i,\xi_i) &=(\eta_0,\xi_0)+(\tau_i,~\tfrac {\ga+1}a (\theta_i +\bar \beta_2)\sqrt{d} )\mbox{    with   }
  (\eta_0,\xi_0)  =(m_1,N-m_1 -1).
\end{aligned}
\label{eta-xi}\ee
With this scaling and after a conjugation, the kernel (\ref{L-kernel'}) of the ${\mathbb L}$-process tends to the  new kernel ${\mathbb L}^{\mbox{\tiny dTac}} $, as in (\ref{Final0}), depending only on the width $\rho$ of the strip $\{\rho\}$, the number $r$ of blue dots on the oblique lines in the strip $\{\rho\}$ and the parameter 
\be\bl \beta&:=-\bar\beta_1-\bar\beta_2
= \lim_{d\to \infty} \bigl({2d(d+c) +(m_1+m_2)(d-c)} \bigr)\tfrac{\sqrt{d^{-1}-c^{-1}}}{d+c};
\el\label{beta}\ee to be precise\footnote{It was noticed earlier that $\eta+\xi=$odd, and thus  the blue dots belonging to levels $\eta=k$ take on even or odd values of $\xi$, implying that the ``discrete differential" $\Dt\xi=2$. In the scaling limit this discrete differential will turn into a continuous differential.},
\be\begin{aligned}
 \lim_{d\to \infty} (-1)^{\tfrac 12 
   (\eta_1+\xi_1-\eta_2-\xi_2)}
&\left(\sqrt{d}\frac{\ga\!+\!1}{2a}\right)^{\eta_2-\eta_1
 }
{\mathbb L}( \eta_1 ,\xi_1;\eta_2,\xi_2)\frac 12 \Dt\xi_2
\\
&={\mathbb L}^{\mbox{\tiny dTac}} (\tau_1, \theta_1;\tau_2, \theta_2)d\theta_2.\end{aligned}
 \label{limit}\ee 
The kernel satisfies the following involution:
$$
{\mathbb L}^{\mbox{\tiny dTac}} (\tau_1, \theta_1;\tau_2, \theta_2)
=
{\mathbb L}^{\mbox{\tiny dTac}} (\rho-\tau_2, \beta-\theta_2;\rho-\tau_1, \beta-\theta_1).
$$
This involution exchanges ${\mathbb L}_1^{\mbox{\tiny dTac}}\leftrightarrow
{\mathbb L}_2^{\mbox{\tiny dTac}}$, with  ${\mathbb L}_k^{\mbox{\tiny dTac}}$ being self-involutive for $k=3,4$. Also ${\mathbb L}_1^{\mbox{\tiny dTac}}$ has support on $\{\tau_1>\rho\}$, ${\mathbb L}_2^{\mbox{\tiny dTac}}$ has support on $\{\tau_2<0\}$ and ${\mathbb L}_4^{\mbox{\tiny dTac}}$ on $\{\tau_1>\rho\}\cap \{\tau_2<0\}$.
\end{theorem}

\remark The condition $1<\ga<3$ just above (\ref{geomscaling}) is essential. When $\ga\sim 1$ moves to $\ga\sim 3$, it turns out that, under the scaling (\ref{geomscaling}), the point $y_d$ moves from $x_{c+1}$ to $x_c$. Thus, outside the range $\ga\in (1,3)$, the geometric condition $x_{c+1}<y_d<x_c$ on the model $\bf P$, as in (\ref{geom}), would be violated.


\medbreak

The main Theorem \ref{Theorem2} assumes equal sizes for the two cuts on the opposite sides. The next statement shows it does not need to be so: \begin{theorem}\label{theorem3}
For a hexagon with edges  $m'_1,m'_2,b_R,c_R\sqrt{2},n'_2,n'_1,b_L,c_L\sqrt{2}$
with two cuts of sizes $d$ and $d'$, as in Fig. 4, satisfying $m'_1+m'_2=n'_1+n'_2$ and $N=b_L+c_L=b_R+c_R$ and $c_L+d=c_R+d'$, the same limit (\ref{limit}) holds. Here also the limiting kernel ${\mathbb L}^{\mbox{\tiny dTac}}$, as in (\ref{Final0}), only depends on the width $\rho'$ of the oblique strip $\{\rho'\}$ formed by the two cuts, the number $r'$ of dots on each oblique line in the strip $\{\rho'\}$, and $\beta$ as in (\ref{beta}),where
$$ r'=b_L-d=b_R-d'\geq 0,~\mbox{and}~\rho'=n'_1-m'_1+r' .$$
\end{theorem}


 \remark The scaling in Theorem \ref{theorem3} is given by the following recipe. Setting $\dt:=d'-d$, we define a new hexagon with edges $m _1,m _2,b ,c\sqrt{2} ,n _2,n _1,b ,c \sqrt{2}$, and with two equal cuts of size $d$, where
\be\bl
&b=b_L+\dt=b_R,~~~c=c_L-\dt=c_R,~~~d=d'-\dt,
\\
&m_1=m'_1,~~ m_2=m'_2,~~n_1=n'_1-\dt, ~~n_2=n'_2+\dt,
\el\label{newHex}\ee
with $r=r'+\dt$ and $\rho'=n_1-m_1+b-d=\rho$. Substituting the formulas (\ref{newHex}) in the scaling (\ref{geomscaling}) gives the correct scaling for the model of Theorem \ref{theorem3}. 

Given the kernel (\ref{Final0}) it is very natural to ask for the (joint) density of blue dots along oblique levels $\tau=\tau_1,~\tau_2 $. A sample of these results is stated here; their proofs will appear elsewhere. For an illustration, see Fig. 5.

\begin{corollary} (\noindent {\bf Density of blue dots along oblique lines}).  Given two oblique lines (levels) $\rho\leq \tau_1<\tau_2$ to the right of the strip $\{\rho\}$ (as in lower Fig.5), each carrying $n_{ \al }=\tau_\al-\rho+r$ blue dots (tiles) at locations $\theta_i=:x_i^{(\al)}$, with $1\leq i\leq n_{\al}$.  
Setting 
${\boldsymbol{\theta}}^{(\al)}=({x}_1^{(\al)}\geq
 x_2^{(\al)}\geq \dots\geq x_{n_\al}^{(\al)})$ for $\al=1,2$, 
  the joint density of blue tiles can be expressed as : 
 \be\bl
 \BP&\left( {\theta}_i^{(\al)}\in d\boldsymbol{\theta}_i^{(\al)},\mbox{   with  }
 \left\{\bl &\theta_i^{(\al)}\in \mbox{level}~\tau_\al
 \\
 &1\leq i\leq n_\al 
 \el \right\},\mbox{   for $\al=1,2$}\right) 
 \\&=
 C \widetilde \Dt_{ n_1 }({\boldsymbol{\theta}}^{(1)}) \Dt_{n_2 }({\boldsymbol{\theta}}^{(2)})
 {\mathbb V}(n_1,{\boldsymbol{\theta}}^{(1)},n_2,{\boldsymbol{\theta}}^{(2)} )e^{-\frac{|\! | {\boldsymbol{\theta}}^{(2)}|\! |^2}{4}}  
  d{\boldsymbol{\theta}} ^{(1)}d{\boldsymbol{\theta}} ^{(2)} ,\el
\label{joint density}\ee
 where $C$ is a normalization constant and where $
 {\mathbb V}$ is the volume of the truncated cone:
  $$
 {\mathbb V}(n_1,{\boldsymbol{\theta}}^{(1)},n_2,{\boldsymbol{\theta}}^{(2)} )
 :=\int_{{\boldsymbol{\theta}}^{(1)}={\bf z}_{\tau_1 }\prec{\bf z}_{\tau_1+1}\prec \dots\prec {\bf z}_{\tau_2-1 }\prec {\bf z}_{\tau_2  }={\boldsymbol{\theta}}^{(2)}
   }
 d{\bf z}_{\tau_1+1}\dots 
 d{\bf z}_{\tau_2-1}.
 $$
 Density (\ref{joint density}) contains, besides a regular Vandermonde $\Dt_{n_2 }$,
 a determinant $\widetilde \Dt_{n_1 }$ of a matrix of size $n_1=\tau_1-\rho+r$:
%
 
 $\mbox{\footnotesize $ 
 \begin{aligned}
&\widetilde \Dt_{n_1 } ({\boldsymbol{\theta}}):=\det\left(\begin{array}{ccccc}
& 1  &\dots &1\\
& {\theta}_1  &\dots &\theta_{n_1}
\\
&\vdots& &\vdots
\\
&\theta_{1}^{\tau -\rho-1}&\dots &
  \theta_{n_1}^{\tau -\rho-1}
\\  \\
&p_{-\tau} ( \tfrac{\beta-\theta_{1}}2 ) &\dots & p_{-\tau} ( \tfrac{\beta-\theta_{n_1}}2 ) \\
&\vdots  &&\vdots\\
& p_{-\tau+\rho-1} ( \tfrac{\beta-\theta_{1}}2 ) &\dots &
p_{-\tau+\rho-1} ( \tfrac{\beta-\theta_{n_1}}2 )
\\
&p_{-\tau+\rho} ( \tfrac{\beta-\theta_{1}}2 )
&\dots &
 p_{-\tau+\rho} ( \tfrac{\beta-\theta_{n_1}}2 )
\\
&\vdots&&\vdots
\\
&p_{r-1-\tau} ( \tfrac{\beta-\theta_{1}}2 )
&\dots &
  p_{r-1-\tau} ( \tfrac{\beta-\theta_{n_1}}2 )
  \end{array}
\right) 
 \!\!\!\!\!\!  
 \vspace*{-6.4cm}  \begin{aligned}
  &    \left.
  \begin{array}{cccc}\\  \\ \\ \\    \end{array}\right\}
   \tau -\rho
    \\  \\
 &\left.\begin{array}{cccc}\\ \\   \\  \\  \end{array}\right\}\rho
  \\ 
 &\left.\begin{array}{cccc}\\ \\ \\   \\  \end{array}\right\}(r-\rho  )_{_{\geq 0}} 
\end{aligned}
\end{aligned}  $}$
 %
 %
 \newline with $\tau=\tau_1$ and $\beta$ as in (\ref{eta-xi}) and (\ref{beta}); see also (\ref{Final0}). The Vandermonde-like $\widetilde \Dt_{n_1 } $ above contains the integrals $p_\al(x) $, taken along the line  $\uparrow L _{0+}$; for $\al\geq 0$ they are Hermite polynomials and for $\al<0$ truncated normal  moments:
 $$\mbox{\footnotesize$p_\al(x)=\int_{\uparrow  L_{0+}}
 \frac{dv}{2\pi\I} v^{\al}e^{v^2+2x v}
 =\tfrac{\Id_{\al\geq 0}}{2^{\al+1}\sqrt{\pi}}
 e^{-x^2}H_\al(-x)+\Id_{\al< 0}
 \int_0^{\infty}\tfrac{\xi^{-\al-1}}{(-\al-1)!}e^{-(\xi-x)^2}d\xi
 $.}$$
 Similar densities can be written down for the case of two levels within and below the strip $\{\rho\}$.  
Given blue tiles at levels $ \tau_1<\tau_2$, the blue tiles along the lines $\tau_1<\tau<\tau_2$ in between are uniformly distributed (extension of Baryshnikov property to truncated cones). 

The one-level density at level $\tau=\tau_1=\tau_2\geq \rho$ is given by formula (\ref{joint density}), with   $n_1=n_2$ and the volume ${\mathbb V}=1$.

  \end{corollary}
  

 Notice that for $\rho=r$, one has $n_\al=\tau_\al$ and the joint density (\ref{joint density}) coincides with the corresponding density for the GUE-tacnode for overlapping Aztec-diamonds \cite{ACJvM,AvM}; its proof will appear elsewhere. For $\rho=r$, the one-level density coincides (visually) with the one-level density (19) in \cite{AvM} for the GUE-tacnode, after some minor change of variables. This is a very strong indication -although no proof- that the GUE-tacnode kernel of \cite{ACJvM,AvM} is  a special instance of the kernel (\ref{Final0}). Visually these two kernels look entirely different. 
%
%










\section{Revisiting the ${\mathbb K}$-process of red dots}

 This section contains a brief summary of the results obtained in \cite{AJvM3} necessary for this paper. We begin with the {\bf two-cut model}. The assumptions (\ref{geom}), (\ref{sg}) and the one formulated just below (\ref{sg}) imply the following inequalities: 
 \be
\max (-n_2,-m_1)<d-b \leq m_2-n_2=n_1-m_1\leq c-d <\min(m_2,n_1).
\label{ineq}
\ee
It will be clear that all these conditions are satisfied given the scaling (\ref{geomscaling}).
 
The line $m=N$ contains three separate sets of contiguous integers~  $\in \widetilde{\bf P} \backslash {\bf P}$:  
a  $\LR$(eft) region, an upper-${\mathcal C}$(ut)-region and the $\RR$(ight) region, containing respectively $b,~d$ and $c$ integers; in total $d+N $ integers; to wit:
\be\begin{aligned}\LR &:=\{ x_{d+c+b},\dots , x_{d+c+1}\}
,
\CR :=\{ x_{c+d},\dots , x_{ c+1}\} ,~
\RR :=\{ x_{c},\dots , x_{ 1}\}.
\end{aligned}\label{LCR}\ee
We define two other sets of  contiguous integers on the line $m=N$, related to the strips $\{\Sg\}$ and $\{\rho\}$ :
$$\begin{aligned}
 \{\bar\Sg\}  :=\{\Sg\}\cap\{m=N\}&=\{x_{c+1}+1,\dots, y_d-1\}=\{n_1-c,\dots,m_1-d-1\}
\\
   \{\bar\rho\} :=\{\rho\}\cap\{m=N\}&=  \{ x_{c+d}-\rho,\dots, x_{c+d}-1\}
   \\
   &=\{ m_1-b-c,\dots,n_1-c-d-1\}.   
\end{aligned}$$
The inequalities (\ref{ineq}) imply that each of the sets $\CR\cup \{\bar\Sg\}$ and $\{\bar\rho\}\cup\CR$ form a contiguous set of integers, such that each of the three sets are completely separated: $\CR<\{\bar\Sg\}<\RR$ and $\LR<\{\bar\rho\}<\CR$.

From the polynomials $P(z)$ and $Q(z)$  of degree $N-d$ and $N+d$, defined in (\ref{P}), we define other polynomials:
\be\begin{aligned}
P(z)&  =(z-y_d+1)_{N-d} 
 = P_{ \rho }(z)Q_{ \CR }(z)P_{ \Sg }(z)
  \\&
  :=(z-x_{d+c}+1)_\rho (z-x_{c+1})_d(z-y_d+1)_\Sg
 \\
Q(z) & =  Q_{ \LR }(z)Q_{ \CR }(z)Q_{ \RR }(z)  
 :=(z-x_{d+c+1})_b(z-x_{c+1})_d(z-x_1)_c,
\end{aligned}\label{PQ}\ee
where $P_{ \rho },~ P_{ \Sg }, ~Q_{ \LR },~ Q_{ \CR },~ Q_{ \RR }$ are monic polynomials whose roots are given by the sets $\{\bar\rho\},~\{\bar\Sg\}, ~ \LR,  \CR$ and $  \RR$, repectively. Later we will represent them as ratios of $\Ga$-functions\footnote{using $(a)_b=a(a+1)\dots (a+b-1)=\frac{(a+b-1)!}{(a-1)!}=\frac{\Ga(a+b)}{\Ga(a)}$.};  see (\ref{int1}) and (\ref{int2}) later. 

  Referring to contour integration in this paper, the notation $\Ga(\mbox{set of points})$ will denote a contour encompassing the points in question and no other poles of the integrands; e.g., contours like  
  \be
  \Ga(\RR), ~\Ga({\LR}), ~\Ga(x+\mathbb N),\dots.
  \label{GaRL}\ee

 %

\bigbreak

 For notational simplicity, we set 
  \be\begin{aligned}
 &R_1(v):=\! \frac{  (v\!-\!x\!+\!1)_{N-m-1}}{ Q _{ \RR }(v)Q _{ \CR }(v)} ,
 ~~R_2^{-1} (z):=\!\frac{   Q _{ \RR }(z)Q _{ \CR }(z)}
 {(z-y)_{N-n+1} },~ h(v):=\!\frac{ Q _{ \RR }(v)  }{P_\rho(v )P_\Sg(v )Q_\LR(v )}
 \\
&(R_2(z)h(z))^{-1} =\frac{P(z)Q_\LR(z)}{(z-y)_{N-n+1}}, ~~~~~~~R_1(v)h(v)=\frac{(v-x+1)_{N-m-1}}{P(v)Q_\LR(v)},\end{aligned}\label{Rih}\ee
which appear crucially in the kernel below and in the following $k$-fold contour integral\footnote{Set $\Om _0(v,z)=1$. A shorthand notation for the Vandermonde is $\Dt_k(u):=\Dt_k(u_1,\dots,u_k)=\prod_{1\leq i<j\leq k}(u_i-u_j)$.} for $k\geq 0$, (see (\ref{GaRL}))
 %
  \be\begin{aligned} 
 \Om^{ }_k (v,z)&:=\left(\prod_{\al=1}^k  \oint_{\Gamma(\LR)}
\frac{du_\al  h(u_\al)   }{2\pi \I ~  }
  ~\frac{  z-u_\al }{ v-u_\al  }
\right) \Dt^2_k (u) .
 %
\\
 \Om^{\varepsilon }_k(v,z)
 &:=\left(\prod_{\al=1}^k  \oint_{\Gamma(\LR)}
\frac{du_\al  h(u_\al)   }{2\pi \I ~  }
    (v-u_\al)^{\varepsilon}(z-u_\al)^{\varepsilon }
\right) \Dt^2_k (u) .
 \end{aligned}\label{Omr}\ee
 Be aware of the slightly different notation, in comparison with \cite{AJvM3}!

  \begin{theorem} \label{Kr}
 For the {\bf two-cut model} and for $(m,x)$ and $(n,y)\in {\bf P}$, the determinantal process of red dots is given by the kernel ${\mathbb K}   (m,x;n,y)$ below, involving at most $r+2$-fold integrals, with $r$, defined in (\ref{r}), being the number of blue dots along the oblique lines $\eta=k$ within the strip $\{\Sg\}$,%
     \be \begin{aligned}
   {\mathbb K}  &(m,x;n,y) =: \BK_0+\tfrac{(N-n)!}{(N-m-1)!}(\BK_1+ \tfrac {1}{r+1}\BK_2) 
   \\&= -\frac{(y-x+1)_{n-m-1}}{(n-m-1)!}\Id_{n>m}\Id_{y\geq x}
 + \frac{(N\!-\!n\! )!}{(N-m-1)!} 
\oint_{ {\Gamma( {x +{\mathbb N}}) }}  \frac{dvR_1(v)}{2\pi \I}
\\
& \left(\oint_{\Gamma_{\infty}} \frac{dz}{2\pi \I(z\!-\!v)R_2(z)}  
 \frac{  \Om^{ }_r (v,z)}{  \Om^{ }_r (0,0)} 
  +\!\tfrac{1}{r\!+\!1}\!\oint_{ \Ga_{ -\tau
  }}  
  \frac{dz   }{2\pi \I R_2(z)h(z) }   
 \frac{ \Om^{-}_{r+1} (v,z)}{  \Om  _r (0,0)} \right),
  \end{aligned}\label{Kernlim5}\ee
%
 \be\begin{aligned}
\Gamma(x+{\mathbb N})&:=\mbox{contour containing the set}~  x+{\mathbb N}=\{x,x+1,\ldots\}
\\
\Gamma_{\infty}&:=\mbox{very large contour containing all the poles of the $z$-integrand}
\\
\Ga_{-\tau }&:=\Ga(y+n-N,\dots, \min(y_1 -N,y) )\Id_{ \tau<0},
\end{aligned}
\label{cont0}\ee
 where\be \label{tau'}\tau:= y+n- m_1.
  \ee

\end{theorem}


   
   The relationship between the ${\mathbb K}$ and ${\mathbb L}$-kernels, stated in Theorem \ref{Theo:L-kernel} for {\bf the multi-cut model}, will be established by  showing that the kernel ${\mathbb K}$ is the inverse of the Kasteleyn matrix; this will be done in sections \ref{section3} and \ref{KKastinv}.   To do so, we use the very general form of the ${\mathbb K}$-kernel, given in (\ref{Kernlim''}) for the multi-cut model; see \cite{AJvM3}.  
 Incidentally, obtaining  the $d\to \infty$ asymptotics for the kernel (\ref{Kernlim''}) would be awkward, in view of  expression (\ref{Kernlim''}) involving $d+2$-fold integrals. In the two-cut case formula (\ref{Kernlim''}) reduces to (\ref{Kernlim5}).


 The set $\RR$ in the multi-cut model is to be defined as $\RR=\{x_i~~\mbox{such that} ~~x_i\geq y_d\}$. It reduces to the definition (\ref{LCR}) of $\RR$ for the two-cut model, using the geometric condition (\ref{geom}). %
Setting for any $k\geq 0$, \be S_x^{(k)}(w):=w^{x+1}(1-w)^k,\label{S} \ee 
we have 
for $x,y \in \BR$ and integer $k\geq 1$ the following identity, 
%
 \be
\oint_{\Gamma_0}\frac{w^{y }dw}{2\pi \I S_{x }^{(k)}(w)} 
=\frac{(x -y  +1)_{k-1}}{(k-1)!},~\mbox{   for   }x-y+k-1\geq 0.\label{ascfact}\ee
Setting for brevity $u=(u_1,\dots,u_d)$), define  $D_{ d}^{({\bf y}_{\mbox{\tiny cut}})}(w):=\det(w_k^{y_\ell})_{ 1\leq k,\ell\leq d}$ and  

\noindent$\mbox{\footnotesize$ %
 \Dt_{ d}^{({\bf y}_{\mbox{\tiny cut}})}( u_1,\dots, u_d)
   :=\det\!\left( \!\frac{(u_\al -y_\beta+1)_{N-1 }  }{(N-1)!}\!\!\right)_{\!\!1\leq \alpha,\beta \leq d} 
   =\left(\prod_{\al=1}^d\oint_{\Gamma_0} \frac {dw_\al 
 }{2\pi \I S^{(N)}_{u_\al}(w_\al)} \right)
 D_{ d}^{({\bf y}_{\mbox{\tiny cut}})}(w),$}$
 
 \noindent$\mbox{\footnotesize$ \widetilde \Dt_{ d}^{({\bf y}_{\mbox{\tiny cut}})}( w;u_2,\dots, u_d)
  := \det\left(\begin{array}{cccccccc}
w^{ y_1}\!\!\!\!\!\!&\dots\!\!\!\!\!\!&\!\!\!\!\!\!w^{y_d}       \\ \\
 &   \left(  \frac{(u_\al -y_\beta+1)_{N-1 }  }{(N-1)!} \right)_{ {2\leq \alpha  \leq d}\atop{1\leq \beta\leq d}}
 \end{array}
\right)  
$}$
 \hspace*{3.4cm}$\mbox{\footnotesize$=   
\left(
\prod_{\al=2}^d\oint_{\Gamma_0} \frac {dw_\al 
 }{2\pi \I S^{(N)}_{u_\al}(w_\al)} \right)
 D_{ d}^{({\bf y}_{\mbox{\tiny cut}})}(w)\Bigr|_{w_1=w}.$}$
 \vspace*{-1cm}
 \be \label{Deltacut}\ee
%
%
 %
%
Indeed, using (\ref{ascfact}), the integral expressions are valid, since for all $u_\al\in \RR$ and $y_\beta\in$ the lower-cut, we have $u_\al-y_\beta+N-1\geq y_d-y_1+N-1\geq 0$. 

\bigbreak


\begin{proposition} \label{Kernlim}The kernel for the {\bf multi-cut model} has the following form for $d\geq 0$, and $(m,x)$, $(n,y)\in {\bf P}$:
\be \begin{aligned}
   {\mathbb K} &(m,x;n,y)   =:(\widetilde{\mathbb K}_0+ \widetilde{\mathbb K}_1+\widetilde{\mathbb K}_2)(m,x;n,y)
\\  =& -\frac{(y-x+1)_{n-m-1}}{(n-m-1)!}\Id_{n>m}\Id_{y\geq x}
+  
\oint_{ {\Gamma( {x +{\mathbb N}}) }}  \frac{dv}{2\pi \I}
 \left(\frac{  (v-x+1)_{N-m-1}}{ (N-m-1)! Q(v)}\right)
\\
&\times\left(\oint_{\Gamma_{\infty}} \frac{dz }{2\pi \I(z\!-\!v)}\left(\frac{(N\!-\!n\! )!Q(z)}{(z-y)_{N-n+1}}\right) 
\frac{\Om_\RR(v,z)
 }{\Om_\RR(0,0)}\!\! \right.
 \\& \left.~~~~~~~~~~~~~~~~~~~~~~~~~~~~~~+d\oint_{\Ga_0}\frac{dw}{2\pi \I S_y^{(n)}(w) } 
 \frac{ {\widetilde{\Om}^{(1)}}_\RR(v,w)}{\Om_\RR(0,0)}\right), 
 \end{aligned}\label{Kernlim''}\ee
%
where $\Om_\RR(v,z)$ and $ {\widetilde\Om}^{(1)}_\RR(v,w)$ are multiple integrals\footnote{where $ \Om_\RR(v,z)=1$ for $d=0$, as before, and $\widetilde\Om^{(1)}_\RR(v,w)=w^{y_1}$ for $d=1$}, 
\be\begin{aligned}
\Om_\RR(v,z):= & \left(\prod_{\al=1}^d  \oint_{\Gamma(\RR)}
\frac{du_\al  }{2\pi \I    Q(u_\al )}
   \frac{v-u_\al}{z-u_\al}
    \right)
  \Dt_d (u ) \Dt_{ d}^{({\bf y}_{\mbox{\tiny cut}})}(u) 
%
\\ {\widetilde\Om}^{(1)}_\RR(v,w):=&\left( 
\prod^{d }_{\al=2} \oint_{\Gamma(	\RR)}   {du_\al \over 2\pi \I Q(u_\al)}
\right)\Dt_{d }   (  v, {u_2 },\ldots, {u_d} )
\widetilde \Dt^{({\bf y}_{\mbox{\tiny cut}})}_{d }(w;u_2,\dots, u_d)
%
,\end{aligned}\label{OmR'}\ee
 containing the expressions $\Dt_{ d}^{({\bf y}_{\mbox{\tiny cut}})}$ and $\widetilde \Dt^{({\bf y}_{\mbox{\tiny cut}})}_{d }$ in (\ref{Deltacut}) and the polynomial $Q(z)$ as in (\ref{P}).

\end{proposition}



\section{An identity along boundary points of $\bf P$}\label{section3}

In this section, we prove an identity valid near boundary points ({\bf multi-cut model}). Using the notation (\ref{OmR'}) in Proposition \ref{Kernlim}, we state the following:
 
  
 \begin{proposition} \label{CorK2K3zero}
For red dots $(n,y)\in \BZ^2$ in the sets \be \begin{aligned}&{\cal S}:=\left\{\!\!
\begin{aligned}& \{(n,y)\in   (\widetilde {\bf P} \backslash {\bf P})~\bigr|~(n,y)\in 
\left( \left\{ {\mbox{upper oblique boundaries}  
  } \right\} +(0,\tfrac 12) \right)  \}
     \\
     &  \{(n,y)\in   (\widetilde {\bf P} \backslash {\bf P})~\bigr|~(n,y)\in 
\left( \left\{ {\mbox{upper vertical boundaries}  
  } \right\} -(0,\tfrac 12) \right)  \}
    \end{aligned} \right.,\end{aligned}\label{locus}\ee   
we have\footnote{E.g., in $(  \eta,\xi)$-coordinates, the upper oblique lines in the two-cut case (as in Figure 3) correspond to $\eta=m_1+m_2+b+\tfrac 12$ and $\eta=m_1+\rho+\tfrac 12$.}
\be\label{BdryIdentity}
\oint_{\Ga({y,\dots,y-N+n})}\frac{dz~\Om_{\RR}(v,z)}{2\pi \I(z-v)}\left(\frac{(N-n)!Q(z) }{ (z-y)_{N-n+1}}\right)+ d\oint_{\Gamma_0}\frac{dw ~\widetilde \Om^{(1)}_\RR(v,w)}{2\pi \I S_y^{(n)}(w)}=0.
 \ee
 
 \end{proposition}


\bigbreak  

\noindent Refering to the kernel ${\mathbb K}=\widetilde{\mathbb K}_0+\widetilde{\mathbb K}_1+\widetilde{\mathbb K}_2$ in the notation of (\ref{Kernlim''}) ({\bf multi-cut model}), this proposition leads to the following:
\begin{corollary}\label{CorK2K3}
For $(n,y)\in {\cal S}$, we have
\be\label{K2K3}
\left( \widetilde\BK_1+ \widetilde \BK_2\right)(m,x;n,y)=\Id_{x\leq y}\frac{(n-m)_{y-x}}{(y-x)!}.
\ee
\end{corollary}

 %
 %
 We will need the following:
 \begin{lemma}\label{LemmaS}
 For $(n,y)\in {\cal S}$, the sum over  the roots of $(z-y)_{N-n+1}$ equals:
 \be \label{Sidentity}
 \sum_{z=y-N+n}^{y}\frac{(N-n)!Q(z)}{(z-y)_{N-n+1}Q'(z)S_z^{(N)}(w)}~
 = \frac{1}{S_y^{(n)}(w)} .\ee
 \end{lemma}
\proof Using the fact that $(n,y)\in {\cal S}$ and consequently that the roots of $(z-y)_{N-n+1}$ are roots of $Q(z)$, implies that the left hand side equals:$$ 
\begin{aligned}
   \sum^{N-n}_{r=0} & {(N-n)! (z-(y-r)) \over  \dis \prod^{N-n}_{s=0} (z-y+s) S^{(N)}_z (w)} \Bigl |_{z=y-r}
 \\
& = \sum^{N-n}_{r=0} {(N-n)!  \over  \dis \prod^{N-n}_{s=0 \atop s\neq r} (s-r) S^{(N)}_{y-r} (w)}
\\
&= {1\over w^{ y +1}(1-w)^N} \sum^{N-n}_{r=0} (-1)^r {(N-n)! \over r! (N-n-r)!} w^r
 = {1\over S^{(n)}_y (w)}.
\end{aligned} 
$$  \qed

\noindent{\em Proof of Proposition \ref{CorK2K3zero}}: The form (\ref{Kernlim''}) of the kernel will be most convenient here: it contains the expressions $ \Om_\RR$ and and an integral of $\widetilde\Om_\RR^{(1)}$; so, inserting (\ref{Deltacut}) into them, one finds:
\be\label{OmBdry1}
\begin{aligned}
  \Om_\RR&(v,z)  \\
  :=& 
\left(\prod^d_{\al =1} \oint_{\Ga(x_1,\ldots,x_c)}  {du_\al \over2\pi \I Q(u_\al )} \oint_{\Ga_0} {dw_\al \over 2\pi \I S^{(N)}_{x_{i_\al }}(w_\al )}\right)  
\prod^d_{\al =1} {v-u_\al \over z-u_\al } \Delta_d  (u) D_{ d}^{({\bf y}_{\mbox{\tiny cut}})}(w)    
\\
  =&\sum_{1\leq i_1,\ldots i_d  \leq c
}
  {\Delta_d  (x_{i_1},\ldots,x_{i_d })\over \displaystyle \prod^d_{\al=1}Q'(x_{i_\al })} 
 \prod^d_{\al =1} {v-x_{i_\al }\over z - x_{i_\al }}
\left(\prod^d_{\al =1} \oint_{\Ga_0} {dw_\al \over 2\pi \I S^{(N)}_{x_{i_\al }}(w_\al )}\right) D_{ d}^{({\bf y}_{\mbox{\tiny cut}})}(w).
\end{aligned}
\ee
and
\be\begin{aligned}\label{OmBdry2}
\oint_{\Gamma_0} \!\!\frac{ dw \widetilde\Om_\RR^{(1)}(v,w)}{2\pi \I S^{(n )}_y(w)} 
=& \oint_{\Gamma_0} \frac{ dw_1}{2\pi \I S^{(n )}_y(w_1)}    \left(\prod^{d }_{\al=2} \oint_{\Gamma	(x_1,\ldots,x_c)} \!\!  {du_\al    \over 2\pi \I Q(u_\al)}
\oint_{\Gamma_0} \frac {dw_\al   }{2\pi \I  S^{(N)}_{u_\al }(w_\al)}
  \right)
 \\&~~~ \times \Dt_{d  }   (  v,  {u_2 },\ldots, {u_d} )D_{ d}^{({\bf y}_{\mbox{\tiny cut}})}  ( w )
\\
=&  d  \oint_{\Ga_0} {dw_1 \over 2\pi i S^{(n)}_y (w_1)} 
\sum_{1\leq i_2,\ldots,i_d \leq c}  \Dt_{d-1} (x_{i_2},\ldots,x_{i_d})
\\
& \times\prod^d_{\al=2} \left({(v-x_{i_\al}) \over Q' (x_{i_\al})} \oint_{\Ga_0} \frac{d\om_\al }
 { 2 \pi i S^{(N)}_{x_{i_\al}}(\om_\al)}\right) D_{ d}^{({\bf y}_{\mbox{\tiny cut}})} (w).
\end{aligned}\ee
Notice that in $ \Om_\RR(v,z)$ all $x_{i_\al}$ can be taken distinct, because of the presence of the Vandermonde $\Dt_d(x_{i_1},\ldots,x_{i_d })$. 

 The $z$-integral in the ${\mathbb K}_2$-part of the kernel ${\mathbb K}(m,x;n,y)$ contains an expression of the following type:
 \be\label{ratio0}
\frac{Q(z) }{(z-y)_{N-n+1} (z-x_{i_\al})}, \mbox{    for  } x_{i_\al}\in \RR.
\ee
 For any $(n,y)\in \cal S$, 
 the roots $y,y-1,\dots,y-N+n$ of the $z$-polynomial $(z-y)_{N-n+1}$ are roots of the polynomial $Q(z)$ as well. So, for $(n,y)\in {\cal S}$, the expression (\ref{ratio0}) has simple poles at those $z=x_{i_\al}$, for which $ x_c\leq x_{i_\al}\leq y$ and no poles at the  $z=x_{i_\al}$'s such that  $ y+1 \leq x_{i_\al} $. Therefore we shall distinguish between the case $y\geq x_c$ and $y<x_c$. 
 %
     
%



{\bf At first, assume $y\geq x_c$}. The multiple sum in $ \Om_\RR$ can be decomposed as follows and rewritten, taking into account the fact above and removing the noncontributing part of the sum,
$$\begin{aligned}
\sum_{1\leq i_1,\dots,i_d\leq c} &=  \sum^d_{\beta=1} \sum_{
 \begin{array}{c}
  \textrm{{\scriptsize $1\leq i_\beta \leq c$}}      \\
 \textrm{{\scriptsize such that}}      \\
  \textrm{{\scriptsize $y-N+n \leq x_{i_\beta} \leq x_1 $}} 
 \end{array}}
 \sum_{1\leq i_1,\ldots,\widehat{i_\beta},\ldots,i_d \leq c} 
\\&= \sum_{\beta=1}^d
\sum_{
 \begin{array}{c}
  \textrm{{\scriptsize $1\leq i_\beta \leq c$}}      \\
 \textrm{{\scriptsize such that}}      \\
  \textrm{{\scriptsize $y-N+n \leq x_{i_\beta} \leq y$}} \end{array}
  }
\sum_{1\leq i_1,\dots,\hat {i_\beta},\dots, i_d\leq c} .
\end{aligned}$$
It is then useful to re-express the following  Vandermonde's in terms of lower-degree Vandermonde's; namely, for every $1\leq \beta\leq d$, we have
$$
\begin{aligned}
 \Dt_d (x_{i_1},\ldots,x_{i_d}) &D_{ d}^{({\bf y}_{\mbox{\tiny cut}})}(w_1,\ldots,w_d) = \prod_{\alpha \neq \beta} (x_{i_\beta}-x_{i_\alpha}) 
\\
&\times \Dt_{d-1} (x_{i_1},\ldots,\widehat{x_{i_\beta}},\ldots,x_{i_d}) D_{ d}^{({\bf y}_{\mbox{\tiny cut}})} ({w_\beta},w_1,\ldots,\widehat{w_\beta},\ldots,w_d)
\end{aligned}
$$
Then
\be \label{D10}
\begin{aligned}
& \oint_{\Ga({y,\dots,y-N+n})}\frac{dz}{2\pi \I(z-v)}\left(\frac{(N-n)!Q(z) }{ (z-y)_{N-n+1}}\right)\Om_{\RR}(v,z)
\\&= \sum^d_{\beta =1} \oint_{\Ga_0} {dw_\beta\over 2\pi i} \sum_{y-N+n\leq x_{i_\beta}\leq y}
\\
& \times \textrm{Res}_{z=x_{i_\beta}} \left[{(N-n)! Q(z) (v-x_{i_\beta})  \over  (z-y)_{N-n+1}Q'(x_{i_\beta})S^{(N)}_{x_{i_\beta}}(w_\beta) (z-v)(z-x_{i_\beta})} \prod^d_{\alpha=1 \atop \alpha\neq \beta} {x_{i_{ \beta}}-x_{i_\al}  \over  z-x_{i_\al}}\right]
\\
&\times   \sum_{\begin{array}{c}{    \textrm{{\scriptsize $  1\leq i_1,\ldots,\widehat i_\beta,\ldots, i_d  \leq c$}}    }
\\  
\mbox{\tiny all distinct}\\
\mbox{\tiny with all}~~   \textrm{{\scriptsize $x_{i_\al} \neq  x_{i_\beta}$}} \end{array}}  \Dt_{d-1} (x_{i_1},\ldots,\widehat{x_{i_\beta}},\ldots,x_{i_d})
\\
&\times \prod^d_{\al = 1 \atop \al \neq \beta} \left({v-x_{i_\al} \over Q'(x_{i_\al})} \oint_{\Ga_0}\frac{dw_\al 
 }{  2\pi i S^{(N)}_{x_{i_\al}}(w_\al)}\right) D_{ d}^{({\bf y}_{\mbox{\tiny cut}})} (w_\beta,w_1,\ldots,\widehat{w_\beta},\ldots,w_d).
\end{aligned}
\ee
For fixed $1\leq \beta \leq d$, and using identity (\ref{Sidentity}) in Lemma \ref{LemmaS}, the sum of the residues on the second line above equals
$$\begin{aligned}
\sum_{y-N+n \leq x_{i_\beta} \leq y} \textrm{Res}_{z=x_{i_\beta}} \left[ ~~ \right] &=-\!\!\!\sum_{z=y-N+n}^y \frac{(N-n)!Q(z)}{(z-y)_{N-n+1}Q'(z)S_z^{(N)}(w_\beta)} 
\\&=
 - {1\over S^{(n)}_y (w_\beta)}.
\end{aligned}$$
Observe that all $u$ such that $y-N+n\leq u\leq y$, with $(n,y)\in {\mathcal S}$, figure in the list of $x_{i_\beta}\in \RR$, since the upper-cuts do not contain $y_d$. So, the above sum is valid.

Notice that the new expression for the right hand side of  (\ref{D10}), is independent of the value of $x_{i_\beta}$.   Moreover the multiple sum in the expression above is, in reality, a free sum over $d-1$ indices $1\leq i_1,\ldots,\widehat{i_\beta},\ldots,i_d \leq c$, with $\beta$ playing no role in neither of the Vandermonde's.
 Also, we can remove the constraint in the sum, because if two or more $x_{i_\al}  =x_{i_\beta}$, then we get zero by the Vandermonde  $\Dt_{d-1} (x_{i_1},\ldots,\widehat{i_\beta},\ldots,  {i_d})$ and if exactly one $x_{i_\al}  =x_{i_\beta}$, then the skew-symmetry of $w_\al,~w_\beta$ in $D_{ d}^{({\bf y}_{\mbox{\tiny cut}})} (w_\beta,w_1,\ldots,\widehat{w_\beta},\ldots,w_d)$ in (\ref{D10}) will kill expression (\ref{D10}).  So its right hand side can be rewritten:
\be \label{D10'}
\begin{aligned}
 &=-\sum^d_{\beta =1} \oint_{\Ga_0} {d\zeta \over 2\pi i S^{(n)}_y (\zeta)} \sum_{1\leq i_1,\ldots,\widehat{ {i_\beta}},\ldots,i_d \leq c} \Dt_{d-1} (x_{i_1},\ldots,\widehat{x_{i_\beta}},\ldots,  x_{i_d})
\\
& \left(\prod_{\al=1 \atop \al\neq \beta} {(v-x_{i_\al}) \over Q'(x_{i_\al})}  \oint_{\Ga_0}     {dw_\al  \over 2\pi i S^{(N)}_{x_{i_\al}}(w_\al)} \right)  D_{ d}^{({\bf y}_{\mbox{\tiny cut}})} (\zeta,w_1,\ldots,\widehat{w_\beta},\ldots,w_d)
\end{aligned}
\ee
Each one of terms in the sum $   \sum^d_{\beta=1}$ is independent of $\beta$; so we may rename the variables in each of the terms:
\[
(\xi_{i_2},\ldots, \xi_{i_d})  := (x_{i_1},\ldots, \widehat{x_{i_\beta}},\ldots,x_{i_d}) 
\]
and accordingly
$$
(\om_2,\ldots,\om_{d}) := (w_1,\ldots,\widehat{w_{\beta}},\ldots,w_d).
$$
With this renaming each term in the sum will be the same. So, we conclude that, the expression (\ref{D10}), upon using (\ref{D10'}) and (\ref{Deltacut}), equals 
 $$
\begin{aligned}
 & \oint_{\Ga({y,\dots,y-N+n})}\frac{dz}{2\pi \I(z-v)}\left(\frac{(N-n)!Q(z) }{ (z-y)_{N-n+1}}\right)\Om_{\RR}(v,z)
 \\ =& -d  \oint_{\Ga_0} {d\zeta \over 2\pi i S^{(n)}_y (\zeta)} 
\sum_{1\leq i_2,\ldots,i_d \leq c}  \Dt_{d-1} (\xi_{i_2},\ldots,\xi_{i_d})
\\
& \times\prod^d_{\al=2} \left({(v-\xi_{i_\al}) \over Q' (\xi_{i_\al})} \oint_{\Ga_0} {d\om_\al   \over 2 \pi i S^{(N)}_{\xi_{i_\al}}(\om_\al)}\right)  D_{ d}^{({\bf y}_{\mbox{\tiny cut}})} (\zeta,\om_2,\ldots,\om_d)
 \\
 =&-d  \oint_{\Ga_0} {d\zeta \over 2\pi i S^{(n)}_y (\zeta)} 
 \left(\prod_{\al=2}^d\oint_{\Ga(\RR)}\frac{du_\al}{2\pi \I Q(u_\al)}\right)\Dt_d(v,u_2,\dots,u_d)\Dt_{ d}^{({\bf y}_{\mbox{\tiny cut}})} (\zeta,u_2,\dots ,u_d)
 \\
 =&
    - d\oint_{\Gamma_0}\frac{d\zeta ~\widetilde \Om^{(1)}_\RR(v,\zeta)}{2\pi \I S_y^{(n)}(\zeta)}
,\end{aligned}
$$
This establishes identity (\ref{BdryIdentity}) for $y\geq x_c$. 

{\bf The case $y<x_c$} is easier, because each of the terms in (\ref{BdryIdentity}) vanishes. Indeed, in that case the expression (\ref{ratio0}) has no poles and thus the first integral in (\ref{BdryIdentity}) vanishes. The points $(n,y)\in {\cal S}$ such that $y<x_c$ belong automatically to the inside of an upper-dent, and so $y\leq x_{c+1}$. But $x_{c+1}<y_d$ by assumption in the multi-cut case; see the paragraph before (\ref{S})) and so $y<y_d$. This implies that $y-y_\beta+1\leq 0$ for $1\leq \beta\leq d$, implying in turn that 
$$\oint_{\Gamma_0} \frac{ dw_1 ~w_1^{y_\beta}}{2\pi \I S^{(n )}_y(w_1)} =\oint_{\Gamma_0} \frac{ dw_1 ~ }{2\pi \I w_1^{y-y_\beta+1}(1-w_1)^n  }=0 \mbox{     for   }  1\leq \beta\leq d.
$$

Thus performing the $w_1$-integration on the first row of the determinant $D_{ d}^{({\bf y}_{\mbox{\tiny cut}})}  ( w )$ leads a zero integral, showing that the second integral in 
(\ref{BdryIdentity}) vanishes. This ends the proof of Proposition \ref{CorK2K3zero}.


\bigbreak

 \noindent{\em Proof of Corollary \ref{CorK2K3}}: The expression $\widetilde{\mathbb K}_1$ in the kernel (\ref{Kernlim''}) contains a $z$- integration along the contour $\Ga_\infty$; in view of the integrand, this contour can be decomposed into $\Ga_\infty=\Ga_v+\Ga(y,\dots,y-N+n)$. Identity (\ref{BdryIdentity}) then tells us that for $(n,y)\in {\cal S}$,
 $$\begin{aligned}
 & \left( \widetilde\BK_1+ \widetilde \BK_2\right)(m,x;n,y)
\\&=
\oint_{ {\Gamma( {x +{\mathbb N}}) }}  \frac{dv}{2\pi \I}
 \left(\frac{  (v-x+1)_{N-m-1}}{ (N-m-1)! Q(v)}\right)
 \oint_{\Gamma_{v}}\!\! \frac{dz }{2\pi \I(z\!-\!v)}\left(\frac{(N\!-\!n\! )!Q(z)}{(z-y)_{N-n+1}}\right) 
\frac{\Om_\RR(v,z)
 }{\Om_\RR(0,0)} 
\\&=\frac{(N-n)!}{ (N-m-1)!}
\oint_{ {\Gamma( {x +{\mathbb N}}) }}  \frac{dv}{2\pi \I}
  \frac{  (v-x+1)_{N-m-1}}{ (v-y)_{N-n+1}} 
   =\Id_{x\leq y}\frac{(n-m)_{y-x}}{(y-x)!};
 \end{aligned}
 $$
for this last integral, see Petrov \cite{Petrov}. This ends the proof of formula (\ref{K2K3}) in Corollary \ref{CorK2K3}.\qed

\section{The $ {\mathbb K}_{ }$-kernel is the inverse of the Kasteleyn matrix and deriving the $ {\mathbb L}_{ }$-kernel of blue dots}
\label{KKastinv}

In this section we show that, {\bf for the multi-cut model},  the inverse of the Kasteleyn matrix equals the $ {\mathbb K}_{ }$-kernel (Theorem \ref{ThKastInv} below), from which the proof of Theorem \ref{Theo:L-kernel} is an easy consequence.

\newpage

\newcommand{\z}{\setlength{\unitlength}{0.015in}\begin{picture}(8,11)
 \linethickness{.001mm} \put(   0,   0){\line(-1,  1){2}}
  \put(   2,   0){\line(-1,  1){4}} \put(   4 ,   0){\line(-1,  1){6}}
  \put(   6,   0){\line(-1,  1){8}}\put(  8,   0){\line(-1,  1){10}}
 \end{picture}}

\newcommand{\za}{\setlength{\unitlength}{0.030in}\begin{picture}(8,11)
 \linethickness{.0001mm} \put(   0,   0){\line(-1,  1){2}}
  \put(   2,   0){\line(-1,  1){4}} \put(   4 ,   0){\line(-1,  1){6}}
  \put(   6,   0){\line(-1,  1){8}}\put(  8,   0){\line(-1,  1){10}}
   \put(  10,   0){\line(-1,  1){12}} \put(  12,   0){\line(-1,  1){14}}
   \put(  14,   0){\line(-1,  1){16}} \put(  16,   0){\line(-1,  1){18}}
   \put(  18,   0){\line(-1,  1){20}} \put(  20,   0){\line(-1,  1){22}}
 \end{picture}}

 \vspace*{.9cm}


 \begin{picture}( 0,0)
  \put(-40,-400)
   {{\rotatebox{0}
   {\includegraphics[width=160mm,height=225mm] {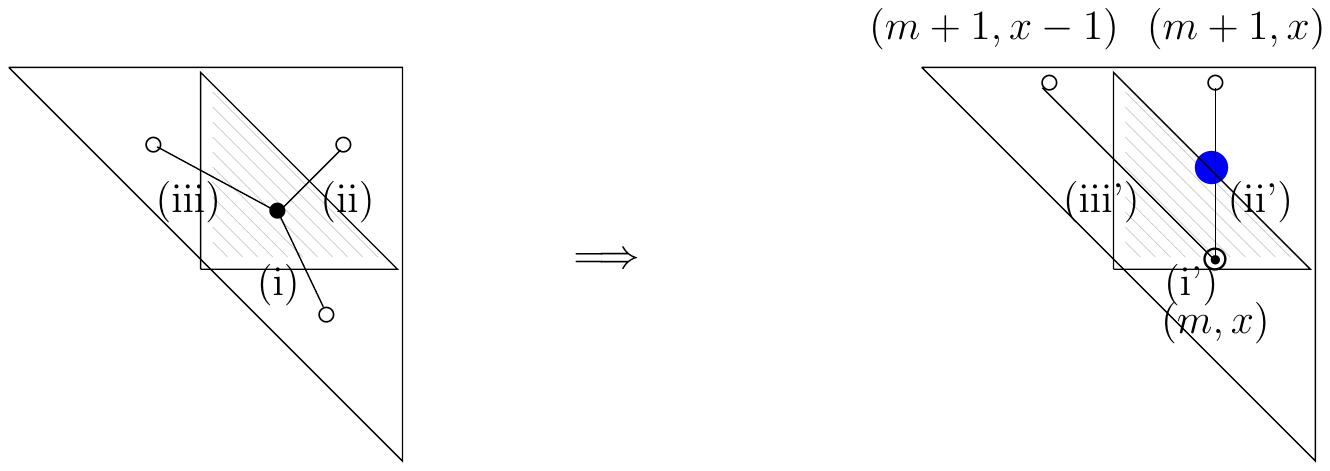} }} }
 

\end{picture}

\vspace*{.3cm}

\noindent Fig. 7 \& 8.: The three dimers on the honeycomb lattice corresponding to the three types of lozenges. Changing the coordinates by moving the white circles to the middle upper-part of the 
   triangle and the black dots to the middle of the lower-part of the shaded triangle.

\vspace*{.2cm}

\begin{picture}( 0,0)
  \put(-80,-400)
   {{\rotatebox{0}
   {\includegraphics[width=170mm,height=200mm] {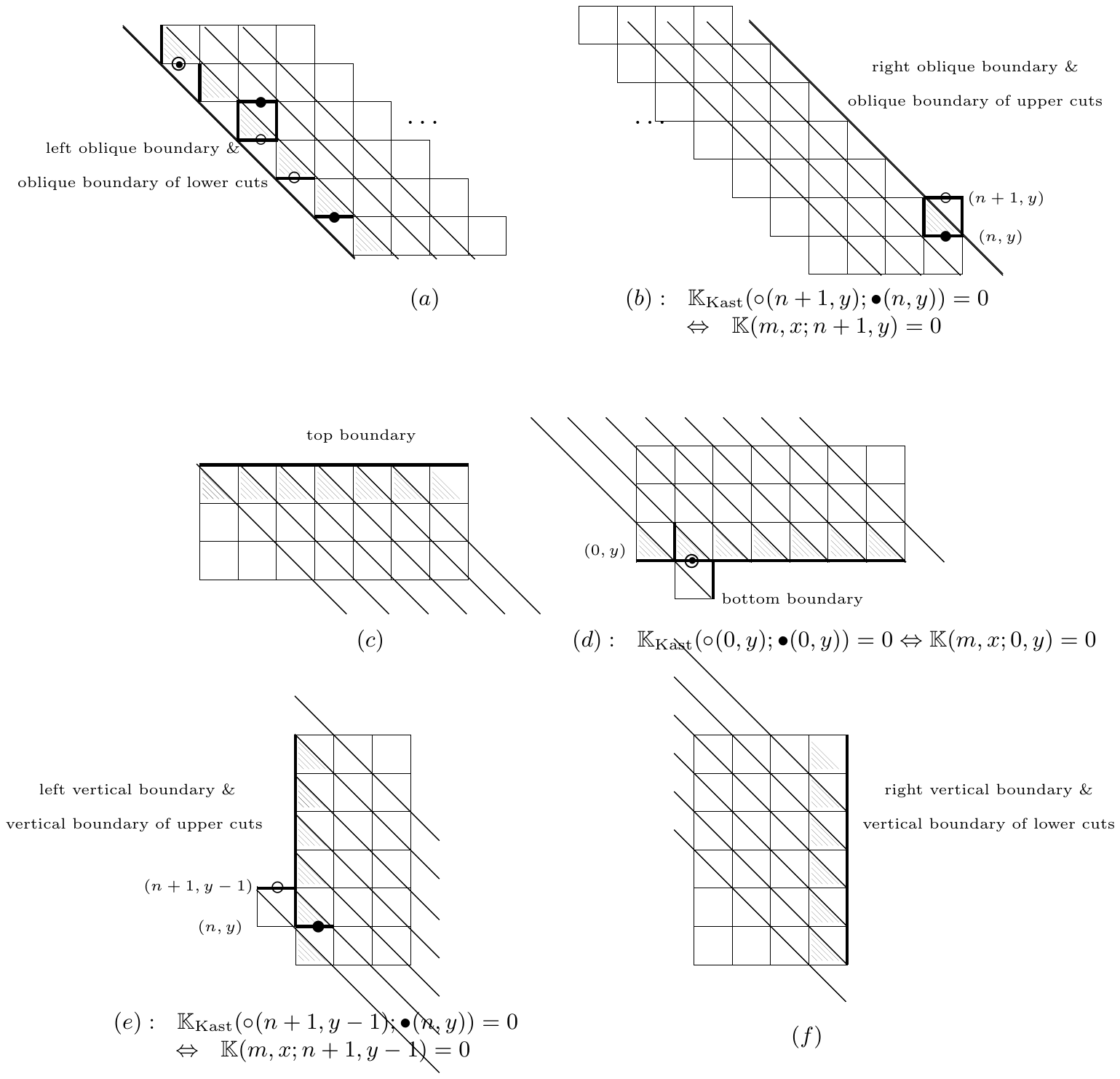} }} }
 

\end{picture}

\vspace*{12.5cm}

\noindent Fig. 9. The boundaries at which the $\mathbb K$-kernel vanishes.

 \newpage

\newpage

 The lozenge tilings of the hexagon is equivalent to dimers on its dual graph,  a honeycomb graph (bipartite), consisting of black dots, alternating with white circles; see Fig. 7   above. The black dots correspond to the shaded triangles and the white circles to white triangles. Only one segment emanates out of each vertex, as shown in Figs. 7 and 8. The Kasteleyn matrix for this honeycomb graph is the adjacency matrix and is given by:
  $$\begin{aligned}
  {\mathbb K}_{\mbox{\tiny Kast}}(\circ \to \bullet)&=1, ~~\mbox{for (i),~ (ii) and (iii) only}
 \\& =0, \mbox{~  otherwise,}
 \end{aligned} $$
 in terms of the dimers  (i),~(ii)  and (iii) given in Fig. \!8. Typically the vertices would belong to the middle of the black and white triangles, as in the left Fig. \!8.
 However, the $(m,x)$-coordinates of our problem correspond to the integer points coinciding with the middle of the lower-edge of the squares, which is the middle of the lower-edge of the shaded triangles in the right Fig. \!8 above. In order to make the two figures match, one moves the dots in the white triangles to the top-edge and the circles in the shaded triangles to the lower-edge. Thus the Kasteleyn matrix in the new coordinates (i'), (ii'), (iii') reads 
  \be\begin{aligned}
  {\mathbb K}_{\mbox{\tiny Kast}}(     \circ (n,y);   \bullet (m,x))&=1, ~~\mbox{if $(m,x)=(n,y)$, as in (i') }
 \\& =1, ~~\mbox{if $(m,x)=(n-1,y)$, as in (ii') }
 \\& =1, ~~\mbox{if $(m,x)=(n-1,y+1)$, as in (iii') }
 \\& =0, ~~\mbox{otherwise},
 \end{aligned} \label{Kast}\ee
 with {\em vanishing boundary conditions} along three of the six types of boundaries appearing in the hexagon ${\bf P}$. The Kasteleyn matrix vanishes there, as indicated in Fig. \!9, since the corresponding dimer traverses the boundary in (b),~(d) and (e), whereas in (a), (c) and (f), the shaded triangles are adjacent to white triangles, and thus no dimer crosses the boundary, leading to the non-vanishing of the Kasteleyn matrix.
 %

Before proceeding, we prove a three-step relation for the kernel $\mathbb{K} (m,x;n,y)$:
\begin{lemma}\label{Lemma3Step} The following formal identity involving integrals, is valid for all $(m,x), (n,y)\in {\bf P}$, including $(n,y) \in $ boundary\footnote{This is to say that the white triangles, corresponding to $(n,y)$, $(n+1,y)$ and $(n+1,y-1)$, can be outside  ${\bf P}$.}:
\begin{equation}
\Dt\mathbb{K} := \mathbb{K} (m,x;n,y) - \mathbb{K}(m,x;n+1,y) + \mathbb{K} (m,x;n+1,y-1) 
= \Id _{\{(m,x)=(n,y)\}.}\label{3Step}
\end{equation}
%
\end{lemma}
 \proof To do so, use version (\ref{Kernlim''}) of the kernel ${\mathbb K} (m,x;n,y)=\sum_{i=0}^2 \widetilde{\mathbb K} (m,x;n,y)=:\sum_{i=0}^2 
\mathbb{K}_i (n,y) $, each piece being abbreviated as $\mathbb{K}_i (n,y)$, with the $(m,x)$-variables being implicit. So we check:  \be
\Dt \mathbb{K}_i:=\mathbb{K}_i ( n,y)  - \mathbb{K}_i( n+1,y) + \mathbb{K}_i ( n+1,y-1) 
=\left\{\begin{aligned} &\Id _{(m,x)=(n,y)}\mbox{,~~for $i=0$}\\
&0\mbox{,~~~~~~~~for $i=1,~2$.}\end{aligned}\right. 
\ee
At first:
$$\begin{aligned}
\Dt \mathbb{K}_0
 =&\displaystyle - {(y-x+1)_{n-m-1}  \over (n-m-1)! }
\\
& ~~~~~\displaystyle \left(\mathbbm{1}_{n>m} \mathbbm{1}_{y\geq x} - {y-x+n-m \over n-m} \mathbbm{1}_{n+1>m} \mathbbm{1}_{y\geq x} + {y-x\over n-m} \mathbbm{1}_{n+1>m}\mathbbm{1}_{y\geq x+1}\right)
\\=& \Id _{(m,x)=(n,y)},
\end{aligned}
$$
as for $n=m$ and $y=x$, only the middle term in the equation above counts and contributes 1; otherwise the expression above vanishes.

To show that
$ \Dt \mathbb{K}_1=0$, we consider the part of the integrand containing $(n,y)$ only and check that 
\begin{eqnarray*}
&&\displaystyle {(N-n)! \over  (z-y)_{N-n+1}} - 
{(N-n-1)! \over  (z-y)_{N-n }} +
{(N-n-1)! \over  (z-y+1)_{N-n }}
\\
&&\displaystyle   = {(N-n-1)! \over  (z-y+1)\ldots (z-y+N-n-1)}
\\
&& \displaystyle    \left[{N-n  \over  (z-y) (z-y+N-n  )} 
- {1\over z-y} + {1\over z-y+N-n  }\right] = 0,
\end{eqnarray*}
whereas checking $ \Dt \mathbb{K}_2=0$ reduces to showing
$$
\frac1{S_y^{(n)}(z)} -\frac1{S_y^{(n+1)}(z)}+\frac1{S_{y-1}^{(n+1)}(z)}=
 \frac1{z^y(1-z)^n}\left(\frac1z-\frac1{z(1-z)}+\frac1{1-z}\right)=0.
$$
ending the proof of Lemma \ref{Lemma3Step}.\qed

\medbreak

 \noindent We now prove the first statement of this section:
 \begin{theorem} \label{ThKastInv}For the multi-cut case, the $ {\mathbb K}$-kernel is the inverse of the ${\mathbb K}_{\mbox{\tiny Kast}}$. To be precise:
  \be \label{KastInv}\begin{aligned}
  {\mathbb K}_{\mbox{\tiny Kast}}^{-1}(  \bullet (m,x);   \circ (n,y))&=
  (-1)^{x-y+ m-n}
   {\mathbb K}_{ } (        m,x ; n,y ).
   \end{aligned} \ee
 \end{theorem}
 
 \proof It suffices to prove that
 \be \label{Kinv}
 \sum_{(n' ,y')\in {\bf P} }(-1)^{x-y' +m-n' }
  {\mathbb K}_{ } (        m,x ; n' ,y'  )
  {\mathbb K}_{\mbox{\tiny Kast}} (     \circ (n' ,y' );   \bullet (n,y))
  =\Id_{_{(m,x)=(n,y)}}.
 \ee
 {\bf I}. {\bf For a fixed black triangle $\bullet (n,y)$ belonging to the interior of the hexagon ${\bf P}$ or to any of the boundary types (a),~(c),~(f) } of Fig. \!9 (i.e., lower cuts, lower boundaries or top horizontal boundary) the Kasteleyn matrix takes on three non-zero values as in (\ref{Kast}), and thus formula (\ref{Kinv}) reduces to the three-step relation (\ref{3Step}) in Lemma \ref{Lemma3Step}.
  %


 \medbreak
 
\par\noindent
\textbf{II. Bottom boundary of ${\bf P}$, away from the lower-cuts of type  (d) }: Putting the boundary condition ${\mathbb K}_{\mbox{\tiny Kast}} (     \circ (0,y);   \bullet (0,y))=0$ into identity (\ref{Kinv}) reduces to proving
 \be\begin{aligned}
-\mathbb{K} (m,x;1,y) + \mathbb{K} (m,x;1,y-1) = \Id_{_{\{(m,x)=(0,y)\}}},\label{bottom}
\end{aligned}\ee
and, upon subtracting (\ref{3Step}) from this equation, it suffices to show :
$$
\mathbb{K} (m,x;0,y)=0 \ \textrm{for all $x$ and }\ m\geq 0. 
$$ 
%
$ \mathbb{K}_0 (0,y)=0 $ is automatic, since $\Id_{n>m}=0$ for $n=0$, since  
both $m,n\geq 0$. 
To show
 $\mathbb{K}_1 (0,y)=0 $, we evaluate the behavior of the $z$-integrand at $z=\infty$; i.e., 
$$\frac{dzQ (z)}{(z-v)(z-y)_{N+1}} \Om_\RR(v,z)
 \simeq\frac{z^{d+N}dz}{z^{N+2}z^d} \simeq z^{-2}dz,
$$
which has zero residue at $z=\infty$. Finally, $\mathbb{K}_2 (0,y)=0$ because the $w$-integration in  $\mathbb{K}_2 (0,y)$ acting on the first row of $\widetilde \Dt^{({\bf y}_{\mbox{\tiny cut}})}_{d }(w,u_2,\dots, u_d)$, as given in (\ref{Deltacut}) (see  (\ref{OmR'}) and (\ref{S})), involves the following integral, which is $=0$ for all $1\leq \beta\leq d$,
$$
\oint_{\Gamma_0} \frac{dw}{2\pi \I w^{y+1}}w^{y_\beta}=0, \mbox{  if  } y\notin \{y_1,\dots,y_d\},
 $$ thus establishing (\ref{bottom}).

\medbreak

\par\noindent
\textbf{III. Upper-vertical boundaries of ${\bf P}$ of type (e)},  as in Fig. 9 : Inserting ${\mathbb K}_{\mbox{\tiny Kast}}(     \circ (n+1,y-1);   \bullet (n,y))=0$ into equation (\ref{Kinv}), it suffices to prove that for the points $(n,y)\in {\bf P}$ adjacent to the upper-vertical boundaries, we have:
 \be\begin{aligned}
 \mathbb{K} (m,x;n,y) - \mathbb{K} (m,x;n+1,y) = \Id_{_{(n,y)=(m,x)}}.
\end{aligned}\label{Kinv(e)}\ee
 Again subtracting the three-step relation (\ref{3Step}), it thus suffices to prove $\mathbb{K} (m,x;n+1,y-1)=0$ for the $(n,y)\in {\bf P}$ as above; then, we have that $(n+1,y-1)\in {\cal S}$ and thus from Corollary \ref{CorK2K3} we have the first equality below, at these points,
\be \label{K123} \begin{aligned}(\mathbb{K}_1+\mathbb{K}_2 )(m,x;n+1,y-1)&= \mathbbm{1}_{x\leq y-1} {(n-m+1)_{y-x-1} \over (y-x-1)!} 
\\
&\stackrel{\ast}{=} \frac{(y-x)_{n-m} }{ (n-m)!}  \Id_{n+1> m}\mathbbm{1}_{x\leq y-1}\\
&=-\mathbb{K}_0 (m,x;n+1,y-1).
\end{aligned}\ee
Equality $\stackrel{\ast}{=}$ is automatic for $n+1>m$. 
  For $m\geq n+1 $, we consider two cases:
  
 \noindent (i) $(n,y)\in$ left-vertical boundary$+(0,\tfrac 12)$. Then it must be that 
 $x\geq y$, if $(m,x)$ is to remain within the polygon; but then both sides of equality $\stackrel{\ast}{=}$ in (\ref{K123}) equal  $0$.
 
  \noindent (ii) $(n,y)\in$ vertical boundary of upper-cut$+(0,\tfrac 12)$. Then  if $(m,x)$ is to remain within the polygon, then either $y-x\leq 0$, or $y-x>m-n\geq 1$. In the first case, both sides of $\stackrel{\ast}{=}$ in (\ref{K123}) vanish, for the same reason as (i), and in the second case, the left hand side vanishes, because: 
  $ (n-m+1)_{y-x-1} =0$,
  since $n-m+1\leq 0$ and $n-m+y-x-1\geq 0$ and so does the right hand side of $\stackrel{\ast}{=}$, because $m-n\geq 1$. Then one notices that the expression in $\stackrel{\ast}{=}$ equals $\mathbb{K}_0 (m,x;n+1,y-1)$. To conclude, we have  $\mathbb{K} (m,x;n+1,y-1)=0$   along the upper-vertical boundaries.
  
  

%

\bigbreak

\par\noindent
\textbf{IV. Upper-oblique boundaries of ${\bf P}$ of type (b)}, as in Fig. \!9 :  Inserting ${\mathbb K}_{\mbox{\tiny Kast}} (     \circ (n+1,y);   \bullet (n,y))=0$  in equation (\ref{Kinv}), it suffices to prove that for the points $(n,y)\in {\bf P}$ adjacent to the upper-oblique boundaries, one has:
\be\begin{aligned}
 \mathbb{K} (m,x;n,y) + \mathbb{K} (m,x;n+1,y) = \Id_{_{(n,y)=(m,x)}}.
\end{aligned}\label{Kinv(b)}\ee
 Again subtracting the three-step relation (\ref{3Step}), one needs  to prove $\mathbb{K} (m,x;n+1,y )=0$. For $(n,y)$ in the locus above, we have that $(n+1,y)\in {\cal S}$, and thus applying again Corollary \ref{CorK2K3},
\be \label{K123'} \begin{aligned}(\mathbb{K}_1+\mathbb{K}_2 )( n+1,y )&= \mathbbm{1}_{x\leq y } {(n-m+1)_{y-x } \over (y-x )!} 
\\
&\stackrel{\ast}{=} \frac{(y-x+1)_{n-m} }{ (n-m)!}  \Id_{n+1> m}\mathbbm{1}_{x\leq y } 
 =-\mathbb{K}_0 (n+1,y ).
\end{aligned}\ee
 It remains to prove equality $\stackrel{\ast}{=}$, which is automatic for $n\geq m$, whereas for $n+1\leq m$, the right hand side vanishes. Then for $(m,x)$ to belong to the hexagon, we must have, if $n+k=m $ for $k\geq 1$, that $x\leq y-k$. Then $(n+1-m)_{y-x}=(1-k)_{y-x}=0$, since $y-x\geq k$, implying the left hand side of $\stackrel{\ast}{=}$ vanishes. This ends the proof that  $\mathbb{K} (m,x;n+1,y)=0$ and the proof of Theorem \ref{ThKastInv}. 
 \qed

Before proving Theorem \ref{Theo:L-kernel}, we need Kenyon's proposition:
 
 \begin{proposition}\label{Kenyon} (Kenyon, '97)
Suppose that $E=\{e_i\}_{i=1}^n$ are a collection of distinct dimer edges of a bipartite graph $G$, with $e_i=(w_i,b_i )$, where $b_i$ and $w_i$ denote black and white vertices and let $K$ be the associated Kasteleyn matrix. 
    The dimers form a determinantal point process on the edges of $G$ with correlation kernel given by
    \begin{equation}
	  L(e_i,e_j) = K(w_i,b_i) K^{-1} (b_i,w_j) 
  ,  \end{equation}
 where $K(w,b) = K_{ wb}$ and $K^{-1}(b,w )= (K^{-1})_{bw}$
 \end{proposition}
 

 \noindent {\em Proof of Theorem \ref{Theo:L-kernel}}: At first observe that for $(m,x)\in (\BZ+\tfrac12)\times \BZ$,
 $$\bl
 &\left\{ {\mathbb L}\mbox{-dot at }(m,x)\right\}
 \\
 &\Leftrightarrow
 \left\{\mbox{dimer connects $ \bullet$-dot $( m-\tfrac 12,x)$ and $\circ$-circle $( m+\tfrac 12,x)$}  \right\},
 \el $$
   as illustrated in Fig. \!8, where the ${\mathbb L}\mbox{-dot}$ is represented by a blue dot. 
 Applying Kenyon's  Theorem, 
for $(w,b)=e_1=((m+\tfrac12,x),(m-\tfrac12,x))$ and $(w',b')= e_2=((m'+\tfrac12,x'),(m'-\tfrac12,x'))$ with $ (m,x),~(m',x') \in \BZ^2$, we find for the kernel $\widetilde{\mathbb L}$ of blue dots, expressed in the $(m,x)$-coordinates, using Proposition \ref{Kenyon}, the following:
%
   %
   \be\label{Kenyon'}\begin{aligned}\widetilde{\mathbb L}(m,x ; m',x') =&{\mathbb K}_{\mbox{\tiny Kast}}\Bigl(    \circ (m+\tfrac 12 ,x ),  \bullet (m -\tfrac 12 ,x  )   \Bigr)
   \\
  &~~~~~ \times {\mathbb K}_{\mbox{\tiny Kast}}^{-1}\Bigl(    \bullet (m-\tfrac 12 ,x);   \circ (m'+\tfrac 12 ,x')  \Bigr)\\
 =&{\mathbb K}_{\mbox{\tiny Kast}}^{-1}
\Bigl(        \bullet (m -\tfrac 12 ,x  ); \circ (m' +\tfrac 12 ,x') \Bigr),~~~\mbox{using (\ref{Kast}) }\\
 =&(-1)^{m-m'+x -x' -1}
   {\mathbb K}_{ } (       m -\tfrac 12,x; m'+\tfrac 12,x' ).
   \end{aligned} \ee
  It will be more convenient to reexpress the kernel of blue dots $\widetilde{\mathbb L}(m,x ; m',x')$  in the $(\eta,\xi )$-coordinates\footnote{\label{foot7}using the change of coordinates (\ref{Lcoord}) on both $(m,x)$ and $(m',x')$; namely $m=\tfrac{\eta+\xi}{2},~~ x=\tfrac{\eta-\xi-1}{2}$ and $m'=\tfrac{\eta'+\xi'}{2},~~ x'=\tfrac{\eta'-\xi'-1}{2}$.
      }, with $(\eta,\xi )=(k,2\ell-k-1 )$ and $k,\ell\in \BZ$ and performing a conjugation. The $ {\mathbb L}$-process of blue dots is thus expressed by the following kernel: 
   \be \label{L-kernel}\begin{aligned}
    {\mathbb L} (\eta,\xi  ; \eta',\xi')
    &:=(-1)^{\eta -\eta' }
   \widetilde{\mathbb L}((m,x ; m',x') )
   =-{\mathbb K}_{ } (       m -\tfrac 12,x; m'+\tfrac 12,x' ),
   \end{aligned} \ee
    leading to the result in Theorem \ref{Theo:L-kernel}. \qed

\section{Scaling and moving contours in the ${\mathbb L}$-kernel
 }


In this section, entirely devoted to {\bf the two-cut model}, we will prepare the ${\mathbb L}$-kernel (\ref{L-kernel}) of blue dots in order to do the  asymptotics in the neighborhood of the strip $\{\rho\}$. While the lines $\{\eta=\mbox{integer}\}$ carrying the blue dots will be kept as such, the (discrete) running variable $\xi$ along those lines will be rescaled into a continuous variable. In view of the integrands in the ${\mathbb L}$-kernel and the related ${\mathbb K}$-kernel as in (\ref{Kernlim5}), it will be necessary to move the integration contours into an appropriate position vis-\`a-vis the saddle points, to be given later in Proposition \ref{PropKernlim6}.

From (\ref{L-kernel}) and using the usual map $(m,x)\mapsto (\eta,\xi)$ as in footnote \ref{foot7}, the ${\mathbb L}$-kernel of blue dots along oblique lines $\{\eta=\mbox{integer}\}$ reads in some new notation:
 \be \begin{aligned}
   {\mathbb L}&( \eta_1 ,\xi_1;\eta_2,\xi_2)\\
&=-{\mathbb K}\left(\tfrac12 (\eta_1+\xi_1-1),\tfrac12 (\eta_1-\xi_1-1);
\tfrac12 (\eta_2+\xi_2+1),\tfrac12 (\eta_2-\xi_2-1)\right),
\\&=-{\mathbb K}(m',x';n',y'), \end{aligned}\label{Lkernel}\ee
giving rise to a map
$\MR: (m',x';n',y')\to (\eta_1 ,\xi_1;\eta_2,\xi_2)$, defined by 
  \be\begin{aligned}
 (m',x';n',y')&= \left(\tfrac12 (\eta_1+\xi_1-1),\tfrac12 (\eta_1-\xi_1-1);
\tfrac12 (\eta_2+\xi_2+1),\tfrac12 (\eta_2-\xi_2-1)\right).
\end{aligned}
\label{xy-to-xi eta}\ee
In (\ref{Lkernel}), we use the kernel ${\mathbb K}   (m,x;n,y)$ of red dots along horizontal lines, as given by (\ref{Kernlim5}):
\be \begin{aligned}
   {\mathbb K}  &(m',x';n',y') 
   = \left[{\mathbb K}_0+ \frac{(N-n' )!}{(N-m'-1)!} \left({\mathbb K}_1+\tfrac{1}{r+1}{\mathbb K}_2\right)\right](m',x';n',y').
 \end{aligned}\label{Kkernel'}\ee
It is worth paying attention to the parameter $\tau$ appearing in the contour $\Ga_{-\tau}$ in formula (\ref{Kernlim5}) for ${\mathbb K}   (m',x';n',y') $, which here is given by 
\be\tau=y'+n'-(y_1+1)=\eta_2-m_1.
\label{tau}\ee
In the rest of this section we {\bf omit the primes} in (\ref{Lkernel}) and (\ref{xy-to-xi eta}); that is to say that henceforth $(m ,x ;n ,y )$ will be defined by the map (\ref{xy-to-xi eta}), without the primes.

%
%
%

\medbreak

\noindent {\bf Proposing a scaling}.  Assuming $N$ even, the asymptotics will be performed about the black dot in the middle of Fig. \!3, namely the point
$$\bl(m_0,x_0)&=(\mbox{halfway point along the left boundary of the $ \{\rho\}$-strip })-( \tfrac 12, -\tfrac 12)\\
&=(\tfrac {N-1}2 ,m_1  -\tfrac N2)
\\&\Updownarrow
\\ (\eta_0,\xi_0) &=(m_1,N-m_1 -1 ).
\el
$$  
%
%
We now {\em propose} the following scaling of the geometric variables of the figure, with $r,\rho\geq 0$ fixed and $b,c,d>0 $ getting large, and still satisfying $n_1+n_2=m_1+m_2$, 
\be \label{scalinggeom}
\begin{array}{lllll}
b=d+r&& c=\ga d
\\
n_1 = m_1+(\rho-r) &&  m_1=\al_1 d+\beta_1\sqrt{d}+\ga_1
\\
n_2 = m_2-(\rho-r) &&  m_2=\al_2 d+\beta_2\sqrt{d}+\ga_2,
\end{array} \ee
together with the following scaling of the coordinates $(\eta,\xi)\in \BZ\times\BZ$ and the integration variables $v,z,u$ in the kernel ${\mathbb K}$, as in (\ref{Kernlim5}), leading respectively to new variables $(\tau, \widetilde\sg)\in \BZ\times\BR$ and new integration variables $\om,\ze,U$ :
\be\begin{aligned}
 (\eta_i,\xi_i) &=(\eta_0,\xi_0)+(\tau_i,\widetilde\sigma_i\sqrt{d} )\\
  &=
 (m_1+\tau_i, N-m_1-1+\widetilde\sg_i\sqrt{d} )
\\&=(\al_1d+\beta_1\sqrt{d}+\ga_1+\tau_i,    (\ga-\al_1+1)d+(\widetilde\sg_i-\beta_1)\sqrt{d}+r-\ga_1 -1)
\\  v&= \om d,~~z= \zeta d ,~~u=Ud;
\end{aligned}\label{scaling}\ee
throughout we set $d=t^2$.

For future use, consider a map ${\mathcal T} $ on the variables $(m,x;n,y)$ in (\ref{xy-to-xi eta}) as below; it translates into a map on the $\eta_i,\xi_i $-variables and further into a map on the new $\tau_i,\widetilde\sg_i$ variables:
\be{\mathcal T}: ~~\left\{\begin{aligned}  m&\to n-2\\
x&\to y+1
\end{aligned}\right\}\Leftrightarrow 
\left\{\begin{aligned}
\xi_1&\to \xi_2-2\\
\eta_1&\to \eta_2
\end{aligned}\right\}\Leftrightarrow 
\left\{\begin{aligned}
\widetilde\sg_1&\to \widetilde\sg_2-\frac{2}{t}\\
\tau_1&\to \tau_2
\end{aligned}\right\}.
\label{xymap}\ee

\noindent{\bf Constraints on the geometry}. 
Given the scaling (\ref{scalinggeom}) of the geometric variables, the inequalities (\ref{ineq}) lead to the positivity of the following expressions for large $t$,
\be\begin{aligned}
m_2-(c-d)&=(\al_2-\ga+1)t^2+\beta_2t+\ga_2>0
\\
n_1-(c-d)&=(\al_1-\ga+1)t^2+\beta_1t+\ga_1+\rho-r>0
\\
d-b+m_1&= \al_1t^2+\beta_1t+\ga_1-r >0
\\
d-b+n_2&= \al_2t^2+\beta_2t+\ga_2-\rho >0
\\
\Sg&=c-d-n_1+m_1 = (\ga-1)t^2+r-\rho\geq 0,\end{aligned}\label{ineq'}\ee
and thus we must assume for $i=1,2$, 
\be\begin{aligned}
&
 ~~\al_i>0,~  \ga> 1 \mbox{    and   }  \al_i-\ga+1> 0 .
\end{aligned}\label{ineq''}\ee
Using the map (\ref{xy-to-xi eta}) and the formula (\ref{tau}) for $\tau$, combined with the scaling (\ref{scalinggeom}) and (\ref{scaling}), and using the identities in Fig. \!3, leads to the following behavior for the variables involved:
\be  \footnotesize
   \bl
x_1&=(\al_1+\al_2)t^2+(\beta_1+\beta_2)t+\ga_1+\ga_2-1
\\
x_c&=(\al_1+\al_2-\ga)t^2+(\beta_1+\beta_2)t+\ga_1+\ga_2 
\\
m_0&=\tfrac{\ga+1}{2}t^2+\tfrac{r-1}{2},~~~x_0 =(\al_1-\tfrac{\ga+1}{2})t^2+ \beta_1 t+\ga_1-\tfrac r2
\\
N-m-1&=\tfrac{\ga+1}{2}t^2-\tfrac{\widetilde \sg_1}{2}t+\tfrac 12(r-\tau_1), ~~~N-n =\tfrac{\ga+1}{2}t^2-\tfrac{\widetilde \sg_2}{2}t+\tfrac 12(r-\tau_2)
\\
x_{c+1}&=(\al_1 -\ga)t^2+ \beta_1 t+\ga_1-r+\rho-1
\\
 x_{c+d} &=(\al_1 -\ga-1)t^2+ \beta_1 t+\ga_1-r+\rho
\\
y_d&=(\al_1-1)t^2+\beta_1 t+\gamma_1t
\\
 x-1 &=x_{c+1}+\tfrac 12\{(\ga-1)t^2-\widetilde\sg_1t+  r-2\rho+\tau_1\}
\\
y_1-N&=x_{c+d}-\rho-1
\\
x+m-N&=x_{c+d}+\tau_1 -\rho-1,~~
y+n-N  =x_{c+d}+\tau_2  -\rho ~~(*)
\\
x+m-m_1&=\tau_1-1,~~~~~~~~~~~~~~
\tau=y+n-m_1  = \tau_2   ~~(**)
\\
x_{c+d+1}&=  -(\ga +1)t^2-1
,~~x_{c+d+b} =  -(\ga +2)t^2-r
\\
\left.{{x}\atop{y}}\right\}&=
(\alpha_1 -\tfrac{\gamma  +1}2  ){t}^{2}+(\beta_1  
-\tfrac 
{\widetilde\sigma_{1\atop2}}2) t+\gamma_1+\tfrac{\tau_{ {1\atop2} }-r}2 
    \\
 x-x_0  &=-\tfrac12 ( \widetilde\sg_1 t-\tau_1)
 ,~ ~~~~y- x_0  =-\tfrac12 ( \widetilde\sg_2 t-\tau_2)
 \\
 m- m_0  &= \tfrac12 ( \widetilde\sg_1 t+\tau_1-1),~ ~~n-m_0  = \tfrac12 ( \widetilde\sg_2 t+\tau_2+1)\el.
\label{CoordAs}\ee
From (\ref{ineq''}), it follows that for large enough $t$, the $(m,x )$- and $(n,y)$-variables  are asymptotically related to the geometrical points of the polygon as follows  (on the first line the respective order of the $m,n$ and $x,y$ does not matter):
\be \begin{aligned}x_{c+d+1}+N-\left\{{m\atop n}\right\}<& x_{c+1}<\left\{{x\atop y}\right\}<y_d<x_c.
\el\label{ineq2}\ee

The kernel ${\mathbb K}$ as in (\ref{Kernlim5}) involves three contours (\ref{cont0}). In particular, for the contour $\Ga_{-\tau}$ and from (\ref{CoordAs} $(\ast)$) and (\ref{CoordAs}$(\ast\ast)$), we have, asymptotically, that $y>y_1-N$, that $\tau=\tau_2$   and that $\Ga_{-\tau}=\Ga_{-\tau_2}$ can be written as
$$
\Ga_{-\tau_2}=\Ga
\{x_{c+d}-(\rho-\tau_2),\dots,x_{c+d}-(\rho+1)\}\Id_{\tau_2<0}.
$$
Also, from (\ref{CoordAs}($\ast$)), it follows that the left-most roots of  the polynomials $(v-x+1)_{N-m-1}$ and $(z-y)_{N-n+1}$  are given by $v=x+m-N +1=x_{c+d}+\tau_1-\rho$ and $z=y+n-N=x_{c+d}+\tau_2-\rho $, where $x_{c+d}$ is the left-most integer in the upper-cut; see Fig. \!3.

\medbreak

\noindent{\bf Poles of the integrands of the kernel ${\mathbb K}$: } 
%
%
Using the identities (\ref{CoordAs}) and inequalities (\ref{ineq2}), the functions (\ref{Rih}), defined in terms of the polynomials $P,Q$ 
as in (\ref{PQ}), have the following poles, where $\Ga$, appearing under the bracket, denotes a contour containing exactly those poles, written in ascending order,

%

\be\begin{aligned}
& \mbox{Poles of}~~h(u) 
 \stackrel{(i)}{=}\{\underbrace{x_{c+d}\!-\!\rho,\dots, x_{c+d}\!-\!1}_{ \Ga_{\bar\rho} }\}\Id_{\rho>0}
\cup \{\underbrace{x_{c+1}\!+\!1,\dots,y_d\!-\!1}_{ \Ga_{\bar\Sg} }\}\cup \underbrace{\LR}_{\Ga_\LR} \\
 &\mbox{Poles of}~~ R_1(z)  \stackrel{(ii)}{=}\left(\mbox{Poles of}~~ \frac{(z-x+1)_{N-m-1}}{Q_{\CR}(z)}\right)\cup \RR
\\&=\underbrace{\{ x_{c+d},\dots, x_{c+d}+\tau_1-\rho-1 \} \Id_{\tau_1>\rho}}_{\Gamma_{\tau_1-\rho}} \cup \RR
\\ 
&\mbox{Poles of}~~ R_2^{-1}(z)  \stackrel{(iii)}{=}\mbox{Poles of}~~\frac{Q_{\CR}(z)}{(z-y)_{N-n+1}}
\\&
 =\underbrace{\{ x_{c+d}-(\rho-\tau_2),\dots,x_{c+d}-1 \}\Id_{\tau_2< \rho}}_{\Ga_{\rho-\tau_2}}\cup
\underbrace{\{x_{c+ 1}+1,\dots,y\}}_{\Ga_{y-x_{c+1}}}
\\ 
&\mbox{Poles of}~~R_2^{-1}(z)h^{-1}(z) \stackrel{(iv)}{=} \mbox{Poles of}~~ \frac{P(z)}{(z-y)_{N-n+1}}
\\&=\underbrace{\{x_{c+d}-(\rho-\tau_2),\dots,x_{c+d}-(\rho+1)\}\Id_{\tau_2<0}}_  { \Ga_{_{\!-\tau_2}} }
,
 \\  
&\mbox{Poles of}~~\frac{R_1(z)}{R_2(z)} \stackrel{(v)}{=}\mbox{Poles of}~~\frac{ (z-x+1)_{N-m-1}}{(z-y)_{N-n+1}}
\\&=\{y-N+n,\dots, x-N+m\}\Id_{y +n\leq x +m }\cup
\{x,x+1,\dots,y\}\Id_{x\leq y}
\\
&= \underbrace{\{ x_{c+d}-(\rho-\tau_2), \dots,x_{c+d}+(\tau_1-\rho)-1 
\}\Id_{\tau_1> \tau_2}}_{\Ga_{\tau_1-\tau_2}  }
\cup
\underbrace{\{x,x+1,\dots,y\}\Id_{x\leq y}}_{\Ga_{y-x}}
%
\end{aligned}\label{poles}\ee
 
\noindent {\bf Two useful Lemmas} will be needed: 

\begin{lemma} \label{LemOmtOm}(\cite{AJvM3}, section 8)
Given a rational function $R(u)$ with possibly poles within a contour $\Gamma$ and a point $z$ not within $\Gamma$, not a pole of $R(u)$. Then for $0\leq k-1\leq \ell$, we have (the notation $\Gamma\cup z$ refers to the contour $\Gamma$, deformed so as to contain $z\in \BC$)
\be\begin{aligned}
\Bigl(\prod_{\al=1}^\ell \oint_{\Ga \cup z}&\frac{du_\al R(u_\al)}{2\pi \I( u_\al-z)}\Bigr)\Dt_\ell ^2(u_1,\dots,u_\ell)\\
=& (k-1)R(z)\left(\prod_{\al=1}^{k-2}\oint_{\Ga} \frac{du_\al}{2\pi \I} R(u_\al) 
(u_\al-z)\right) \Dt_{\ell-1}^2(u_1,\dots,u_{\ell-1})
\\
&+\prod_{\al=1}^{k-1}   \oint_{\Ga} \frac{du_\al R(u_\al)}{2\pi \I( u_\al-z)}
\prod_{\al=k }^{\ell } \oint_{\Ga\cup z}\frac{du_\al R(u_\al)}{2\pi \I( u_\al-z)}\Dt_\ell^2(u_1,\dots,u_\ell).
\end{aligned}\label{OmtOm}\ee
In particular for $k=\ell+1$,
\be\begin{aligned}
\Bigl(\prod_{\al=1}^{\ell} &\oint_{\Ga \cup z} \frac{du_\al R(u_\al)}{2\pi \I( u_\al-z)}\Bigr) \Dt_\ell ^2(u_1,\dots,u_\ell) 
-\Bigl(\prod_{\al=1}^{\ell }   \oint_{\Ga} \frac{du_\al R(u_\al)}{2\pi \I( u_\al-z)}\Bigr)\Dt_\ell^2(u_1,\dots,u_\ell)  
\\
=& \ell R(z)\Bigl(\prod_{\al=1}^{\ell-1}\oint_{\Ga} \frac{du_\al}{2\pi \I}R(u_\al) 
(u_\al-z)\Bigr)\Dt_{\ell-1}^2(u_1,\dots,u_{\ell-1}).
\end{aligned}\label{OmtOm'}\ee


\end{lemma}

\begin{lemma} \label{LemOmrv} The $r$-fold $u$-integral $ \Om _r(v,z)$ about $\Gamma_\LR$, not containing $v$, can be expressed as the same $r$-fold integral $ \Om^{(v)} _r(v,z)$, with $\Gamma_\LR$ containing $v$, by adding the  $r-1$-fold integral $\Om^+_{r-1}$ and $\Om_r(z,v)$, to wit:
\be \begin{aligned}
 \Om _r(v,z) &= \Om^{  (v) }_r( v,z)+r(z-v) h(v) \Om^+_{r-1}(v,z)
  \\
 \Om^- _{r+1}(v,z) &= \Om^{-(v)} _{r+1}(v,z)+(r+1)\frac{h(v)}{(z-v) } \Om _{r }(z,v).
 \end{aligned}\label{reduce2}\ee

\end{lemma}
\proof Apply (\ref{OmtOm'}) with $\ell=r,~r+1$, with $z\to v$ and $R(u)=h(u)(u-z),~\frac{h(u)}{u-z}$.\qed


\noindent{\bf Preparing the kernel, by moving contours: }  It will turn out that the saddle points for the various integrands will all coincide and will belong to the upper-cut, namely to the interval $[x_{c+d}, x_{c+1}]$; this will enable us to perform steepest descent analysis on the kernel ${\mathbb L}$.   Before doing this analysis, we will need to  move contours, so as to tailor the contours to the integrands. Of course, moving those contours will be at the expense of adding extra-terms; that will depend on the pole structure (\ref{poles}) of the functions (\ref{Rih}). We now rewrite the $\mathbb L$-kernel as follows, remembering the notation in (\ref{Kernlim5}):


\begin{proposition} \label{PropKernlim6} Using the map ${\mathcal M}$ as in (\ref{xy-to-xi eta}), the kernel ${\mathbb L}$ (as in (\ref{Lkernel}) and (\ref{Kkernel'})) in the range $x_{c+1}<x,y<y_d<x_c$, can now be written as\footnote{For $a,b\in \BZ$ and $b\geq 0$, the symbol $\left(a\atop b\right)=\frac{a(a-1)\dots(a-b+1)}{b!}$, which coincides with the usual definition for $a\geq b$.} 
 %
\be
 \begin{aligned}
 {\mathbb L} &(\eta_1,\xi_1;\eta_2,\xi_2)
  =-  {\mathbb K}    (m,x;n,y)   =
-  \BK_0-\tfrac{(N-n)!}{(N-m-1)!}(\BK_1+ \tfrac {1}{r+1}\BK_2)  
 \\&= \Id_{x>y} \Id_{x+m-y -n\geq 0} (-1)^{n-m}
 \left({x-y-1}\atop{x+m-y-n}\right) +\frac{(N - n )!}{(N-m-1)!}
\\   \times&\left[  
   \oint_{ {\tau_1-\rho 
  }}  \frac{dv  R_1(v)}{2\pi \I}
  \oint_{ \Ga_{\rho-\tau_2}+ \Ga^{(v)}_{\tau_1-\rho}+\Ga_{y-x_{c+1}} }
    \frac{dz}{2\pi \I (z-v) R_2(z)} 
   \frac{  \Om_r(v,z)}{  \Om_r (0,0)}\right.
 \\
   & 
   +\oint_{ {\Ga_{\LR}   }}  \frac{dv R_1(v)}{2\pi \I}
 \oint_{ \Ga_{\rho-\tau_2}+ \Ga^{}_{\tau_1-\rho}+\Ga_{y-x_{c+1}}  } \frac{dz }{2\pi \I(z\!-\!v)R_2(z)} \frac{\Om^{(v)}_r(v,z) }{\Om_r(0,0)}
   \\
    &  + \oint_{ {\Ga  _{\LR} 
  }}  \frac{dv h(v)  R_1(v)}{2\pi \I}
  \oint_{  
\Ga_{_{\!-\tau_2}} 
 }
    \frac{dz}{2\pi \I (z-v)h(z) R_2(z)} 
   \frac{  \Om_r(z,v)}{  \Om_r (0,0)}
   \\
    & +r\oint_{ \Ga_\LR
  }   \frac{dv h(v)R_1(v)}{2\pi \I}
  \oint_{\Ga_{\rho-\tau_2}+ \Ga^{ }_{\tau_1-\rho}+\Ga_{y-x_{c+1}} }
   \frac{dz}{2\pi \I R_2(z)} 
  \frac{  \Om^+_{r-1} (v,z)}{  \Om_r (0,0)}  
 \\
 &   + \frac{1}{r+1}  \oint_{ {\Ga_{\tau_1-\rho}  
 }}  \frac{dv R_1(v)}{2\pi \I}
 \oint_{ \Ga_{_{\!-\tau_2}} }
 \frac{dz }{2\pi \I h(z)R_2(z)} \frac{ \Om^{-  }_{r+1}(v,z) 
  }{\Om_r(0,0)}
  \\
 & 
 +\left.\left.  \frac{1}{r+1}  \oint_{ {\Ga_{\LR}  
 }}  \frac{dv R_1(v)}{2\pi \I}\oint_{  
\Ga_{_{\!-\tau_2}} 
 } \frac{dz }{2\pi \I h(z)R_2(z)} \frac{ \Om^{-(v) }_{r+1}(v,z) 
  }{\Om_r(0,0)} 
\right]  \right |_{\MR}
\\
&=:\left[{\mathbb L}_0 +\tfrac{(N - n  )!}{(N-m-1)!}
\left({\mathbb L}_1+{\mathbb L}'_1+{\mathbb L}_2+r{\mathbb L}_3+\tfrac{1}{r+1}({\mathbb L}_4+{\mathbb L}'_4)\right)\right](\eta_1,\xi_1;\eta_2,\xi_2),
\end{aligned}\label{Kernlim6}\ee
where the $\Om_k$ and $\Om^\vr_k$ were given in (\ref{Omr}). The contours are all spelled out in (\ref{poles}); see also Fig. \!10. The contour $\Ga_{\tau_1-\rho}^{(v)}$ refers to $\Ga_{\tau_1-\rho}^{ }$ containing the point $v$ as well; i.e., $\Ga_{\tau_1-\rho}$ sits inside $\Ga_{\tau_1-\rho}^{(v)}$. The functions $\Om_k(v,z), ~\Om_k^{\pm}(v,z)$ involve integrations along $u_\al \in \Ga_\LR$. Similarly, the functions $\Om^{(v)}_k(v,z), ~{\Om_k}^{-  (v) }$ refer to integrating along $u_\al \in \LR^{(v)}$-contours.

\end{proposition}

\proof Referring to the kernel ${\mathbb K} $ in (\ref{Kernlim5}), each  ${\mathbb K}_i$  will be prepared separately:
\medbreak

 \noindent{\bf (i) Preparing ${\mathbb K}_1$ in (\ref{Kernlim5})}, 
$$\bl   
{\mathbb K}_1\stackrel{I}{=}&\oint_{ {\Gamma( {x +{\mathbb N}}) }}   \frac{dvR_1(v)}{2\pi \I}
\oint_{\Gamma_{\infty}} \frac{dz}{2\pi \I(z\!-\!v)R_2(z)}
 \frac{  \Om_r (v,z)}{  \Om_r (0,0)} ,
\el $$

\noindent with $v$-contour (blue in Fig. \!11.I), $z$-contour (red) and $u$-contour (green). Since the degree in $v$ of the integrand of the first expression above (using (\ref{Rih}) and (\ref{Omr})) equals 
 $$
 (N-m-1)-c-d-r-1=-m-2,
 $$ one can move the (blue) $v$-contour $\Gamma( {x +{\mathbb N}})$ on the Riemann sphere across $\infty$, leading to  equality $\stackrel{II}{=}$ below; indeed given the poles (\ref{poles}{\em (ii)}) of $R_1(v)$, the $v$-contour in Figure \!11.I gets replaced by three other $v$-contours $\Ga_{\tau_1-\rho} ,~ \Gamma^{(u)}_\LR$ (contour about $\LR$, containing the $u=(u_1,u_2,\dots)$-variables of $\Om_r$) and a little contour about $z\in \Ga_{\infty}$, as in Fig. \!11.II. The latter can be evaluated using the Residue Theorem, yielding: 
$$\bl 
{\mathbb K}_1\stackrel{II}{=}&  \oint_{\Ga_\infty}\!\!  \frac{dz R_1(z)}{2\pi \I R_2(z)}
-\oint_{ {\Ga^{(u)}_{\LR}+\Ga_{\tau_1-\rho}
 }}  \frac{dv R_1(v)}{2\pi \I}
 \oint_{ \Ga_{\infty}}
   \frac{dz}{2\pi \I(z\!-\!v)R_2(z)} 
  \frac{  \Om_r (v,z)}{  \Om_r (0,0)}.
  \el $$
   Next, due to the poles (\ref{poles}({\em v})) of $R_1(z)/R_2(z)$, we can shrink the $z$-contour $\Gamma_\infty$ of the first integral to $\Ga_{\infty}=\Ga_{y-x}+\Gamma_{\tau_1-\tau_2} $. Also, in view of the poles (\ref{poles}({\em iii})) of $R_2^{-1}(z)$, and since $\Om_r(v,z)$ has no $z$-poles, the $z$-contour $\Gamma_\infty$ of the double integral in $ \stackrel{II}{=}$ can be reduced to $\Ga_{y-x_{c+1}}+\Ga_{\rho-\tau_2}$ and a small circle about $v\in 
 \Gamma_\LR^{(u)}+\Gamma_{\tau_1-\rho}$; the latter leading to another residue term in $z$, which integrated over $v$ gives the second single integral in $ \stackrel{III}{=}$, with $v,~z,~u$-contours in the double integral as in Fig. \!11.III,
$$\bl  
  {\mathbb K}_1\stackrel{III}{=}&  \oint_{\Ga_{y-x}+\Gamma_{\tau_1-\tau_2} }\!\!  \frac{dz R_1(z)}{2\pi \I R_2(z)}
-\oint_{\Ga^{(u)}_\LR+\Ga_{\tau_1-\rho}}\!\!  \frac{dz R_1(z)}{2\pi \I R_2(z)}  ~
\\&-\oint_{ {\Ga^{(u)}_{\LR}+\Ga_{\tau_1-\rho}
 }}  \frac{dv R_1(v)}{2\pi \I}    
 \oint_{  
\Ga_{ \rho-\tau_2} +\Ga_{y-x_{c+1}} 
 } \frac{dz}{2\pi \I(z\!-\!v)R_2(z)} 
  \frac{  \Om_r (v,z)}{  \Om_r (0,0)} .
  \el $$



 \newpage

\vspace*{-.8cm}

   \setlength{\unitlength}{0.017in}\begin{picture}(100,170)
\put(159,-80){\makebox(0,0) {\rotatebox{0}{\includegraphics[width=220mm,height=300mm]
 {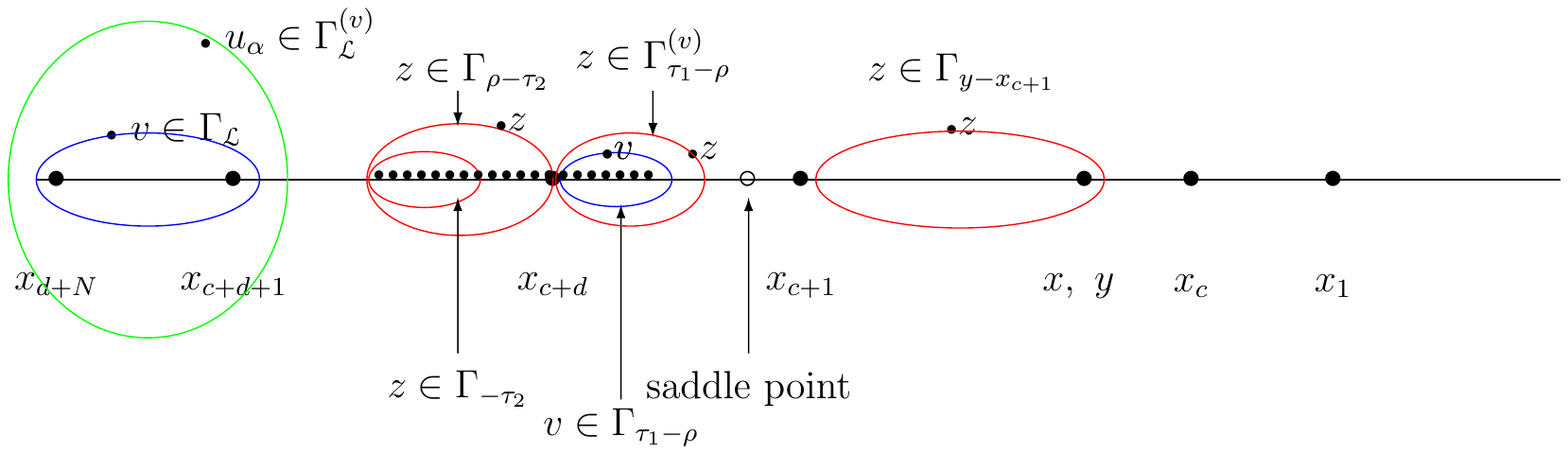} }}}

\end{picture}

 \vspace{-4cm}
 
\noindent Fig. 10. The contours for the ${\mathbb L}$-kernel, with blue $v$-, red $z$- and green $u$-contours. 

 

  
  \vspace*{ 2.5cm}

 \setlength{\unitlength}{0.017in}\begin{picture}(0,170)
\put(125,-30){\makebox(0,0) { \rotatebox{0}{\includegraphics[width=150mm,height=225mm]
 {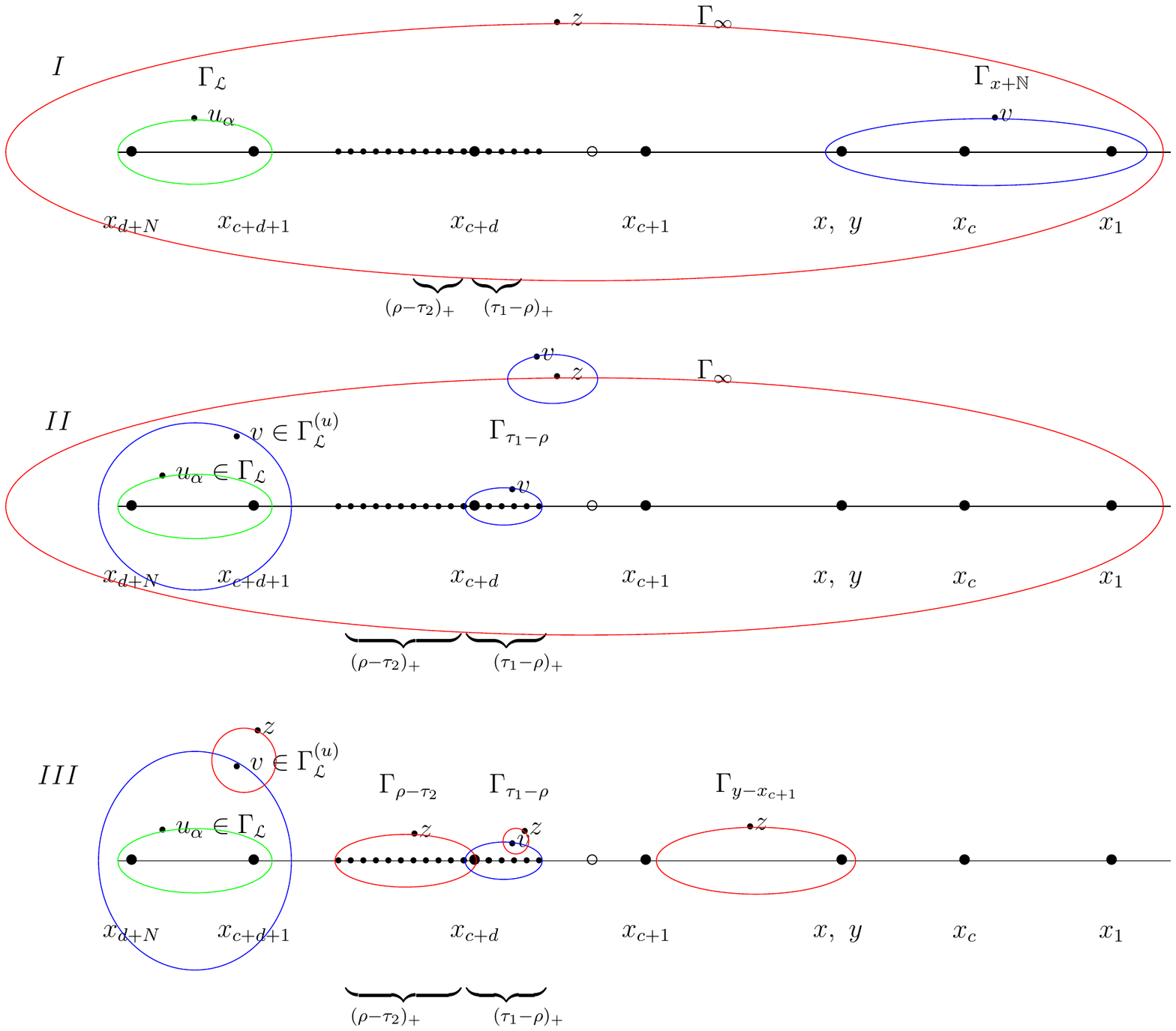} }}}
\end{picture}

\vspace*{4cm}
\noindent Fig. 11. The changes of contours for integral ${\mathbb K}_1$.

\newpage 

%
The two single $z$- and $v$-integrals can be combined in one, with $ \Ga^{(u)}_{\LR}$ removed, because of (\ref{poles}{\em (v)}). Also the $v$-integral in the double integral can be split as a sum of two, yielding:  
 %
%
%
%
  $$\bl
  {\mathbb K}_1 \stackrel{IV}{=}&  \oint_{\Ga_{y-x_{ }}+\Ga_{\tau_1-\tau_2}-\Ga_{\tau_1-\rho}}  \frac{dz R_1(z)}{2\pi \I R_2(z)}
\\&-\oint_{ {\Ga^{(u)}_{\LR} 
 }}  \frac{dv R_1(v)}{2\pi \I}    
 \oint_{  
\Ga_{ \rho-\tau_2} +\Ga_{y-x_{c+1}} 
 } \frac{dz}{2\pi \I(z\!-\!v)R_2(z)} \frac{  \Om_r (v,z)}{  \Om_r (0,0)}
\\&-\oint_{ { \Ga_{\tau_1-\rho}
 }}  \frac{dv R_1(v)}{2\pi \I}    
 \oint_{  
\Ga_{ \rho-\tau_2} +\Ga_{y-x_{c+1}} 
 } \frac{dz}{2\pi \I(z\!-\!v)R_2(z)} \frac{  \Om_r (v,z)}{  \Om_r (0,0)}.
  \end{aligned} $$
 %
 In $\stackrel{IV}{=}$ we interchange the $v$-contour with all $u_\al$-contours in the first double integral only, which leads to an extra-contribution as given by identity (\ref{reduce2}) in Lemma \ref{LemOmrv}, leading to: 
  %
%
  $$
  \begin{aligned}  {\mathbb K}_1 \stackrel{V }{=}&  \oint_{\Ga_{y-x_{ }}+\Ga_{\tau_1-\tau_2}-\Ga_{\tau_1-\rho}}  \frac{dz R_1(z)}{2\pi \I R_2(z)}
 \\&
-\oint_{ {\Ga_{\LR} 
 }}  \frac{dv R_1(v)}{2\pi \I}    
 \oint_{  
\Ga_{ \rho-\tau_2} +\Ga_{y-x_{c+1}} 
 } \frac{dz}{2\pi \I(z\!-\!v)R_2(z)}  \\&\qquad
 \qquad\qquad\qquad\qquad\qquad\qquad \times  
 \frac{\Om^{(v)}_r(v,z) 
 +r(z-v)h(v) \Om^+ _{r-1}(v,z)}{\Om_r(0,0)}
\\&-\oint_{ { \Ga_{\tau_1-\rho}
 }}  \frac{dv R_1(v)}{2\pi \I}    
 \oint_{  
\Ga_{ \rho-\tau_2} +\Ga_{y-x_{c+1}} 
 } \frac{dz}{2\pi \I(z\!-\!v)R_2(z)} \frac{  \Om^{ }_r (v,z)}{  \Om_r (0,0)} .
  \end{aligned}
 $$
Moreover, in view of the poles (\ref{poles}({\em iii})) of $R_2^{-1}(z)$, it is legitimate to add to the $z$-integration in the first double integrals in $\stackrel{V}{=}$ the contour $\Ga^{}_{\tau_1-\rho}$, since $v\in \Gamma_\LR$, i.e., far away from $\Ga^{}_{\tau_1-\rho}$. In the last double integral, one may also add to the $z$-integral a contour smaller than $\Ga_{\tau_1-\rho}$, (i.e., not containing $v$); that contour can be enlarged so as to contain the $v$-contour $\Gamma_{\tau_1-\rho}$, at the expense of an extra-residue term. The last double integral in the formula below comes from the second part in the double integral of (V) above, thus yielding:
\be\begin{aligned}
  {\mathbb K}_1 \stackrel{\tiny\mbox{ }}{\stackrel{VI}{=}} &  \oint_{\Ga_{y-x_{ }}+\Ga_{\tau_1-\tau_2}-\Ga_{\tau_1-\rho}}   \frac{dz R_1(z)}{2\pi \I R_2(z)}
  +\oint_{ {  \Ga_{\tau_1-\rho}}}\frac{dvR_1(v)}{2\pi \I R_2(v)}    
\\&-\oint_{ {\Ga_{\LR}  
 }}  \frac{dv R_1(v)}{2\pi \I}    
 \oint_{  
\Ga_{ \rho-\tau_2} +\Ga_{y-x_{c+1}}+\Ga^{ }_{\tau_1-\rho} 
 } \frac{dz}{2\pi \I(z\!-\!v)R_2(z)} 
 \frac{\Om^{(v)}_r(v,z) 
  }{\Om_r(0,0)}
  \\&-\oint_{ {  \Ga_{\tau_1-\rho}
 }}  \frac{dv R_1(v)}{2\pi \I}    
 \oint_{  
\Ga_{ \rho-\tau_2} +\Ga_{y-x_{c+1}}+\Ga^{(v)}_{\tau_1-\rho}
 } \frac{dz}{2\pi \I(z\!-\!v)R_2(z)} 
 \frac{\Om^{}_r(v,z) 
  }{\Om_r(0,0)} 
\\&-r\oint_{ { \Ga_{\LR}
 }}  \frac{dv R_1(v)h(v)}{2\pi \I}    
 \oint_{  
\Ga_{ \rho-\tau_2} +\Ga_{y-x_{c+1}}+ \Ga_{\tau_1-\rho}
 } \frac{dz}{2\pi \I R_2(z)} \frac{   \Om^+_{r-1} (v,z)}{  \Om_r (0,0)} ,
  \end{aligned}
 \label{PrepKernel1} \ee
 with an obvious cancellation of the two single $\Gamma_{\tau_1-\rho}$- integrals. The sum of the three double integrals equals $-({\mathbb L}'_1+{\mathbb L}_1+r{\mathbb L}_3)
  )$ in that order, as in (\ref{Kernlim6}); the contours are as in Fig. \!10. The single integrals will be dealt with in {\bf (iii)} below.
 
 \medbreak
 
 \noindent{\bf (ii) Preparing ${\mathbb K}_2$ in (\ref{Kernlim5})}: At first, since the $v$-degree of the integrand of  ${\mathbb K}_2 $ is also $-m-2$, one can move the $\Gamma( {x +{\mathbb N}})$-contour of the $v$-integral across $\infty$, giving $\stackrel{I}{=}$. This case is simpler, since the $z-v$ term is now missing in ${\mathbb K}_2$. Then one takes the $u_\al$'s outside the contour, yielding an extra-contribution in $\stackrel{II}{=}$, again using formula (\ref{reduce2}):
%
\be\begin{aligned}   
\tfrac{1}{r\!+\!1}{\mathbb K}_2 =&\tfrac{1}{r\!+\!1}\oint_{ {\Gamma( {x +{\mathbb N}}) }}   \frac{dvR_1(v)}{2\pi \I}
 \oint_{ \Ga_{_{\!-\tau_2}} } \frac{dz}{2\pi \I h(z)R_2(z)}
 \frac{  \Om^-_{r+1} (v,z)}{  \Om_r (0,0)} 
 \\
 \stackrel{I}{=}&  
  -\tfrac{1}{r\!+\!1}\oint_{ {\Ga^{(u)}_{\LR}
 }}  \frac{dv R_1(v)}{2\pi \I}
  \oint_{ \Ga_{_{\!-\tau_2}} }
   \frac{dz}{2\pi \I h(z)R_2(z)} 
  \frac{   \Om^-_{r+1} (v,z)}{  \Om_r (0,0)}
\\&
-\tfrac{1}{r\!+\!1}\oint_{ {\Ga_{\tau_1-\rho}
 }}  \frac{dv R_1(v)}{2\pi \I}
 \oint_{ \Ga_{_{\!-\tau_2}} }
   \frac{dz}{2\pi \I h(z)R_2(z)} 
  \frac{  \Om^-_{r+1} (v,z)}{  \Om_r (0,0)}  
       \\
 \stackrel{II}{=}&  
   -\tfrac{1}{r\!+\!1}\oint_{ \Ga_\LR
 }   \frac{dv R_1(v)}{2\pi \I}
  \oint_{ \Ga_{_{\!-\tau_2}} }
   \frac{dz}{2\pi \I h(z)R_2(z)} 
  \frac{  \Om^{-{(v)}}_{r+1} (v,z)}{  \Om_r (0,0)}  
  \\
  &
    - \oint_{ {\Ga _{\LR} 
  }}  \frac{dv h(v)R_1(v)}{2\pi \I}
   \oint_{ \Ga_{_{\!-\tau_2}} }
    \frac{dz}{2\pi \I (z-v)h(z)R_2(z)} 
   \frac{  \Om_r(z,v)}{  \Om_r (0,0)}.
 \\
 &-\tfrac{1}{r\!+\!1}\oint_{  \Ga_{\tau_1-\rho}
 }   \frac{dv R_1(v)}{2\pi \I}
  \oint_{ \Ga_{_{\!-\tau_2}} }
   \frac{dz}{2\pi \I h(z)R_2(z)} 
  \frac{  \Om^-_{r+1} (v,z)}{  \Om_r (0,0)} .
  \\ \end{aligned}
  \label{PrepKernel2} \ee
  This leads to $-\tfrac{1}{r\!+\!1}  {\mathbb L}'_4-{\mathbb L}_2-\tfrac{1}{r\!+\!1}{\mathbb L}_4 $ in (\ref{Kernlim6}), in that order.
%


  \medbreak
 
 \noindent{\bf (iii) Combining ${\mathbb K}_0$ in (\ref{Kernlim5}) with the sum of the single integrals in (\ref{PrepKernel1})}. Remember ${\mathbb K}_0$ and use the notation,
   \be \bl
 {\mathbb K}_0 & =-\frac{(y-x+1)_{n-m-1}}{(n-m-1)!} \Id_{n>m}\Id_{y\geq x}
 \\
 {\mathcal I}_{\Ga_{y-x}}+{\mathcal I}_{\Ga_{\tau_1-\tau_2}}&:= \frac{(N-n)!}{(N-m-1)!}\oint_{\Ga_{y-x}+\Ga_{\tau_1-\tau_2}}
\frac{dz}{2\pi \I}\frac{R_1(z)}{R_2(z)}
.
\el \ee
 Using (\ref{CoordAs}(**)), we set
\be \label{th}\theta:=x+m-y-n=\tau_1-\tau_2-1,\ee
we now prove the following {\em formulas:}
{\em \be\bl
{\mathcal I}_{\Ga_{y-x}} =\frac{(n-m)_{y-x}}{(y-x)!}\Id_{x\leq y}&\mbox{   ~~~,~~~~~~~  } 
 {\mathcal I}_{\Ga_{\tau_1-\tau_2}} =
\Id_{\theta\geq 0}(-1)^{m-n-1} 
 \frac{(n-m)_{\theta}}{\theta !}
 \\{\mathbb K}_0&=-\Id_{n>m}\Id_{x\leq y }\frac{(n-m)_{y-x}}{(y-x)!}.\label{Int2}\el\ee
Moreover
\be \label{indterm}
{\mathbb K}_0+{\mathcal I}_{\Ga_{y-x}}+ {\mathcal I}_{\Ga_{\tau_1-\tau_2}}=
(-1)^{m-n-1}\left( {x-y-1}\atop{\theta}\right)
\Id_{\theta\geq 0} \Id_{x>y }.
\ee
}

 %
 
 {\em Indeed, } the formula (\ref{Int2}) of ${\mathbb K}_0$ is straightforward; for the proof of the identity (\ref{Int2}) involving ${\mathcal I}_{\Ga_{y-x}} $, see Petrov\cite{Petrov}. We now evaluate the second integral ${\mathcal I}_{\Ga_{\tau_1-\tau_2}} $ for $\tau_1>\tau_2$; otherwise (\ref{Int2}) is trivially satisfied.  Given the definition (\ref{th}) of $\theta$, we have that $0\leq \theta \leq N-n$, since on the one hand $\tau_1 > \tau_2$, and on the other hand the latter is valid asymptotically; see (\ref{ineq2}). The contour  $\Gamma_{\tau_1-\tau_2}$ encloses the points 
$$\{ y - N+n,\ldots,x-N+m\} =\{y - N+n+\al,~~\mbox{with   }  0\leq \al \leq \theta\}.
$$
In order to apply the residue Theorem, we need to evaluate the polynomial numerators of $R_1(v)$ and $R_2(z)$ at those points, using the identity $(a)_k=(-a-k+1)_k (-1)^k$ for $k\geq 0$. Indeed,
\be\bl 
(v-x+1)_{N-m-1}\Bigr|_{v=y-N+n+\al}
  &=(-\theta -(N-m-1)+\al )_{N-m-1}
\\&=(-1)^{N-m-1}(\theta-\al+1)_{N-m-1}
\\
 \frac{(z-y)_{N-n+1}}{z-(y-N+n+\al)}\Bigr|_{z=y-N+n+\al}
 &=(-1)^{N-n-\al} \al!(N-n-\al)!  .
\el \label{eval1}\ee
Using the residue theorem and the evaluations (\ref{eval1}) above, one finds  (the integral only exists for $\tau_1>\tau_2$ or what is the same $\theta\geq 0$):
$$\bl
{\mathcal I}_{\Gamma_{\tau_1-\tau_2}}= & \Id_{\theta\geq 0} (-1)^{N-n} (-1)^{N-m-1} \sum^{\theta}_{\al=0} \frac{(N-n)!}{(N-n-\al)! \al!}\frac{(-1)^\al (\theta -\al+1)_{N-m-1}}{(N-m-1)!} 
\\
= & \Id_{\theta\geq 0}(-1)^{m-n-1} \sum^{\theta}_{\al=0}  (-1)^\al\left(\begin{array}{c} N-n \\ \al\end{array}\right) \left(\begin{array}{c} N-m-1+\theta-\al 
 \\N-m-1\end{array}\right)
\\
=& \Id_{\theta\geq 0}(-1)^{m-n-1}    \left(\frac{(\ell+1)_{\theta}}{\theta !}\right)
=\Id_{\theta\geq 0}(-1)^{m-n-1} 
 \frac{(n-m)_{\theta}}{\theta !},
%
\el  $$
using on the last line the combinatorial identity below, setting $k=N-n$ and $\ell=n-m-1$, valid for $0\leq \theta\leq  k $:
\be \label{combident}
\sum_{\al=0}^{\theta}(-1)^\al\left({k}\atop{\al}\right)
\left({k+\ell+\theta-\al}\atop{k+\ell}\right)
=\frac{(\ell+1)_\theta}{\theta!}.
\ee
%
%
We now prove (\ref{indterm}). Using the identity prior to (\ref{eval1}) again, the following holds :  
\be\label{rel1}\bl
 \Id_{\theta\geq 0} 
 {\cal I}_{\Ga_{y-x}}
 + &\Id_{x\leq y}  {\cal I}_{\Ga_{\tau_1-\tau_2}}
 \\&=\Id_{\theta\geq 0}\Id_{x\leq y} \left(\frac{(n-m)_{y-x}}{(y-x)!} +
 (-1)^{m-n-1} 
 \frac{(n-m)_{\theta}}{\theta !}\right)
 \\&=(-1)^{x-y}\Id_{\theta\geq 0}\Id_{x\leq y}
\left( \frac{(\theta+1)_{y-x}}{(y-x)!} -\frac{(y-x+1)_\theta}{\theta !}\right)=0.
\el\ee
Using the above (\ref{rel1}) in $\stackrel{\ast }{=}$, we have 
$$\bl   {\cal I}_{\Ga_{y-x}}+&{\cal I}_{\Ga_{\tau_1-\tau_2}} 
  =(\Id_{\theta< 0}+\Id_{\theta\geq 0})
 {\cal I}_{\Ga_{y-x}}
 +(\Id_{x\leq y}+\Id_{x>y}) {\cal I}_{\Ga_{\tau_1-\tau_2}}
 \\
 &\stackrel{\ast }{=}\Id_{\theta< 0}(\Id_{n-m\leq 0}+
 \Id_{n-m> 0})
\Id_{x\leq y} \frac{(n-m)_{y-x}}{(y-x)!}
\\&~~~~  + \Id_{x> y}(\Id_{n-m\leq 0}+
 \Id_{n-m> 0}) \Id_{\theta\geq 0}(-1)^{m-n-1} 
 \frac{(n-m)_{\theta}}{\theta !}
\\&\stackrel{\ast\ast}{=}\Id_{\theta< 0}
 \Id_{n>m }
\Id_{x\leq y} \frac{(n-m)_{y-x}}{(y-x)!}
 + \Id_{\theta\geq 0} \Id_{n>m }\Id_{x> y}
  (-1)^{m-n-1} 
 \frac{(n-m)_{\theta}}{\theta !}
\\
&\stackrel{\ast\ast\ast}{=}-{\mathbb K}_0+   \Id_{\theta\geq 0} \Id_{n>m }\Id_{x> y}
  (-1)^{m-n-1} 
 \frac{(n-m)\dots (x-y-1)}{(x-y+m-n) !}.
\el$$
The equality $\stackrel{\ast\ast}{=}$ holds, because the $\Id_{n-m\leq 0}$-terms readily vanish. 
 The last equality $\stackrel{\ast\ast\ast}{=}$ holds; indeed $\Id_{\theta<0}$ can be omitted, since $n>m$ and $y-x\geq 0$ implies $\theta< 0$. Finally, in the second term of $\stackrel{\ast\ast\ast}{=}$, $\Id_{n>m }$ can be omitted, because otherwise that term would automatically vanish, yielding (\ref{indterm}).  

Finally, combining the three contributions (\ref{PrepKernel1}), (\ref{PrepKernel2}) and (\ref{indterm}) leads to formula (\ref{Kernlim6}), ending the proof of Proposition \ref{PropKernlim6}. \qed

 
\section{Formal asymptotics of the main ingredients 
 }

The integrands in the kernel ${\mathbb L}$ as in (\ref{Kernlim6}) contain the functions $R_i,h$ and combinations thereof, besides ascending factorials; see (\ref{Rih}). This section does the formal asymptotics of these functions, given the scaling (\ref{scaling}).

\medbreak

%
%

 
\noindent{\bf Scaling of the integrands: }  The map (\ref{xy-to-xi eta}), combined with the scaling (\ref{scaling}) to the new variables
   $(\tau_1,\sg_1;\tau_2,\sg_2)$, and upon using the map ${\mathcal I} $ as in (\ref{xymap}), leads to the following expressions\footnote{using $(a)_b=a(a+1)\dots (a+b-1)=\frac{(a+b-1)!}{(a-1)!}=\frac{\Ga(a+b)}{\Ga(a)}$.}, 
 $$\begin{aligned}
(v-x+1)_{N-m-1}&=\frac{\Gamma(v-x+N-m )}{\Gamma(v-x+1)}
 \\
 &=\frac{\Ga( (\om+\ga-\al_1+1)t^2-\beta_1 t  +r-\ga_1 -\tau_1+1)}{\Ga( (\om+\tfrac12 (\ga+1)-\al_1)t^2+(\frac {\widetilde\sg_1}2-\beta_1 )t -\ga_1+\frac 12 ( r-\tau_1) +1)}
 \\&
 =:  \frac{\Gamma((\om+A _1)t^2+B _1t+C _1)}{\Gamma((\om+A'_2)t^2+B'_2t+C'_2)}\\
\\
 (z-y)_{N-n+1}&=(v-x+1)_{N-m-1}\Bigr|_{{v\to z}\atop{{\mathcal T}}}
 .\el $$


Notice that, upon using (\ref{xy-to-xi eta}) and (\ref{scaling}) as before, the $R_i(v)$   depend on the geometric variables (see (\ref{PQ})) and the new running variables $\tau_i$ and $\widetilde\sg_i$ as well, whereas $h(u)$ purely depends on the geometry of the polygon; the different $\Ga$-function ratios will be combined appropriately, as follows:
\medbreak
\noindent \hspace*{-1.5cm}$\begin{array}{ ccccccccccccccccccccccccccccc}&&&&&&&&{{ {[x_{c+d},x_{c+d}+\tau_1-\rho-1]}}\atop{\mbox{\tiny finite interval}}}\atop{\downarrow}
& & &&{ [x_{c+1},x]}\atop{\downarrow}
&& &&{ \RR }\atop{\downarrow}
 \end{array}$
\be  \begin{aligned}
 R_1(v)&=\frac{\Ga(v-x-m+N)}{\Ga(v-(x_{c+d}-1)) }
 \frac{\Ga(v-x_{c+1})}{\Ga(v-(x-1))}\frac{\Ga(v-x_{ 1}) }{
 \Ga(v-(x_{ c}-1))}
\\
&= \mbox{\large$ \frac{\Ga( (\om+\ga-\al_1+1)t^2-\beta_1 t  +r-\ga_1 -\tau_1+1)}{\Gamma((\om +\gamma-\al_1+1 )t^2-\beta_1t -\rho+r-\ga_1+1)}
$}
\\&
~~\times \mbox{\large $ \frac{\Gamma(~~(\om +\gamma-\al_1  )t^2-\beta_1t -\rho+r-\ga_1+1))}{\Ga( (\om+\tfrac12 (\ga+1)-\al_1)t^2+(\frac {\widetilde\sg_1}2-\beta_1 )t -\ga_1+\frac 12 ( r-\tau_1) +1)}$}
\\
&
~~\mbox{\large$\times  
\frac{\Gamma(~~(\om-\al_1-\al_2)t^2-(\beta_1+\beta_2)t-(\ga_1+\ga_2)+1)}{ \Gamma( (\om+\gamma-\al_1-\al_2)t^2-(\beta_1+\beta_2)t-(\ga_1+\ga_2)+1)}$}
\\
&=:
 \prod_1^3\frac{\Ga((\om+A_i)t^2+B_it+C_i)}{\Gamma((\om+A'_i)t^2+B'_it+C'_i)}
\\
 R_2(z)&= R_1(v)\Bigr|_{{v\to z}\atop{{\mathcal T}}}
  %
= \prod_1^3\frac{\Ga((\ze+A_i)t^2+B_it+C_i)}{\Gamma((\ze+A'_i)t^2+B'_it+C'_i)}\Bigr|_{\left\{\begin{aligned}
\widetilde\sg_1&\to \widetilde\sg_2-\tfrac{2}{t}\\
\tau_1&\to \tau_2
\end{aligned}\right\} }
\end{aligned}\label{int1}\ee
\noindent
and 

$\begin{array}{ccccccccccccccccccccccccccccccccccc}&&&&&&{\RR}\atop{\downarrow}
&&&&&{\LR}\atop{\downarrow}
&& &&&{\mbox{\tiny$\Sg$-strip}}\atop{\downarrow}
&& &&&{{{\mbox{\tiny$[x_{c+d}-\rho, x_{c+d}-1]$ }}\atop{\mbox{\tiny finite interval}}}}\atop{\downarrow}\end{array}$
\be\begin{aligned}
h(v) 
&=\frac{\Ga(v\!-\!(x_{c}\!-\!1))\Ga(v-x_{ c+d+1}))
\Ga(v-(x_{ c+1}+\Sg)) }{
 \Ga(v- x_1 )\Ga(v-(x_{ d+N}-1))\Ga(v- x_{ c+1 } )}\frac{\Ga(v-(x_{c+d}-1))}{\Ga(v-(x_{c+d}-\rho-1)) }
 \\&= 
 \mbox{\normalsize $ 
\frac{ \Gamma( (\om+\gamma-\al_1-\al_2)t^2-(\beta_1+\beta_2)t-(\ga_1+\ga_2)+1)}{\Gamma(~~(\om-\al_1-\al_2)t^2-(\beta_1+\beta_2)t-(\ga_1+\ga_2)+1)}
\frac{\Gamma((\om +\ga+1) t^2+1)}{\Gamma((\om +\ga+2)t^2+r+1 )}
$}
\\&~~\times
\mbox{\small$  
\frac{\Gamma((\omega  - \alpha_1  +1){t}^{2}-\beta_1 t- 
\gamma_1+1
)}{\Gamma((\omega +\gamma-\alpha_1  ) t ^{2}-  \beta_1 t-\rho+r-\gamma_1+1
)}
\frac{\Gamma((\omega+\gamma -\alpha_1  +1)t^{2}-\beta_1 t-\rho+r-\gamma_1+1)
}{\Gamma((\omega+\gamma -\alpha_1  +1)t^{2}-\beta_1 t+r-\gamma_1 +1)
}$}
\\&=: \prod_3^6\frac{\Ga((\om+A'_i)t^2+B'_it+C'_i)}{\Gamma((\om+A_i)t^2+B_it+C_i)}
. \end{aligned}\label{int2}\ee
One checks the following expressions, useful later on:
 \be\begin{aligned}
 \sum_2^3(A_i-A'_i)& =-\tfrac {\ga+1}2  ,~ 
\sum_2^3(B_i-B'_i) =-\frac 1{2 }\widetilde\sg_1,~ 
\sum_2^3(C_i-C'_i) =\tfrac 1{2 }(r+\tau_1)-\rho
\\
 \sum_3^5(A_i-A'_i) & 
=\sum_3^5(B_i-B'_i) =0,~~
\sum_3^5(C_i-C'_i) =2r-\rho,
\\A_1=A'_1&=A_6=A'_6  = ~\ga- \al_1 + 1, 
  ~~~
 B_1=B'_1 =B_6=B'_6 =-\beta_1\\
  C_1-C'_1&=\rho-\tau_1,~~C_6-C'_6=\rho.
\end{aligned}\label{ABC}\ee
It implies that the first fraction of $R_i(v)$ and the last one of $h(u)$ play a special role, because their numerator and denominator differ by a constant only. This had to be so, because they correspond to finite intervals of width $\rho-\tau_1$ and $\rho$, as seen from (\ref{int1}) and (\ref{int2}).

%
%

 \bigbreak

\noindent {\bf Formal limits and the saddle point}. In view of Stirling's formula $(y-1)!=\Ga(y)= \sqrt{\frac{2\pi} y}e^{y(\log y-1)+O(1/y)}$, we expand the following expression in powers of $1/t$ for $t\to \infty$:
$$\begin{aligned}
(& A t^2+Bt+C)(\log ( A t^2+Bt+C)-1)+O(\frac 1{t^2})
\\&=   t^2A\log A-t^2A+ t B\log A+ (C\log A+ 
\frac {B^2}{2A})+(At^2+  Bt + C)2\log t+O(\frac 1{t }) .
\end{aligned}$$
It follows that
\be\begin{aligned}
& \prod_2^3\frac{\Ga( (\om+A_i) t^2+B_it+C_i)}{ \Ga( (\om+A'_i) t^2+B'_it+C'_i)}=\left(\prod_2^3  \frac{  \om+A'_i }{ \om+  A _i }+O(\frac 1t)\right)^{\tfrac 12}
   \\
   \\
 &
\times\exp  \left\{ \begin{array}{lllll}    t^2 S_1(\om) +t T_1(\om)-t^2\sum_2^3 (A_i-A'_i)
\\ \\+
  \sum_2^3(C_i\log (\om+A_i)+ 
\frac {{B_i}^2}{2(\om+A_i)}-C'_i\log (\om+A'_i)- 
\frac {{B'_i}^2}{2(\om+A'_i)})
\\  \\
 +2 \sum_2^3(t^2(A_i-A'_i)+t (B_i-B'_i)  +   {C_i-C'_i}{} )   \log t  +O(\frac 1{t })
 \end{array}
   \right\}
 \end{aligned}\label{ratio}\ee
 and similarly upon taking the product $\prod_3^5$; in that case the sums get replaced by  $\sum_3^5$. Then $S=S_1,~T=T_1$ goes with $\prod_2^3$, whereas 
 $S=S_2,~T=T_2$ goes with $\prod_3^5$.
 Given an interval $[a,b]$, introduce the functions
 \be S_{[a,b]}(\om):=(\om-b)\log(\om-b)-(\om-a)\log(\om-a).
 \label{Sint}
 \ee
 Given any of the functions $x_k, y_k$ in (\ref{CoordAs}), $x^{(0)}_k, y^{(0)}_k$ refers to their leading terms. The same for the leading terms of the boundary of the intervals, corresponding to $\RR$, $\LR$ and the strip $ \Sigma$.  
 Then the expressions appearing in the asymptotics are as follows:
\be \begin{aligned}
S_1(\om)&:= \sum_2^3((\om+A_i)\log (\om+A_i)-(\om+A'_i)\log (\om+A'_i))
\\
&=  (-S_{[x^{(0)}_{c+1},x^{(0)}_{}]}+S_{\RR^{(0)}})(\om),~~~
\\
 S_2(\om) &:= \sum_3^5((\om+A'_i)\log (\om+A'_i)-(\om+A_i)\log (\om+A_i))
\\& =(- S_{\RR^{(0)}} +S_{\LR^{(0)}}+S_{\Sg^{(0)}})(\om)\\
T_1(\om) &:= \sum_2^3(B_i\log (\om+A_i)-B'_i\log (\om+A'_i))
\\
T_2(\om)& := \sum_3^5(B'_i\log (\om+A'_i)-B _i\log (\om+A_i)).\end{aligned}\label{S-T}\ee
%
%
%
The following derivatives vanish,
 $$\begin{aligned}S_1'(\om)  
=\log \prod_2^3\frac{ \om+A_i }{ \om+A'_i }=0
,~~~~~
 S_2'(\om)  
=\log \prod_3^5\frac{ \om+A'_i }{ \om+A _i }=0,
 \end{aligned}$$
 when the linear  expressions $\prod_2^3(\om+A_i)-\prod_2^3(\om+A'_i )=0 $ and $\prod_3^5(\om+A_i)-\prod_3^5(\om+A'_i)=0  $ respectively; this is so for 
 \be\bl \om &=\om_0:=\frac{-1}{\ga+1}
 (\ga^2+\ga(\al_2-\al_1+1)-\al_1-\al_2) 
 \\ \om&=\om_0+(\al_1-\al_2)\frac{ \al_2\ga(\ga-1) }{(\ga+1)(\ga(\al_2+1)+\al_1+1)} .\el \label{saddle}\ee 
 Considering the kernel (\ref{Kernlim5}), the variables $u_\al$ in the multiple integral  $\Om_r(v,z)$ interact with $v,z$-variables in the first part of the kernel through the ratios $\frac{z-u_\al}{v-u_\al}$. So the scaling for $v$ and $z$ must be, at least to first order, the same as the scaling for the $u_\al$'s. 
 This is to say that the two roots of $S_1'(\om)=0$ and $S_2'(\om)=0$ must be the same, and so we must have $\al_1=\al_2$. One then further computes:
  $$\begin{aligned}
  \tfrac 12 S_1''(\om_0)&=
   -  \frac{(\ga+1)^3}{4\al_1\gamma(\gamma-1)  (\gamma+2\al_1 +1) 
}=:-\frac 1{a ^2}<0
\\
 \tfrac 12 S_2''(\om_0)&=
    \frac 1{a^2}\left(\frac{(\ga+1)(\al_1+1)}{\ga+\al_1+1}\right)   
 =: \frac 1{b^2}>0.
  \end{aligned}$$
  This suggests the change of variables and, for later use, a similar change from $\ze$ to $\tze$, and from $U$ to $\widetilde U$, together with the map in (\ref{scaling})
\be\begin{aligned}
&v=\om t^2,\qquad\qquad~~~~z=\zeta t^2,\qquad\qquad~~~~u_\al=U_\al t^2
\\
&\tom:= \frac{t (\om-\om_0)}{a },~~\tze:= \frac{t (\ze-\om_0)}{a }~~~~\mbox{   and   } ~~~~\widetilde U_\al=\frac{t (U_\al-\om_0)}{a}
\\
&V =\tom-\frac{\beta_1}a
,~~
Z=\tze-\frac{\beta_1}a
,~~
W_\al=\widetilde U_\al -\frac{\beta_1}a 
\mbox{   and   }\sigma_i:= \frac{a\widetilde\sg_i}{\ga+1}.
\label{tomega}\end{aligned}\ee 
 Again, using the scaling (\ref{scaling}) and the further transformation (\ref{tomega}), one checks
  \be\label{diffs}\begin{aligned}
 dvdz&=a^2t^2d\tom d\tze,~~ \frac{dv~dz}{z-v} =\frac{d\om d\zeta}{\zeta-\om}t^2=
  at\frac{d\tom d\tze}{\tze-\tom}~,
  \\ \frac12 \Dt \xi_2&=\frac t2 d\widetilde\sg_2=t\left(\frac{\ga+1}{2a}\right)d\sg\\
  du&=t^2dU=atd\widetilde U,~~~~~ \frac{z-u}{v-u}du=ta\frac{\tze-\widetilde U}{\tom-\widetilde U}d\widetilde U,~~~~
 \\
  (z-u)&(v-u) =a^2t^2(\tze-\widetilde U)(\tom-\widetilde U).
    \end{aligned}\ee%
In this new variable, 
 \be\begin{aligned}
 t^2S_i(\om)&= t^2S_i(\om_0)+(t (\om-\om_0))^2\frac12 S_i''(\om_0)+t^2O((\om-\om_0)^3)
 \\&=\left\{ \begin{aligned}&t^2S_1(\om_0)-\tom^2 + O(\frac 1t)\\
  &t^2S_2(\om_0)+\tom^2\bigl(\frac{a}{b}\bigr)^2 +O(\frac 1t)
  \end{aligned}\right. .
\end{aligned}\label{Taylor0}\ee
%
%
Further, we have 
 $$\begin{aligned}T_1'(\om)\Bigr|_{\om=\om_0}&=\frac{d}{d\om}\sum_2^3(B_i\log (\om+A_i)-B'_i\log (\om+A'_i))\Bigr|_{\om=\om_0}
\\&= 
\sum_2^3 \frac{B_i(\om+A'_i)-B'_i (\om+A_i)}{(\om+A_i)(\om+A'_i)} \Bigr|_{\om=\om_0}
\\&=
 \left(\frac{\ga+1}{2\ga(\ga-1)}\right)
\frac{\ga^2(\beta_1-\beta_2)+2\ga(\widetilde\sg_1\al_1+\beta_1)+\beta_1+\beta_2}{  \al_1 (\ga+2\al_1 +1)}. \end{aligned}$$
Similarly
 $$\begin{aligned}T_2'(\om)\Bigr|_{\om=\om_0}&=\frac{d}{d\om}\sum_3^5(B'_i\log (\om+A'_i)-B_i\log (\om+A_i))\Bigr|_{\om=\om_0}
\\
&=
{\footnotesize \mbox{$-  \frac { ( \gamma+1
  ) ^{2}}{2\ga(\ga-1)}    {\frac {  ( ( { \beta_1}-{\beta_2}){\gamma}^
{2}+(2 \beta_1  +3 \beta_1 \alpha_1  -{\beta_2}{\alpha_1})\gamma+(\beta_1 + \beta_2)(\al_1+1)   )   }  
{ {\alpha_1}   \left(  \gamma +  \alpha_1+1 \right)  \left(  \gamma+2\alpha_1 +1
 \right) }}$}},
 \end{aligned}$$
and so for $i=1,2$, 
$$\begin{aligned}
tT_i(\om)&=tT_i (\om_0)+(t (\om-\om_0)) T_i'(\om_0)+tO((\om-\om_0)^2)
\\
&=tT_i(\om_0)+\tom (aT_i'(\om_0))+ O(\frac 1t).
\end{aligned}$$

\medbreak

 Next we deal with the first factor in (\ref{int1}) and the last factor in (\ref{int2}). Since from (\ref{ABC}), we have $A_1=A'_1=A_6=A'_6$, we fix the saddle $\om_0$ so as to remove the lead-$t^2$-term in the $\Ga$-functions; so $\om_0+A_1=\om_0+\ga-\al_1+1=0$ for $\om_0$ as in (\ref{saddle}); this is so, when $\al_1=\frac{\ga+1}{\ga-1}$. With this substitution and setting $\sg_i=\frac{ a\widetilde \sg_i}{\ga+1}$, all the expressions above simplify to 
\be\begin{aligned}&\om_0=-\frac{(\ga+1)(\ga-2)}{  \ga-1 },~~\frac 1{a^2}=\frac{\ga-1}{4\ga},~~\frac 1{b^2}=\frac 2{a^2}=\frac{\ga-1}{2\ga},~~\al_1=\al_2=\frac{\ga+1}{\ga-1},\\
& aT'_1(\om_0)=
  \sg_1+\kappa,~~~aT'_2(\om_0)=\lb-\kappa,~~~~ a(T'_1(\om_0)+T'_2(\om_0))=\sg_1+\lb,
   \end{aligned}\label{values1}\ee
 with new parameters, expressed in terms of the $\beta_i$ appearing in the scaling for the $m_i$, (see (\ref{scalinggeom}))
 \be\begin{array}{lllll}
  \kappa&:=\frac {a(\ga-1)}{2\ga(\ga+1)} (\beta_1(\ga+1)-\beta_2(\ga-1)) 
~~~ &\mbox{and}&  \lb&:=-\frac{a (\ga-1)}{\ga(\ga+1)}\beta_1.
\end{array}\label{values2}\ee
Later, it will be convenient to define new $\bar\beta_i$, as follows, 
$$
 \bar \beta_i=\frac{2}{a}\tfrac{\ga-1}{\ga+1}\beta_i$$
from which, upon using the value (\ref{values1}) of $a^2$ above, one checks
\be
\bar\beta_1=\lb+\tfrac{2\beta_1}{a} ,~~~-\bar \beta_2=\kappa-\tfrac{2\beta_1}{a} 
,~ ~~~
 ~~\bar\beta_1+\bar\beta_2= \lb-\kappa+\tfrac{4\beta_1}{a} .
\label{newbeta}\ee  
It remains to check the validity of the inequalities (\ref{ineq''}) with the values given in (\ref{values1}); indeed for $1<\ga<3$, we have 
$$\begin{aligned}
\al_i>0 \mbox{  and   } \al_i-\ga+1=\frac{\ga(3-\ga)}{\ga-1}>0
\end{aligned}$$

%

 
  Setting
 $\widetilde\om$ as in (\ref{tomega}),
and taking into account the fact that $\om_0+A_i=0$ for $i=1,6$, we check that the arguments of the $\Ga$-functions for $i=1,6$ (first factor in (\ref{int1}) and last factor in (\ref{int2})) 
$$\begin{aligned}
(\om+A_i)t^2-\beta_1t+C_i&=(a\widetilde\om-\beta_1)t+C_i=:Bt+C_i
\\
(\om+A _i)t^2-\beta_1t+C'_i&=(a\widetilde\om-\beta_1)t+C'_i=:Bt+C'_i,
\end{aligned}$$
and so we have for $i=1,6$, using the $C_i$'s  from (\ref{ABC}):
\be\begin{aligned}
((\om+A_i)t^2&-\beta_1t+C_i)(\log((\om+A_i)t^2-\beta_1t+C_i)-1) 
\\&-((\om+A _i)t^2-\beta_1t+C'_1)(\log((\om+A _i)t^2-\beta_1t+C'_i)-1)
\\
&=
(Bt+C_i)(\log(Bt+C_i)- 1) -(Bt+C'_i)(\log(Bt+C'_i)-1)\\
& =(C_i-C'_i)(\log B+\log t)+O(\frac 1t)
\\
&= \log( (a\widetilde\om-\beta_1)^{C_i-C_i'}) +\log(t^{C_i-C'_i})
 +O(\frac 1t)
 \\
 &=\left\{\bl &\log\left(at (\tom-\tfrac {\beta_1}a)\right)^{\rho-\tau_1}+O(\tfrac 1t)\mbox{    for   }i=1
 \\&\log\left(at (\tom-\tfrac {\beta_1}a)\right)^{\rho}+O(\tfrac 1t)
~~~~ \mbox{        for   }i=6.\el\right.
\end{aligned}\label{simple ratio}\ee
This implies that the {\em ``finite interval"-terms} in $R_1(v)$ and $h(v)$ do not enter into the saddle argument, but merely will give rise to rational functions in the final integrand.
   %


\medbreak

 \noindent {\bf Estimating $R_1(v)$}. $B'_2=\tfrac{\widetilde \sg_1}{2}-\beta_1
 $ and $C'_2=-\ga_1+ \tfrac{r-\tau_1}2+1$ and $C_1=-\gamma_1+r-\tau_1 +1$ are the only terms in $R_1(v)$ depending on the new variables $\widetilde\sg_i,~\tau_i$. Then, taking on the convention that the dots $\dots$ below refer to terms independent of $\tau_i,\widetilde \sg_i$ and $\tom$, but dependent on all the other variables, we have that  {\em the first line} in the exponential of (\ref{ratio}) reads, using (\ref{Taylor0}), (\ref{values1}), the comments right above (\ref{values1}) and $\om_0+A'_2=-\tfrac12(\ga+1)$ in $T_1(\om_0)$:
 \be\bl
  t^2 S_1(\om) &+t T_1(\om)-t^2\sum_2^3 (A_i-A'_i)
  \\&=
  t^2S_1(\om_0)-  \tom^2+tT_1(\om_0)+\tom (\sg_1+\kappa)+t^2(\tfrac{\ga+1}2)+O(\tfrac 1t)
\\&= 
-\tom^2-t(\tfrac{\widetilde \sg_1}{2}-\beta_1)\log(-\tfrac{\ga+1}2)^{}+\tom (\sg_1+\kappa)+t^2(\tfrac{\ga+1}2)+O(\tfrac 1t)+\dots
 \\&= 
-\tom^2+ \log(-\tfrac{\ga+1}2)^{- \widetilde \sg_1t/{2}}+\tom (\sg_1+\kappa) +O(\tfrac 1t)+\dots\el .
\label{ratio1}\ee 
 {\em The second line} of (\ref{ratio}) reads, noticing that $\om-\om_0=O(1/t)$ and using $ \om_0+A'_2=-\tfrac 12 (\ga+1)$,
 \be\bl
 &\sum_2^3(C_i\log (\om+A_i)+ 
\frac {{B_i}^2}{2(\om+A_i)}-C'_i\log (\om+A'_i)- 
\frac {{B'_i}^2}{2(\om+A'_i)})
\\&= \tfrac {\tau_1}{2}\log (\om_0+A'_2)-\frac{\tfrac{{\widetilde \sg_1}^2}{4}-\beta\widetilde \sg_1}{2(\om_0+A'_2)}+O(\tfrac 1t)+\dots
\\&=
\log\ (-\tfrac{\ga+1}{2})^{\tau_1/2} +
 \tfrac{\widetilde \sg_1(\widetilde \sg_1-4\beta)}{4(\ga+1)}+\dots
\el
\label{ratio2}\ee
 and {\em the last line}  reads, using (\ref{ABC}),

 \be\bl
 2\log t \sum_2^3&(t^2(A_i-A'_i)+t(B_i-B'_i)+C_i-C'_i)
 \\&=2t(B_2-B'_2)\log t+2 (C_2-C'_2)\log t
 \\&=
  \log t^{\tau_1-\widetilde \sg_1t} +\dots
\el \label{ratio3}\ee
Finally (\ref{simple ratio}) for $i=1$ reads, 
\be\log\left(a^{-\tau_1}t ^{-\tau_1}(\tom-\tfrac {\beta_1}a)^{ \rho-\tau_1}\right) +O(\tfrac 1t)+\dots\label{ratio4}\ee

\noindent
{\bf Estimating $h(v)$}. Since, as pointed out before,  $h(v)$ is independent of the coordinates, it does not contain any of the variables $(\tau_i,\sg_i)$. Therefore the only contribution will come form the first line of (\ref{ratio}), but calculated for $h$, namely, using again (\ref{Taylor0}), (\ref{values1}) and right above it, compute
$$\bl
  t^2 S_2(\om) &+t T_2(\om)-t^2\sum_3^5 (A_i-A'_i)
  \\&=
  t^2S_2(\om_0)+2  \tom^2+tT_2(\om_0)+\tom (\lb-\kappa)+O(\tfrac 1t)
\\&= 2  \tom^2+ \tom (\lb-\kappa)+O(\tfrac 1t)+\dots,
\el $$
and from (\ref{simple ratio}), 
\be
(\tom-\tfrac {\beta_1}a)^{-\rho}+O(\tfrac 1t)+\dots
.\label{ratio4}\ee
Combining (\ref{ratio1}), (\ref{ratio2}), (\ref{ratio3}), (\ref{ratio4}), one finds
 \be\begin{aligned}
 &\left\{ \begin{array}{lllll}
 R_1(v)
\\ \\
h(v)
\end{array}
\right\} 
 =
 \left\{ {{\prod_1^3 \frac{\Ga( (\om+A_i) t^2+B_it+C_i)}{ \Ga( (\om+A'_i) t^2+B'_it+C'_i)}}\atop{{\prod_3^6 \frac{\Ga( (\om+A'_i) t^2+B'_it+C'_i)}{ \Ga( (\om+A _i) t^2+B_it+C_i)}}}}\right\}
 \\&  {=}
 \left\{   {{ \widetilde f_1(t)a^{-\tau_1} (    \tom -\tfrac{\beta_1}{a}  )^{\rho-\tau_1} t^{-t\widetilde\sigma_1 }\left(-\tfrac{\ga+1}{2}\right)^{ \frac 12 (\tau_1-t\widetilde\sg_1)}}\atop{ \tilde f_2(t)(   \tom-\frac{\beta_1}a )^{-\rho } }}\right\} \left(1+O(\frac 1t)\right)
\\
&~~~\times\exp  \left\{ \begin{array}{lllll}   -\widetilde \om^2\left\{{ {1}\atop{-2}}\right\}+   \widetilde \om \left\{{{\sg_1+\kappa }\atop{\lb-\kappa}}\right\}
 + \left\{{{
 \frac {\widetilde\sg_1(\widetilde\sg_1 -4\beta_1)}{4(\ga+1)}}\atop{0}}\right\}   
 \end{array}
   \right\} , 
 \end{aligned}\label{R1andh}\ee
 where  \be
  \widetilde f_i(t):=\widetilde f_i(t; \beta_1,\beta_2,\ga, \rho,r).
 \label{fg}\ee are functions not depending on the new variables $\sg_i,\tau_i$ and $\om$.

 \newpage
 
 \vspace*{-1cm}
 
 Also, the product of the functions above behaves as
 \be\begin{aligned}
 R_1(v)h(v)=&\tilde f_1(t)\tilde f_2(t)a^{-\tau_1}
 (    \tom -\tfrac{\beta_1}{a}  )^{ -\tau_1} t^{-t\widetilde\sigma_1 }\left(-\tfrac{\ga+1}{2}\right)^{ \frac 12 (\tau_1-t\widetilde\sg_1)}
\\ &\times\exp  \left\{ \begin{array}{lllll}    \widetilde \om^2 +   \widetilde \om (\sg_1+\lb)%
 +  {{
 \frac {\widetilde\sg_1(\widetilde\sg_1 -4\beta_1)}{4(\ga+1)}} }    
 \end{array}
   \right\} \left(1+O(\frac 1t)\right). 
 \end{aligned}\label{R1h}\ee
 Using the map (\ref{xymap}) on formulas (\ref{R1andh}) and (\ref{R1h}), as in the expression (\ref{int2}) for $R_2(z)$, one also checks  \be
 \begin{aligned}
  R_2(z)
    &=
    {{\tilde f_1(t) a^{-\tau_2}(    \tze-\frac{\beta_1 }a )^{\rho-\tau_2} t^{2-t\widetilde\sigma_2 }\left(-\tfrac{\ga+1}{2}\right)^{ \frac 12 (\tau_2-t\widetilde\sg_2)+1}} }  \left(1+O(\frac 1t)\right)
\\
&\times\exp  \left\{ \begin{array}{lllll}   -\tze^2 +   \tze {(\sg_2+\kappa)  }
 +     
 \frac {\widetilde\sg_2(\widetilde\sg_2 -4\beta_1)}{4(\ga+1)}   
+O(\frac 1t) \end{array}\right\}
\\
  R_2(z)h(z)
    &=\tilde f_1(t)\tilde f_2(t)
    {{  a^{-\tau_2}(    \tze-\frac{\beta_1 }a )^{-\tau_2} t^{2-t\widetilde\sigma_2 }\left(-\tfrac{\ga+1}{2}\right)^{ \frac 12 (\tau_2-t\widetilde\sg_2)+1}} }  \left(1+O(\frac 1t)\right)
\\
&\times\exp  \left\{ \begin{array}{lllll}   \tze^2 +   \tze {(\sg_2+\lb)  }
 +     
 \frac {\widetilde\sg_2(\widetilde\sg_2 -4\beta_1)}{4(\ga+1)}   
  \end{array}\right\}.
 \end{aligned}
 \label{R2}\ee
%

 \section{The steepest descent analysis}
 
  Since $\xi-\eta\in 2\BZ+1$, we have for fixed $\eta$ that $\Dt \xi=2$, and so, from the scaling (\ref{diffs}) it follows that \be 1= \tfrac 12 \Dt\xi_2=\tfrac{ t}2 d{\widetilde\sg}_2
=t\frac{\ga+1}{2a}d\sg_2.\label{Delxi}\ee  For convenience, we will denote $d{\mathbb L}_i$ the integrand of $ {\mathbb L}_i$,  $d{\Om}_i$ the integrand of $ {\Om}_i$, etc...

\medbreak



 \noindent {\bf Asymptotics of the ${\mathbb L}_0$-term  }.Then the leading term in  ${\mathbb L}$, as in (\ref{Kernlim6}),  reads (using: $n-m=\tfrac 12 (\tau_2-\tau_1-t(\widetilde\sg_1-\widetilde\sg_2)+1$ according to (\ref{CoordAs})))
\be\begin{aligned}
 {\mathbb L}_0&\frac12 \Dt \xi_2 = \Id_{x> y} \Id_{x-y+m-n\geq 0}(-1)^{ n-m}\left({{x-y-1}\atop{x-y+m-n}}\right)  \tfrac12 \Dt\xi_2
  \\
=& -\Id_{\widetilde\sg_2\geq \widetilde\sg_1}
\Id_{\tau_1> \tau_2}
(-1)^{\tfrac{1}{2}( \tau_1-\tau_2 -t(\widetilde\sg_1-\widetilde\sg_2))}  (-1)^{\tau_2-\tau_1}
 \\
&\times 
\left({\tfrac12((\widetilde\sg_2\!\!-\!\!\widetilde\sg_1)t\!+\! \tau_1\!-\!\tau_2)\!-\!1}\atop{\tau_1-\tau_2-1}\right)  \left(\frac{t(\ga\!+\!1)}{2a}\right)d\sg_2
\\ =& - \Id_{\widetilde\sg_1\geq \widetilde\sg_2}
\Id_{\tau_1> \tau_2}
(-1)^{\tau_2-\tau_1}
(-1)^{\tfrac{1}{2}( \tau_1-\tau_2 -t(\widetilde\sg_1-\widetilde\sg_2))}
\frac{[\tfrac t2(\widetilde\sg_2\!-\!\widetilde\sg_1)]^ {\tau_1-\tau_2-1}}{(\tau_1-\tau_2-1)!}
\left(\frac{t(\ga\!+\!1)}{2a}\right)d\sg_2
\\
=&-(-1)^{\tau_1-\tau_2}\frac{C_t^{(1)}}{C_t^{(2)}}
{\mathbb H}^{\tau_1-\tau_2}(  \sg_2-  \sg_1)d\sg_2,
\end{aligned}\label{L0}\ee
where   \be
 C_t^{(i)}:=C_t(\tau_i,{\widetilde \sg}_i):=(-1)^{\frac12 (\tau_i-t{\widetilde \sg}_i)}
 \left(\frac{t(\ga+1)}{2a}\right)^{\tau_i}.
 \label{conj}\ee

 \noindent {\bf Asymptotics of the factorial}. We  have, using (\ref{CoordAs}) and Stirling's formula, 
 $$\begin{aligned}
 &\frac{(N-n)!}{(N-m-1)! }  
   =\frac{\Ga( \frac{\gamma+1}2 t^2-\frac{\widetilde\sigma_2}2t-\frac{\tau_2}2+\frac r2+1)}{\Ga(\frac{\gamma+1}2 t^2-\frac{\widetilde\sigma_1}2t-\frac{\tau_1}2+\frac r2+1)}\\
   &=(1+O(\frac 1t))\\
   &~~\times\exp\bigl((N\!-\!n\!+\!1) (\log(N\!-\!n\!+\!1)-1)-(N\!-\!m)(\log(N\!-\!m)-1)+O(\frac1{t^2})\bigr)
   \\
   &=\exp\Bigl[\frac{(\widetilde\sg_1-\widetilde\sg_2)t+\tau_1-\tau_2}2\log(t^2\frac{\ga+1}2)
  -\frac{(\widetilde\sg^2_1-\widetilde\sg^2_2)}{4(\ga+1)} \Bigr](1+O(\frac 1t)).\\
  \end{aligned}$$

\noindent {\bf Asymptotics of the $d\Om_r(v,z),~d\Om^\vr_r(v,z)$-integrands}: 
We now define the following integrands: 
$$  \begin{aligned}
d\Xi_k(\tom,\tze )&= \left(\prod_1^k\frac{e^{    2\tU^2+(\lb-\kappa) \tU}}{(   \tU_\al-\frac{\beta_1}a )^{\rho }}
 ~ \frac{\tze-\tU_\al}{\tom-\tU_\al} \frac{d\tU_\al}{2\pi \I}
   \right)
   \Dt_k^2(\tU_1,\dots,\tU_{k})  
 \\
   d\Xi^\vr_k(\tom,\tze ) &=  
  \left( \prod_1^{k}\frac{e^{    2\tU^2+(\lb-\kappa)  \tU}}{(   \tU_\al-\frac{\beta_1}a )^{\rho }}
  ~(\tze\!-\!\tU_\al)^\vr(\tom\!-\!\tU_\al)^\vr \frac{d\tU_\al }{2\pi \I}  \right)\Dt_{k}^2(\tU_1,\dots,\tU_{k}).
\end{aligned}$$
 Then remembering the form (\ref{Omr}) of $ \Om_r(v,z),~\Om^\vr_r(v,z)$ and using the asymptotics (\ref{R1andh}) for $h(z)$, one finds, setting $du_\al= atd\tU_\al$, as in (\ref{diffs}),
$$  \begin{aligned}
d\Om_r(v,z)&=(at)^{r^2}(\widetilde f_2(t))^r d\Xi_r(\tom,\tze) \left(1+O(\tfrac 1t)\right)
\\
d\Om^\pm_{r\mp 1}(v,z)&=(at)^{r^2-1}(\widetilde f_2(t))^{r\mp 1} d\Xi^\pm_{r\mp1} (\tom,\tze)\left(1+O(\tfrac 1t)\right),
\end{aligned}$$



\noindent and so
$$ \begin{aligned}
\frac{d\Om_r(v,z)}{d\Om_r(0,0)}&=
\frac{d\Xi_r(\tom,\tze)}{d\Xi_r(0,0)}\left(1+O(t^{-1})\right)
\\
\frac{d\Om^\pm_{r\mp1}(v,z)}{d\Om_r(0,0)}&=(at)^{-1}\widetilde f_2(t))^{\mp 1}
\frac{d\Xi^\pm_{r\mp1}(\tom,\tze)}{d\Xi_r(0,0)}\left(1+O(t^{-1})\right).
\end{aligned}$$
%


  \noindent {\bf Asymptotics of the $d{\mathbb L}_i$-integrands  }. Using (\ref{conj}), (\ref{R1andh}), (\ref{R2}), (\ref{Delxi}), using the first line of (\ref{diffs}) and $\sg_i=a\widetilde\sg_i/(\ga+1)$, as in (\ref{tomega}), using the translation $(\tom,\tze)\to (V,Z)$ as in (\ref{tomega}) and using the new $\bar \beta_i$ in the last equality below (see (\ref{values2}) and (\ref{newbeta})), the terms $d{\mathbb L}_i$ in (\ref{Kernlim6})  read as follows:
\be  \begin{aligned}  
&\frac{(N-n)!}{(N-m-1)! }\left\{d{\mathbb L}_{1 },d{\mathbb L}'_{1 }\atop d{\mathbb L}_2\right\} \frac12 \Dt \xi_2 \\
&=  \frac{(N-n)!}{(N-m-1)! } 
    \frac{dvdz}{(2\pi\I)^2(z-v)} \frac{R_1(v)}{R_2(z)}      \left\{ { 1\atop{\frac{h(v)}{h(z)}}} \right\}  \frac12 \Dt \xi_2
  \left\{ {\frac{d\Om^{ }_{r}(v,z)}{d\Om^{}_{r}(0,0)}}
  \atop{\frac{d\Om^{}_{r}(z,v)}{d\Om^{}_{r}(0,0)}}
  \right\}%
\\
 &= -  \frac{C^{(1)}_t  }{C^{(2)}_t  }
%
  \frac{e^{-\frac{\beta_1}{a} ( \sg_1- \sg_2) }}    {(2\pi\I)^2} \left\{
{\frac{( \tom-\frac{\beta_1}{a})^{\rho-\tau_1}}{( \tze-\frac{\beta_1}{a})^{\rho-\tau_2}}
\frac{e^{-\tom^2 + (\sg_1+\kappa)\tom}}
{e^{-\tze^2  + (\sg_2+\kappa) \tze  }}
 \frac{d\tom d\tze}{\tze-\tom} ~   \frac{d\Xi^{}_{r}(\tom,\tze)}{d\Xi^{}_{r}(0,0)}}\atop
{   \frac{( \tom-\frac{\beta_1}{a})^{-\tau_1}}{( \tze-\frac{\beta_1}{a})^{ -\tau_2}}
\frac{e^{  \tom^2 +  (\sg_1+\lb) \tom}}
{e^{  \tze^2 + (\sg_2+\lb)\tze  }}  \frac{d\tom d\tze}{ \tze- \tom } ~ 
 \frac{d\Xi^{}_{r}(\tze,\tom)}{d\Xi^{}_{r}(0,0)}
  }
\right\}d{ \sg}_2 (1+O(\tfrac 1t))
\\
&=
-  \frac{C^{(1)}_t  }{C^{(2)}_t  }
%
  \frac{1
   }    {(2\pi\I)^2} \left\{ 
{
\frac{V^{\rho-\tau_1}}{Z^{\rho-\tau_2}}
\frac{e^{-V^2 + (   \sg_1-\bar \beta_2  )V  }}
{e^{-Z^2 + ( \sg_2-\bar \beta_2  )Z  }}
\frac{dVdZ}{Z-V}    \frac{d\Xi_r(V+\tfrac{\beta_1}{a} ,Z+\tfrac{\beta_1}{a} )} {d\Xi_r(0,0)} 
 }
 \atop
{ 
 \frac{V^{ -\tau_1}}{Z^{ -\tau_2}}
\frac{e^{ V^2 + (   \sg_1+\bar \beta_1  )V  }}
{e^{ Z^2 + ( \sg_2+\bar \beta_1  )Z  }}
\frac{dVdZ}{Z-V}    \frac{d\Xi_r(Z+\tfrac{\beta_1}{a} ,V+\tfrac{\beta_1}{a}  )} {d\Xi_r(0,0)} 
  }
\right\}d{ \sg}_2 (1+O(\tfrac 1t))
\end{aligned}\label{Final2}\ee
and
\be \begin{aligned}  
 &\frac{(N-n)!}{(N-m-1)! }\left\{
 d{\mathbb L}_3\atop  d{\mathbb L}_{4}, d{\mathbb L}'_{4}\right\}  \frac12 \Dt \xi_2
 \\&= \frac{(N-n)!}{(N-m-1)! }  
      \frac{dvdz}{(2\pi\I)^2 } \frac{R_1(v)h(v)}{R_2(z)}            \left\{ {  1}\atop{ \frac{1}{h(v)h(z) }}\right\}  \frac12 \Dt \xi_2
  \left\{ {\frac{d\Om^{+}_{r-1}(v,z)}{d\Om^{}_{r}(0,0)}}
  \atop{\frac{d\Om^{-}_{r+1}(v,z)}{d\Om^{}_{r}(0,0)}}
  \right\}
\\&= -  \frac{C^{(1)}_t 
 }{C^{(2)}_t 
  }
  \frac{d\tom d\tze e^{-\frac{\beta_1}{a} ( \sg_1- \sg_2) }  }    {(2\pi\I)^2} %
  \left\{
{   \frac{( \tom-\frac{\beta_1}{a})^{ -\tau_1}}{( \tze-\frac{\beta_1}{a})^{\rho-\tau_2}}
\frac{e^{ \tom^2 + (\sg_1+\lb)\tom}}
{e^{-\tze^2  + (\sg_2+\kappa) \tze }}
  ~ 
 \frac{d\Xi^{+}_{r-1}(\tom,\tze)}{d\Xi^{}_{r}(0,0)}
 }
 \atop{    \frac{( \tom-\frac{\beta_1}{a})^{\rho -\tau_1}}{( \tze-\frac{\beta_1}{a})^{-\tau_2}}
\frac{e^{ -\tom^2 + (\sg_1+\kappa) \tom}}
{e^{ \tze^2 + (\sg_2+\lb)\tze }}  ~  
\frac{d\Xi^{-}_{r+1}(\tom,\tze)}{d\Xi^{}_{r}(0,0)}
} 
\right\}d{ \sg}_2 (1+O(\tfrac 1t))
\\
&=
-  \frac{C^{(1)}_t  }{C^{(2)}_t  }
%
  \frac{dVdZ}{(2\pi\I)^2} \left\{ 
{
 \frac{V^{ -\tau_1}}{Z^{\rho-\tau_2}}
\frac{e^{ V^2 + (   \sg_1+\bar \beta_1  )V  }}
{e^{-Z^2 + ( \sg_2-\bar \beta_2  )Z  }}
     \frac{d\Xi^+_{r-1}( V+\tfrac{\beta_1}{a},Z +\tfrac{\beta_1}{a})} {d\Xi_r(0,0)} 
 }
 \atop
{ 
 \frac{V^{ \rho-\tau_1}}{Z^{ -\tau_2}}
\frac{e^{-V^2 + (   \sg_1-\bar \beta_2  )V  }}
{e^{ Z^2 + ( \sg_2+\bar \beta_1  )Z  }}
      \frac{d\Xi^-_{r+1}(V+\tfrac{\beta_1}{a},Z +\tfrac{\beta_1}{a}  )} {d\Xi_r(0,0)} 
  }
\right\}d \sg_2 (1+O(\tfrac 1t)), \end{aligned}\label{Final3}\ee
   %
%
%
with (using (\ref{newbeta}))
$$  \begin{aligned}
 & d\Xi_r(V\!+\!\frac{\beta_1}a,Z\!+\!\frac{\beta_1}a)
=\left[     
\prod_1^r
 \frac{e^{    2W_\al^2+  (\bar\beta_1+\bar\beta_2)  W_\al}}{    W_\al  ^{\rho }}
 ~\left(\frac{Z\!-\!W_\al}{V\!-\!W_\al}\right) \frac{dW_\al}{2\pi \I}\right]\Dt_r^2(W_1,\dots,W_r)
 \\
  &d \Xi^{\pm}_{r\mp1}(V\!+\!\frac{\beta_1}a,Z\!+\!\frac{\beta_1}a) 
   =\left[     
\prod_1^{r\mp 1}
 \frac{e^{    2W_\al^2+  (\bar\beta_1+\bar\beta_2)  W_\al}}{    W_\al  ^{\rho }}
 ~\left( ({Z\!-\!W_\al} )\ ({V\!-\!W_\al})\right)^{\pm 1} \frac{dW_\al}{2\pi \I}\right]
 \\&\qquad\qquad\qquad\qquad\qquad\qquad\qquad\qquad\qquad\qquad  \qquad\qquad\Dt_{r\mp1}^2(W_1,\dots,W_{r\mp1}).
\end{aligned}$$

\bigbreak

 \noindent{\bf Steepest descent analysis}: 
We will now be working in the $\om -\om_0$ scale. The geometric points $x_i$ in Fig.1 in the new scale $\om-\om_0$ are denoted by  $\widetilde x_i$. That is: $\widetilde x_i=t^{-2}(x_i-\om_0t^2)=\om-\om_0$, using (\ref{tomega}). Fig. \!14. gives a line with the respective positions of the geometric points $\widetilde x_i$, corresponding to the $x_i$. As before set $\widetilde x^{(0)}_i$ for the leading term of $\widetilde x_i$. Then, it follows from the definition (\ref{Sint}) and (\ref{S-T}) that
 \be\begin{aligned}
 S_1(\om_0+\om)&=
 -S _{[\widetilde x^{\mbox{\tiny $(0)$}}_{c+1},\widetilde x^{(0)}]}(\om)+
 S _{[\widetilde x^{\mbox{\tiny$(0)$}}_{c },\widetilde x^{(0)}_1]}(\om)
 \\
 S_2(\om_0+\om)&=
  S _{[\widetilde x^{(0)}_{d+N},\widetilde x^{(0)}_{c+d+1}]}(\om)+
 S _{[\widetilde x^{(0)}_{c+1 },\widetilde y^{(0)}_d]} (\om)
 -S _{[\widetilde x^{(0)}_{c },\widetilde x^{(0)}_1]}(\om)
  \\
 (S_1+S_2)(\om_0+\om)&=
  S _{[\widetilde x^{(0)}_{d+N},\widetilde x^{(0)}_{c+d+1}]}(\om)+
 S _{[\widetilde x^{(0)} ,\widetilde y^{(0)}_d]}(\om) .
 \end{aligned}
 \label{Si}\ee
%


Setting $\om=X+\I Y$, the function (\ref{Sint}),
$$\bl
 S_{[a,b]}(\om)&:=(\om-b)\log(\om-b)-(\om-a)\log(\om-a),
\el $$
has $S'_{[a,b]}(a)=-S'_{[a,b]}(b)=\infty$ and thus it has a local maximum at the mid-point $\tfrac{a+b}2\in [a,b]$, and so the smooth deformation $  S_{[a,b]}(x)+g(x)$ has a local maximum for some $c\in[a,b]$.  It behaves asymptotically for $|X|\to \infty$ as 
$$\bl
 S_{[a,b]}(X)&=-(b-a)(\log |X| +1)+\frac{b^2-a^2}{2X}+O(\tfrac{1}{X^2}).
\el $$
Setting $\om=X+\I Y$, the functions $\Re S_i$ and $\Re(S_1+S_2)(\om_0+\om)$ are symmetric about the $X$-axis and behave, for $|X|\to \infty$ as 
$$\bl \Re S_1(\om_0+X ) &=-\frac{\ga+1}{2}\log|X|+O(\tfrac{1}{|X|}),
\\
\Re S_2(\om_0+X ) &=-\frac{2\ga(\ga+1)}{X(\ga-1)}+O(\tfrac{1}{|X|^2})
\el
$$
using for $S_2$ the fact that (notice $|\LR|+|\Sg|-|\RR|=2r-\rho$)
 $$(\widetilde x^{(0)}_{c+d+1}-\widetilde x^{(0)}_{c+d+b})^\vr+(\widetilde y^{(0)}_d-\widetilde x^{(0)}_{c+1})^\vr-(\widetilde x^{(0)}_1-\widetilde x^{(0)}_c)^\vr=  
\left\{ \begin{array}{ll} &0 ~~\mbox{for}~~\vr=1
 \\&-\frac{4\ga(\ga+1)}{X(\ga-1)}~~\mbox{for}~~\vr=2\end{array}\right. .$$
Using the behavior of the $S_{[a,b]}$ above and the formal Taylor expansion (\ref{Taylor0}), the   properties of the $S_i$, setting  $S_3:=S_1+S_2$, are summarized in Fig. 12\footnote{Set $\widetilde{\RR}:=[\widetilde x_{c },\widetilde x_1]$,~$\widetilde{\LR}:=[\widetilde x_{d+N },\widetilde x_{c+d+1}]$,~$\widetilde\Sg:=(\widetilde x_{c+1},\widetilde y_d)$    
,~ $ [\widetilde x_{c+1} ,\widetilde x]\subset \BR_+$. }. Knowing the type of saddle point for each of the $S_i$ and the location of the maxima and minima, it is easily seen that the $\Re S_i(\om_0+X+\I Y)$ behaves exactly as in the 3-dimensional plots of Fig. 15, with the level profile as indicated just below, at least for $1<\ga<3$. 
Fig. 
\!\!14 contains all the geometric points expressed in the scale $\om-\om_0$; e.g., $\widetilde x=t^{-2}(x-\om_0 t^2)$, $\widetilde x_c=t^{-2}(x_c-\om_0 t^2)$, \dots.
Notice that for $\ga\sim 1$, the saddle point $\om_0$ as in (\ref{values1}) would go to infinity and for $\ga\sim 3$, the geometric condition $y_d<x_c$ on the model would be violated. 

\newpage

\vspace*{-2cm}
 \setlength{\unitlength}{0.017in}\begin{picture}(0,100)
\put(155,-70){\makebox(0,0) {\rotatebox{0}{\includegraphics[width=240 mm,height=250 mm]
 {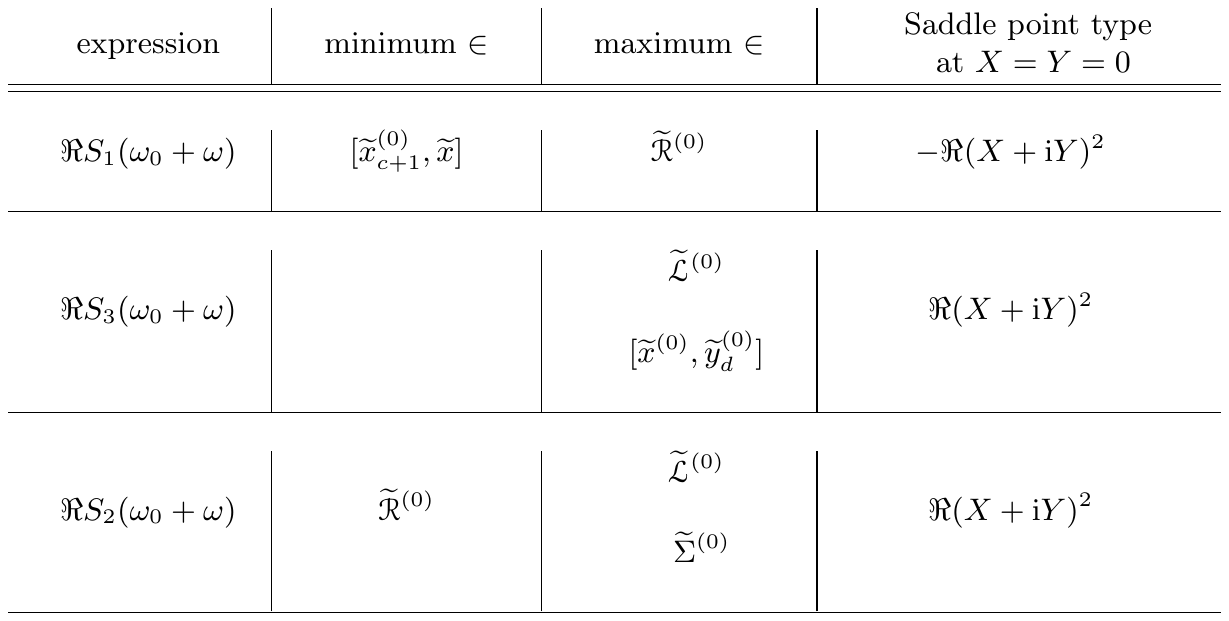} }}}

\end{picture}
 
  \vspace*{.8cm}
Fig. 12. Behavior of $S_1$, $S_3=S_1+S_2$ and $S_2$.
 
 \vspace*{7cm}
 
    \hspace*{ 0.9cm}
 \setlength{\unitlength}{0.017in}\begin{picture}(0,30)
\put(125,-70){\makebox(0,0) {\rotatebox{0}{\includegraphics[width=180 mm,height=300 mm]
 {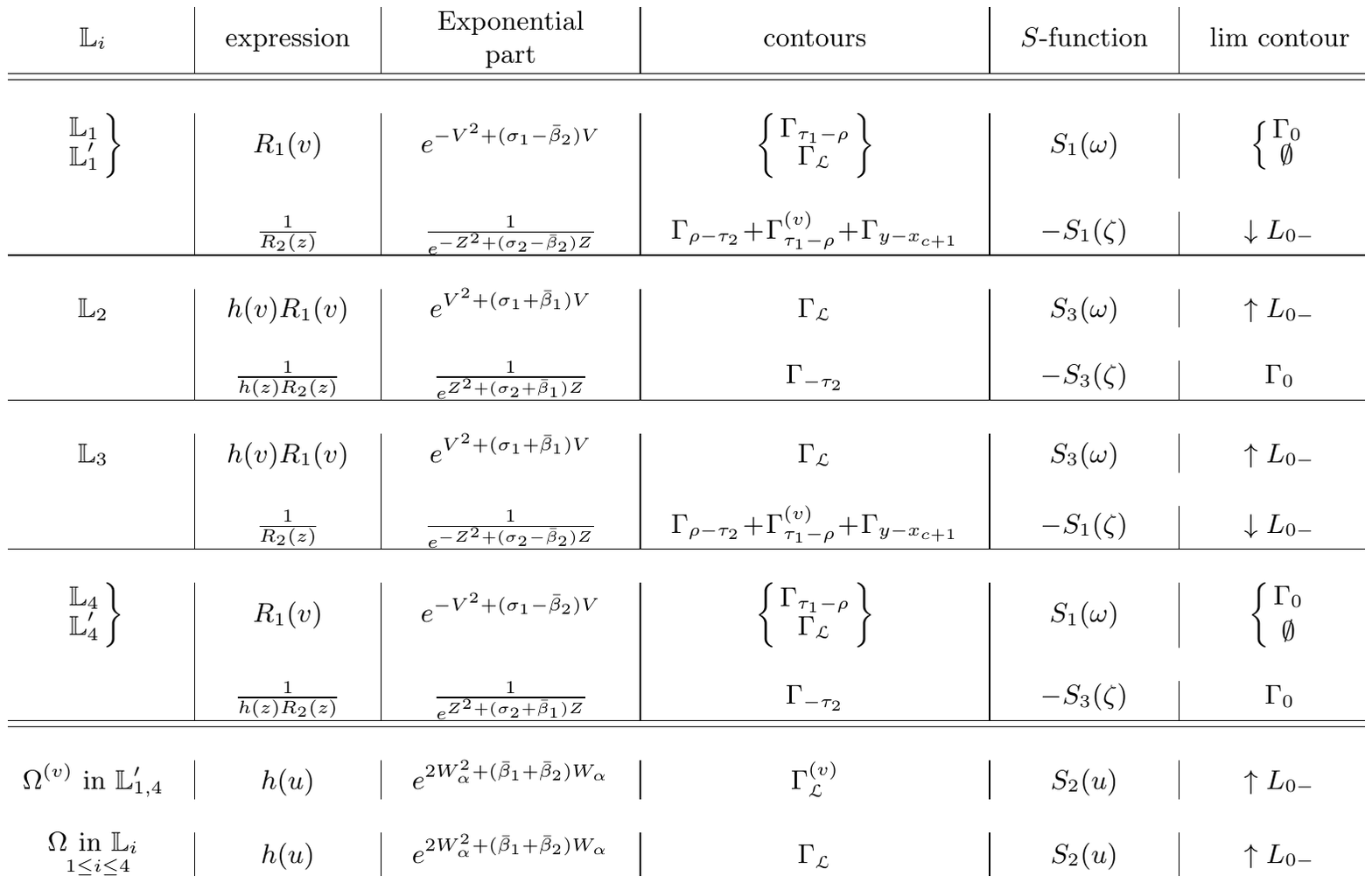} }}}

\end{picture}
 
 \vspace*{4cm}
Fig. \!13. List of expressions in the integrands of the kernel ${\mathbb L}$, their contribution to the exponentials, the corresponding $S_i$-functions and contours, and  the limiting contours in the $V,Z$-variables. Here 
 $\Ga_0 :=\{\mbox{circle about $0$\}, to the right of $ L_{0-}:=0_- +\I \BR$}$.

   \newpage

\vspace*{-4.5cm}

\rotatebox{90}{\includegraphics[width=200mm,height=350mm]{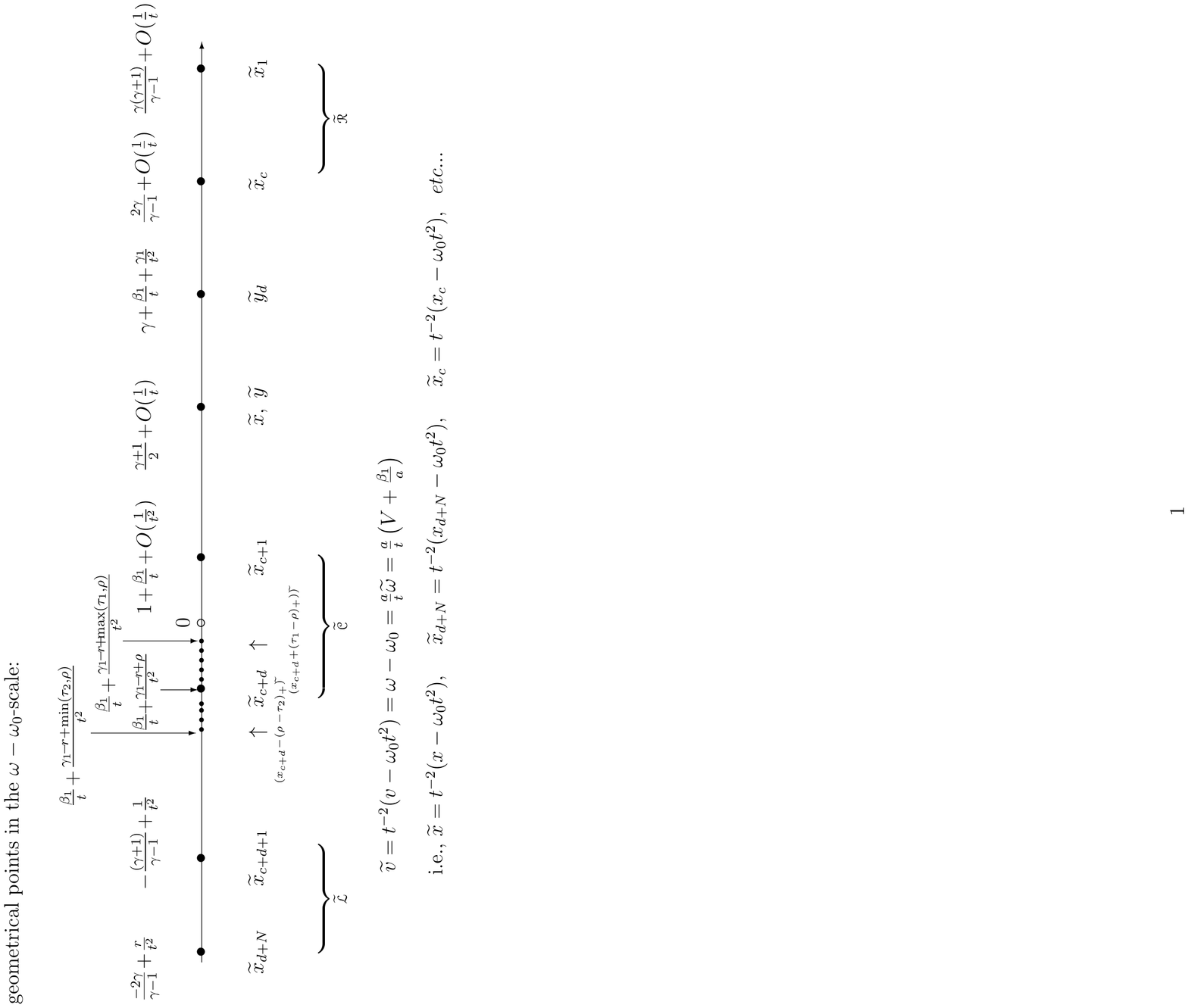}}

\vspace*{-.2cm}
Fig. \!14. The formulas above give the coordinates $\widetilde v=t^{-2}(v-\om_0t^2)=\om-\om_0$    of the points below, with $\circ$ being the origin $\tom=0$, with $x$ and $y$ being the running variables and assuming $\beta_1<0$ and $t$ large enough.

\newpage
  
  \vspace*{-3cm}
  
  \hspace*{-2.9cm}\setlength{\unitlength}{0.017in}\begin{picture}(0,30)
\put(125,-70){\makebox(0,0) {\rotatebox{0}{\includegraphics[width=85 mm,height=95 mm]
 {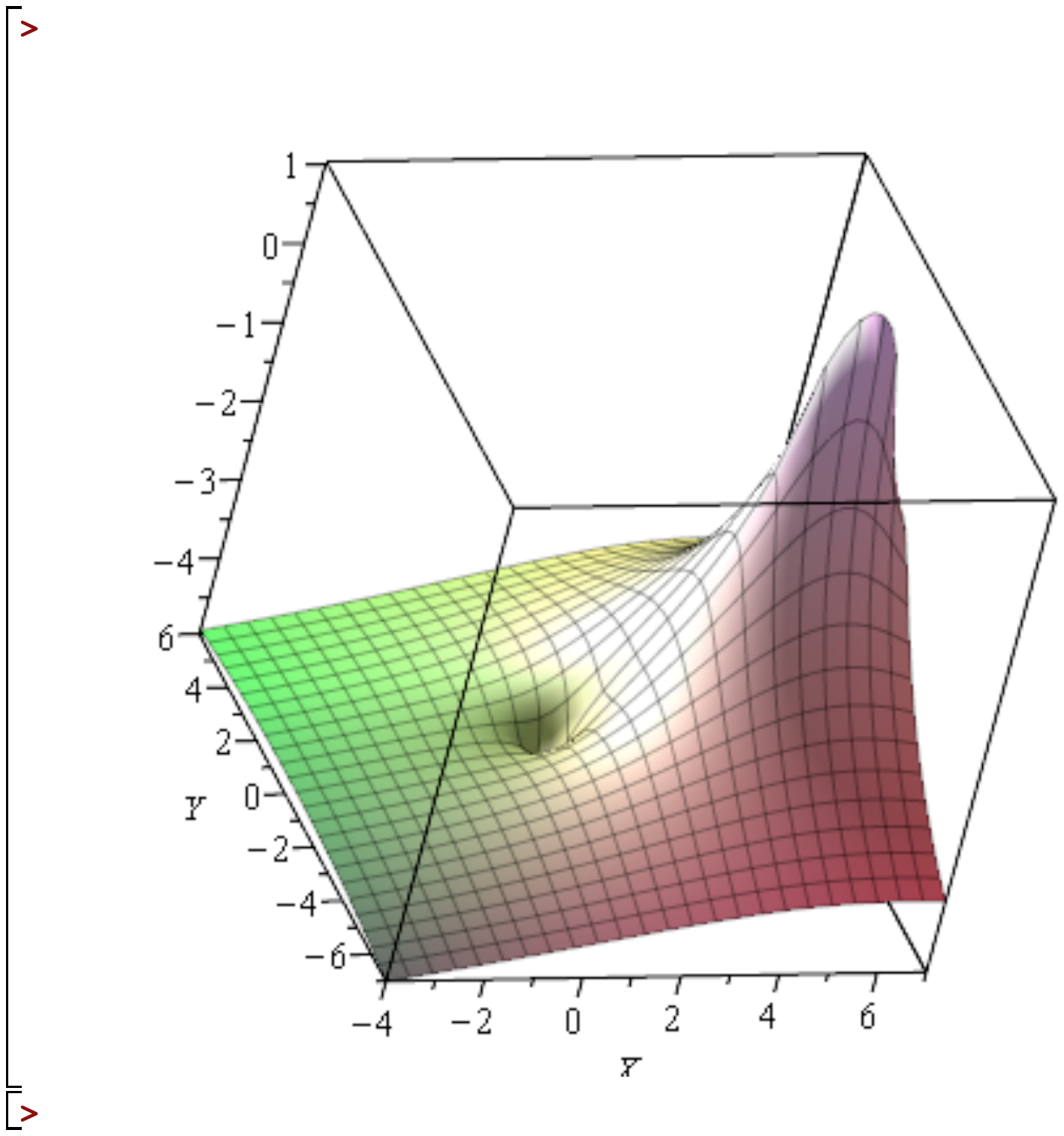} }}}
  \thicklines
  \put(30.69,   -95){\textcolor[rgb]{1.00,1.00,1.00}{\line(0, 1){130}}}

\end{picture}
 \hspace{4.4cm}
  \setlength{\unitlength}{0.017in}\begin{picture}(0,30)
\put(125,-70){\makebox(0,0) {\rotatebox{0}{\includegraphics[width=85 mm,height=95 mm]
 {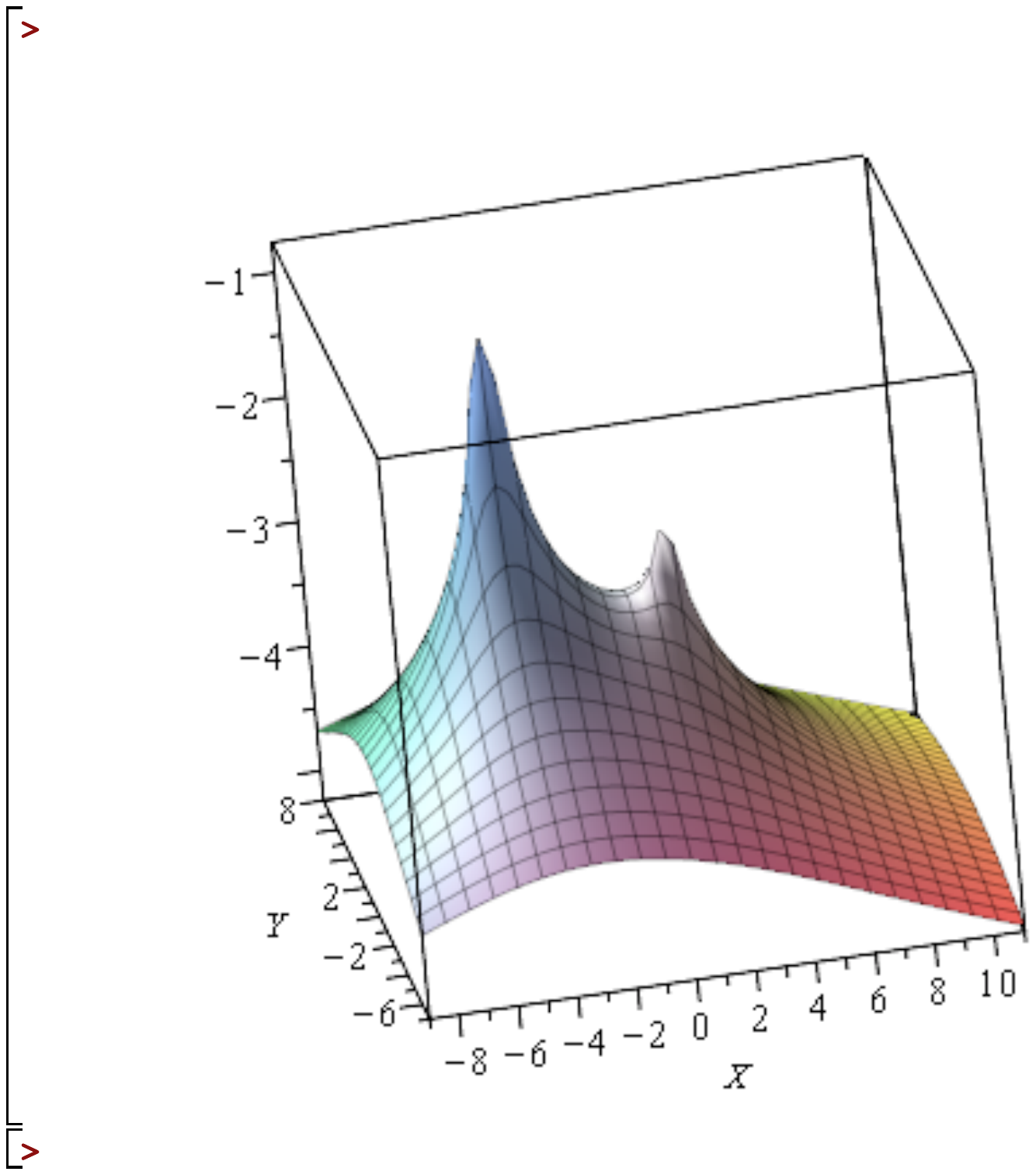} }}}
 \thicklines
  \put(33.90,   -98){\textcolor[rgb]{1.00,1.00,1.00}{\line(0, 1){130}}}

 \end{picture}
\hspace*{-1cm} \setlength{\unitlength}{0.017in}\begin{picture}(0,30)
\put(255,-70){\makebox(0,0) {\rotatebox{0}{\includegraphics[width=85 mm,height=95 mm]
 {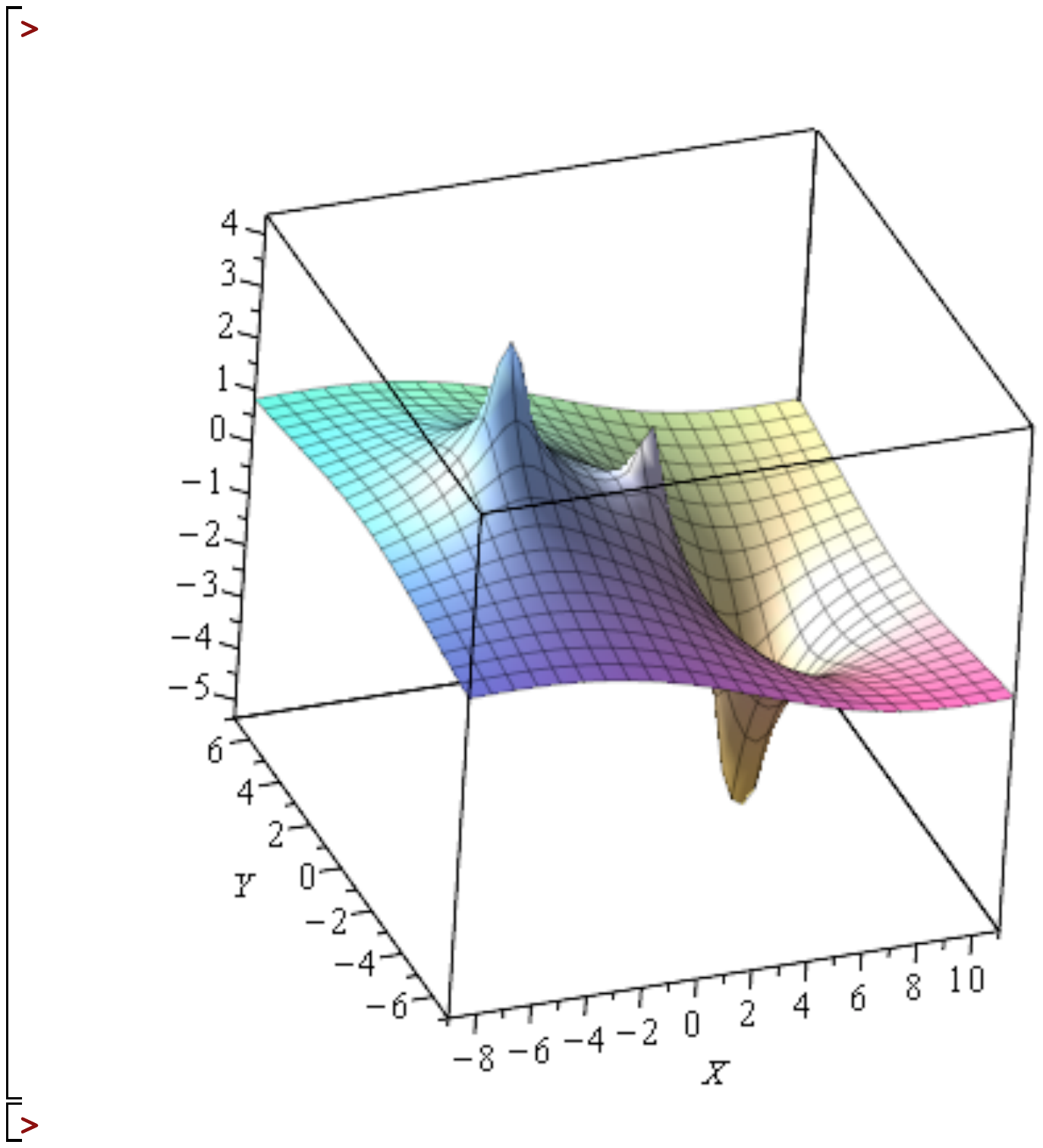} }}}
  \thicklines
  \put(163.90,   -98){\textcolor[rgb]{1.00,1.00,1.00}{\line(0, 1){130}}}

 \end{picture}

 \vspace*{3.5cm}
 
 \setlength{\unitlength}{0.017in}\begin{picture}(0,30)
\put(70,-73){\makebox(0,0) {\rotatebox{0}{\includegraphics[width=70 mm,height=80 mm]
 {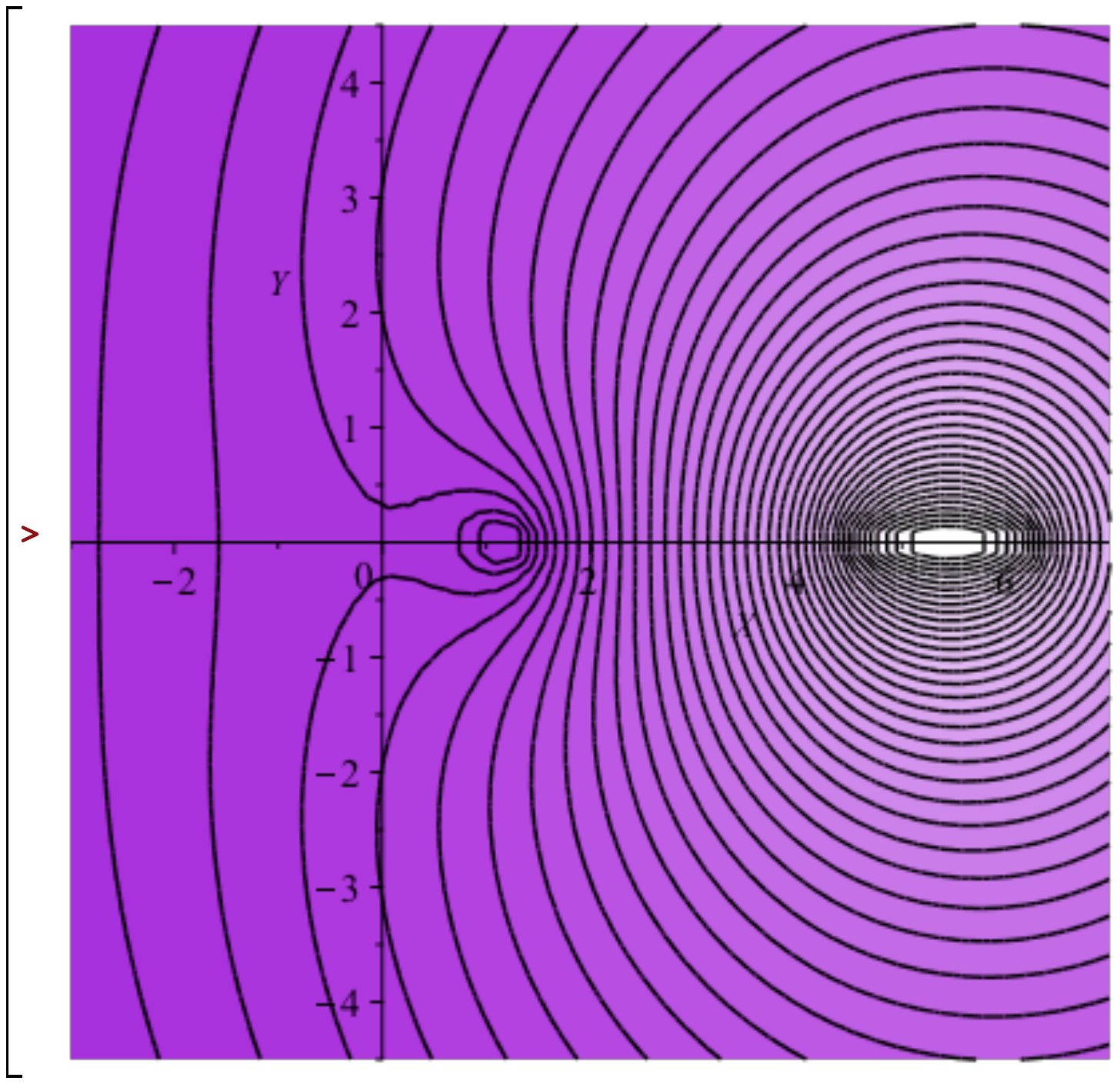} }}}
 
 \put(180,-73){\makebox(0,0) {\rotatebox{0}{\includegraphics[width=70 mm,height=85 mm]
 {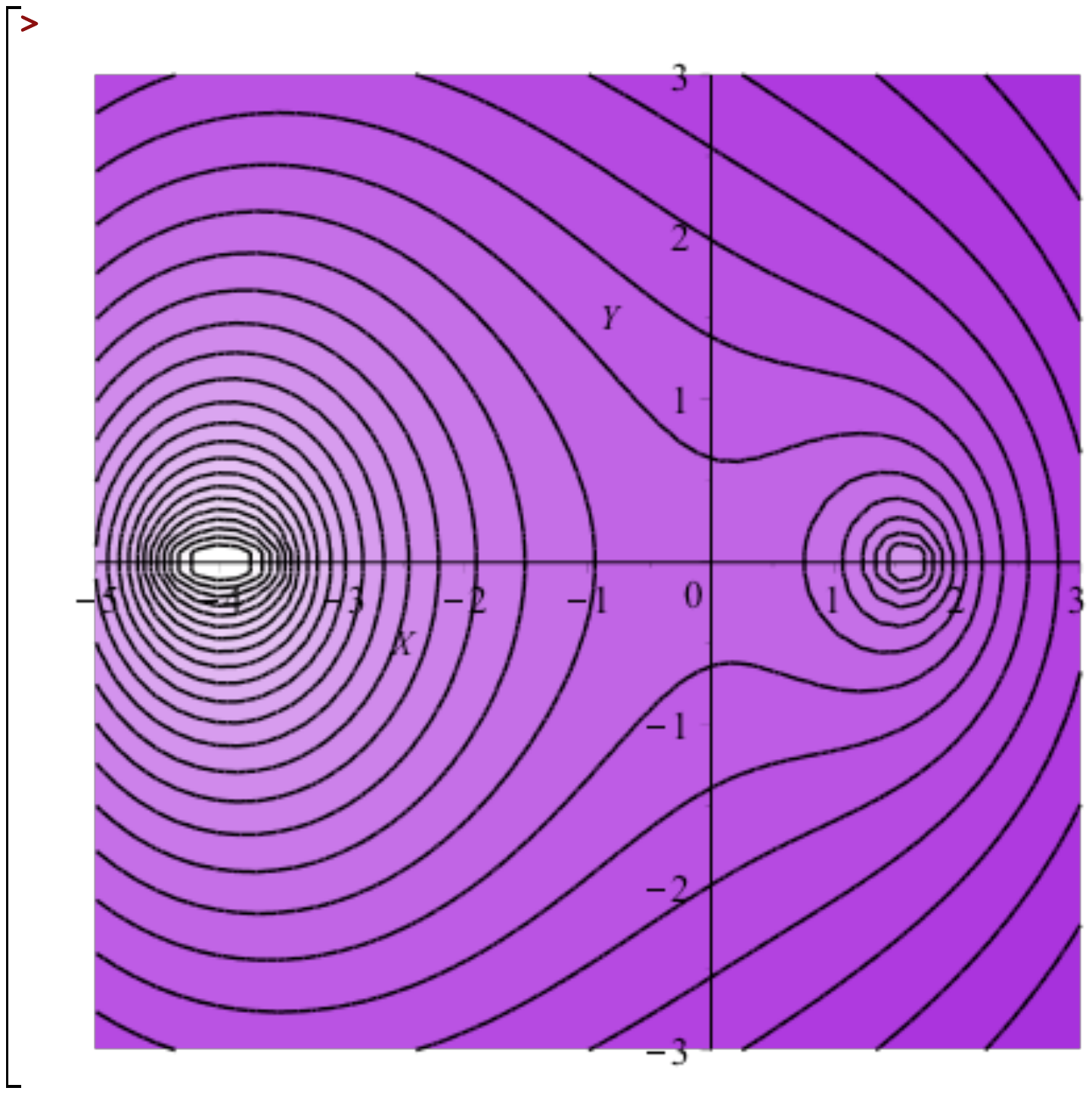} }}}
 
 \put(287,-73){\makebox(0,0) {\rotatebox{0}{\includegraphics[width=70 mm,height=80 mm]
 {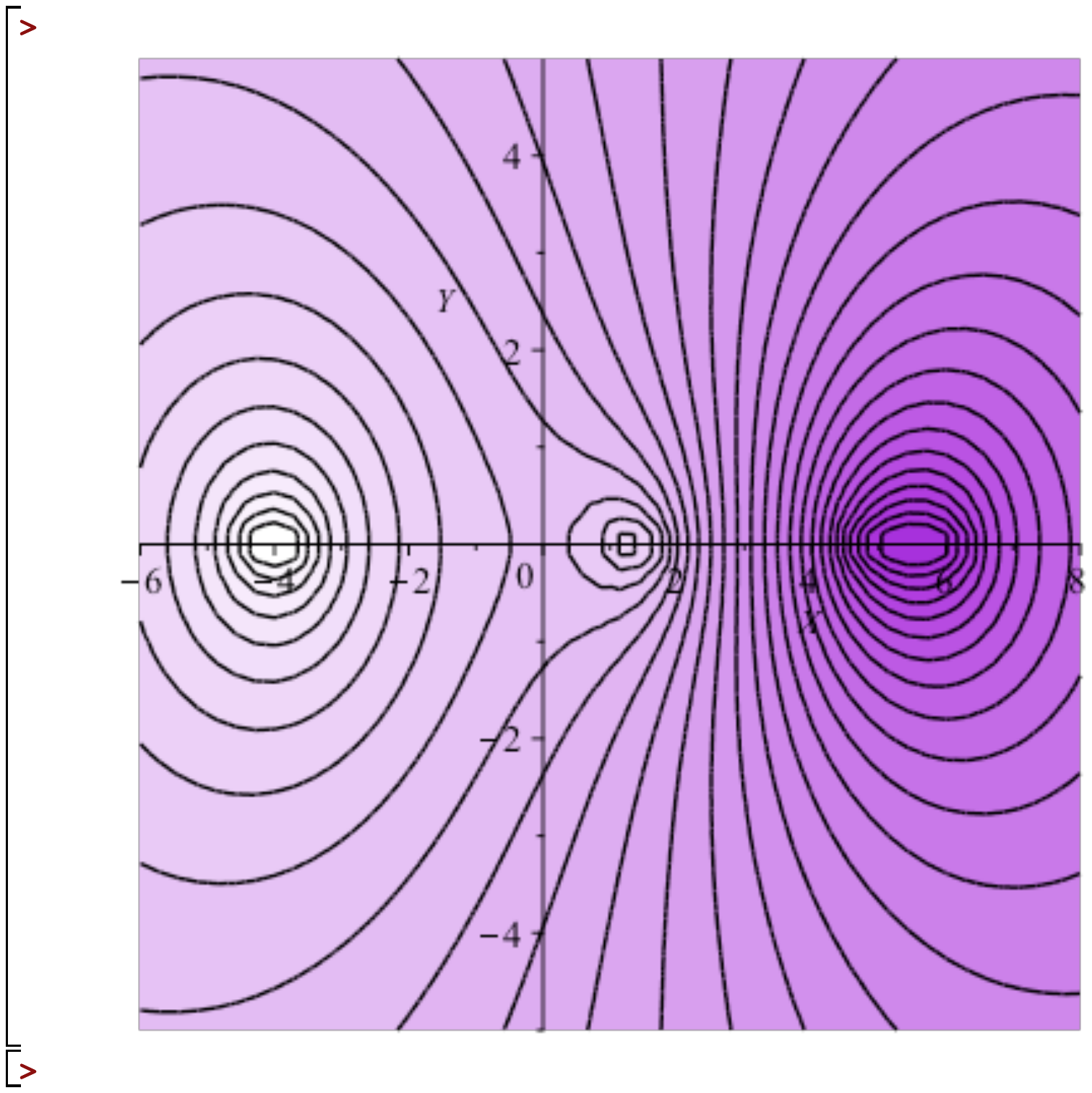} }}}
 
 \end{picture}

\vspace{5cm}


 \hspace*{2.7cm}\setlength{\unitlength}{0.017in}\begin{picture}(0,30)

 \put(1,0){\makebox(0,0)
 {\mbox{$\setlength{\unitlength}{0.017in}\begin{picture}(0,30)
\put(0,0){\makebox(0,0) {\rotatebox{0}{\includegraphics[width=55 mm,height=55 mm]
 {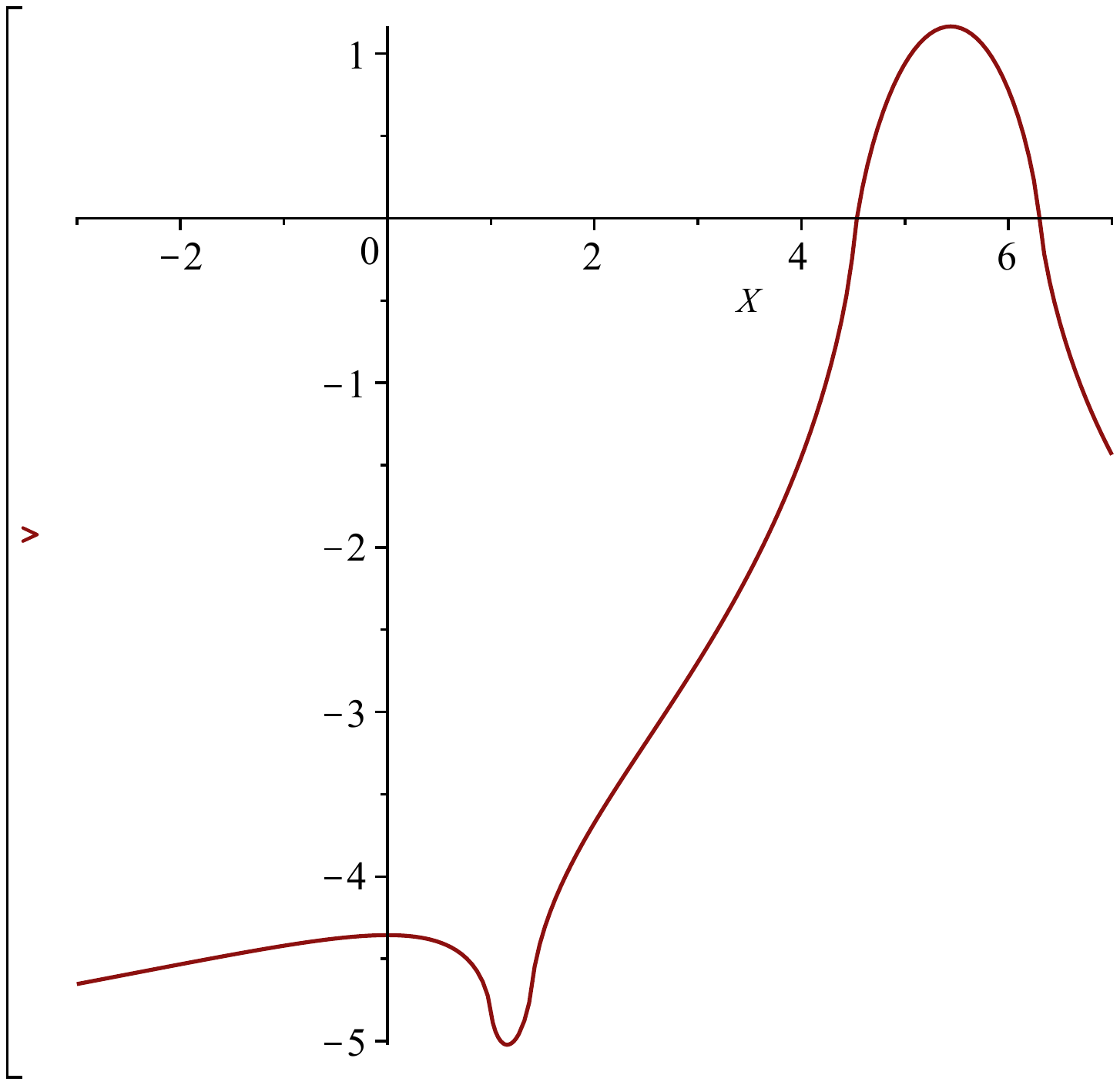} }}}
 
 \thicklines \put(-59.6,  -20){\textcolor[rgb]{1.00,1.00,1.00}{\line(0, 1){80}}}
  \end{picture}$ }
 }}
 
 
\put(100,0){\makebox(0,0) {\mbox{$\setlength{\unitlength}{0.017in}\begin{picture}(0,30)
\put(0,0){\makebox(0,0) {\rotatebox{0}{\includegraphics[width=55 mm,height=55 mm]
 {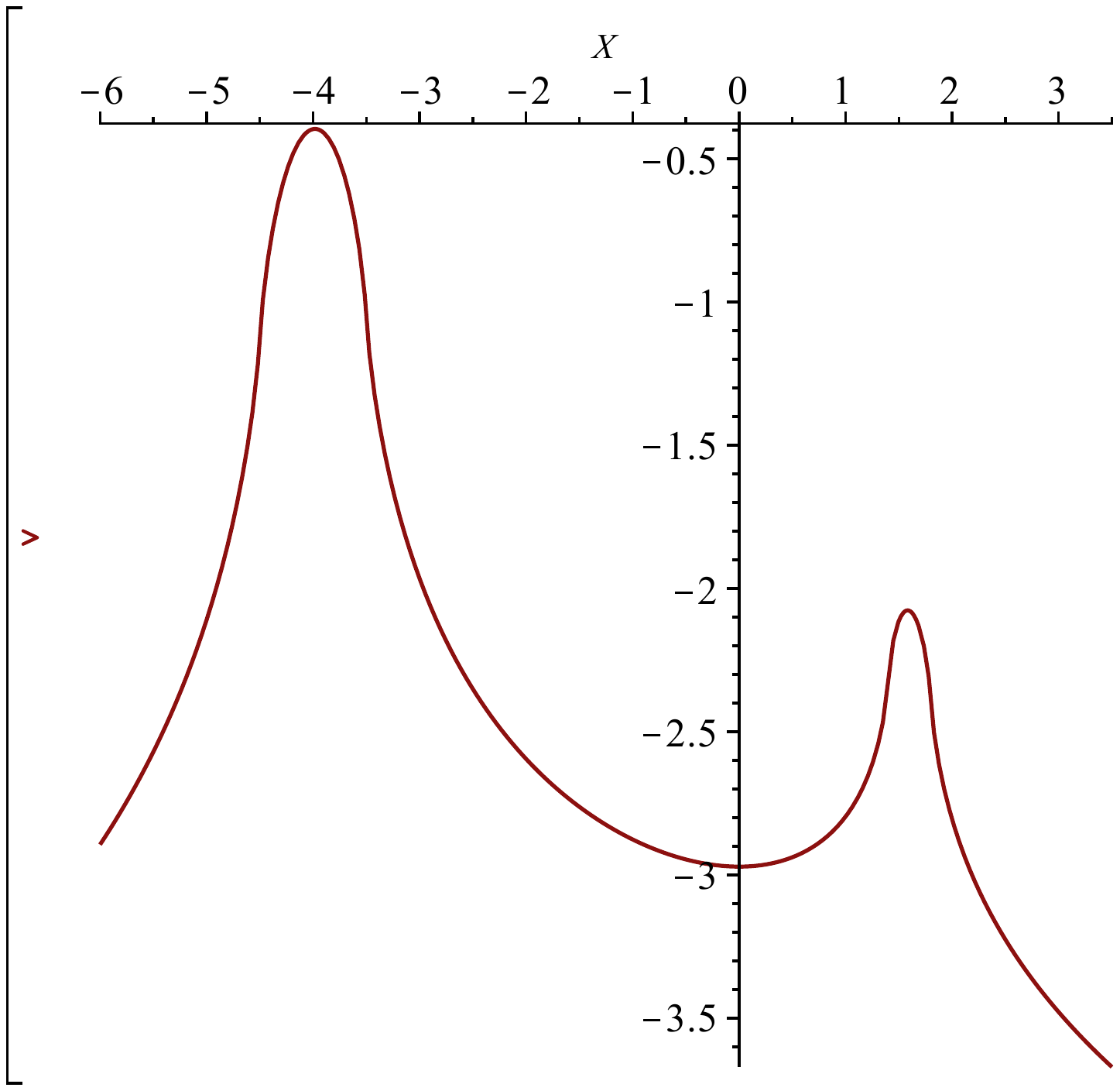} }}}
  \thicklines \put(-59.6,  -20){\textcolor[rgb]{1.00,1.00,1.00}{\line(0, 1){80}}}

\end{picture}$}}
 }

\put(200,0){\makebox(0,0) 
{ \mbox{$\setlength{\unitlength}{0.017in}\begin{picture}(0,30)
\put(0,0){\makebox(0,0) {\rotatebox{0}{\includegraphics[width=55 mm,height=55 mm]
 {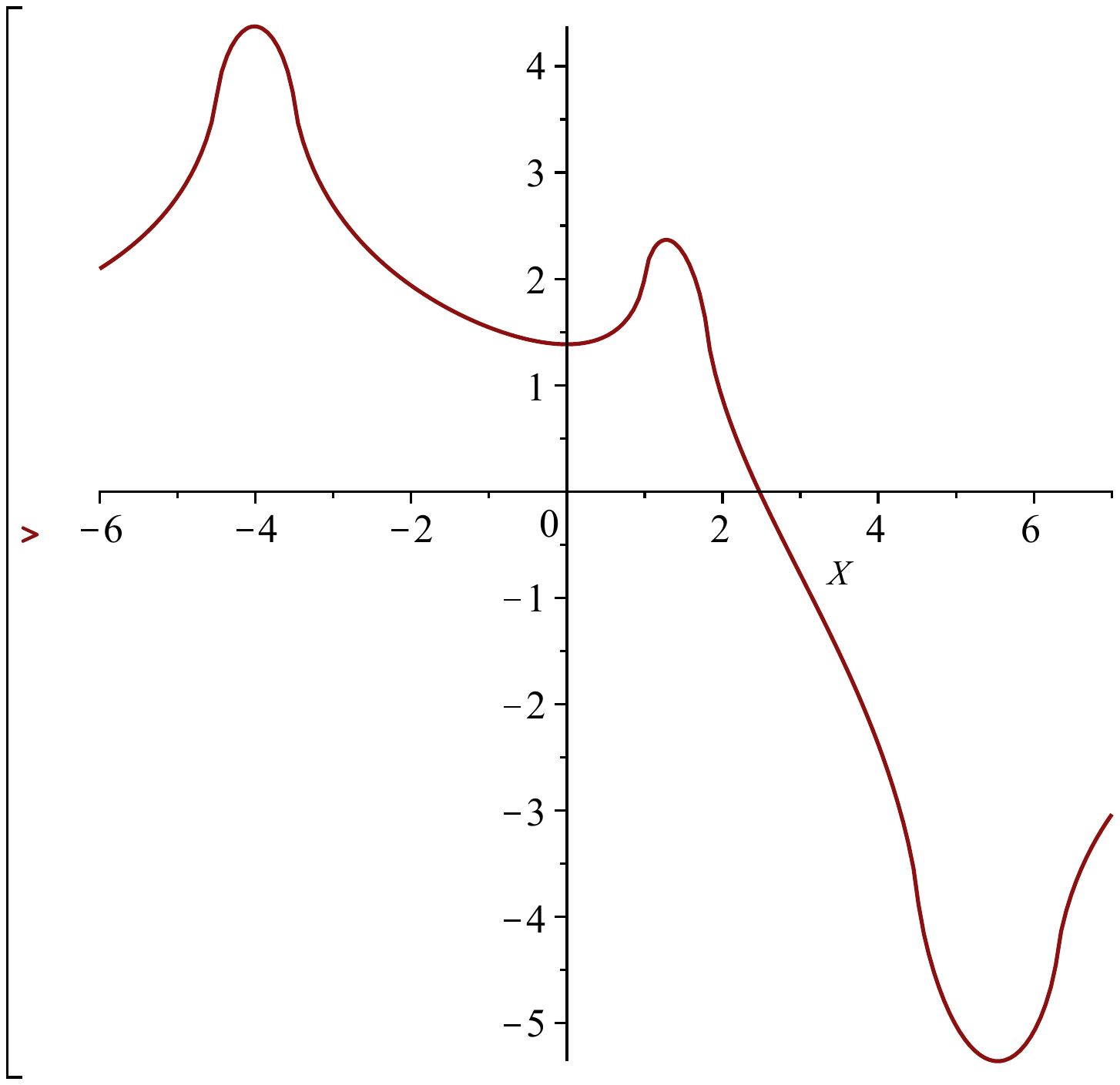} }}}
  \thicklines \put(-59.7,  -20){\textcolor[rgb]{1.00,1.00,1.00}{\line(0, 1){80}}}
\end{picture}$}}
 }

 \end{picture}
 
 \vspace*{2.0cm}

 \hspace*{  -.4cm}
 \noindent Fig. 15. For $\ga=1.8$:  (For the level lines, dark=low, white=high)  
 $$
\hspace*{-.1cm} \begin{array}{llllllllllllllllll}
 \mbox{Plot of}&  \Re S_1(\om_0+X+\I Y)&  
   \Re (S_1+S_2)(\om_0+X+\I Y) 
 &   \Re S_2(\om_0+X+\I Y) 
\\
\mbox{Level lines of}& \Re S_1(\om_0+X+\I Y)&  
   \Re (S_1+S_2)(\om_0+X+\I Y) 
 &  \Re S_2(\om_0+X+\I Y)
  \\
 \mbox{graph of}&  \Re S_1(\om_0+X)&  
 \Re (S_1+S_2)(\om_0+X) 
&  \Re S_2(\om_0+X)
\\
 \end{array}
 $$


  \newpage
  \vspace*{ -1cm}
  
  \begin{picture}(-50,0)
  
  \put(125,-70){\makebox(0,0) {\rotatebox{0}{\includegraphics[width=100 mm,height=180 mm]
 {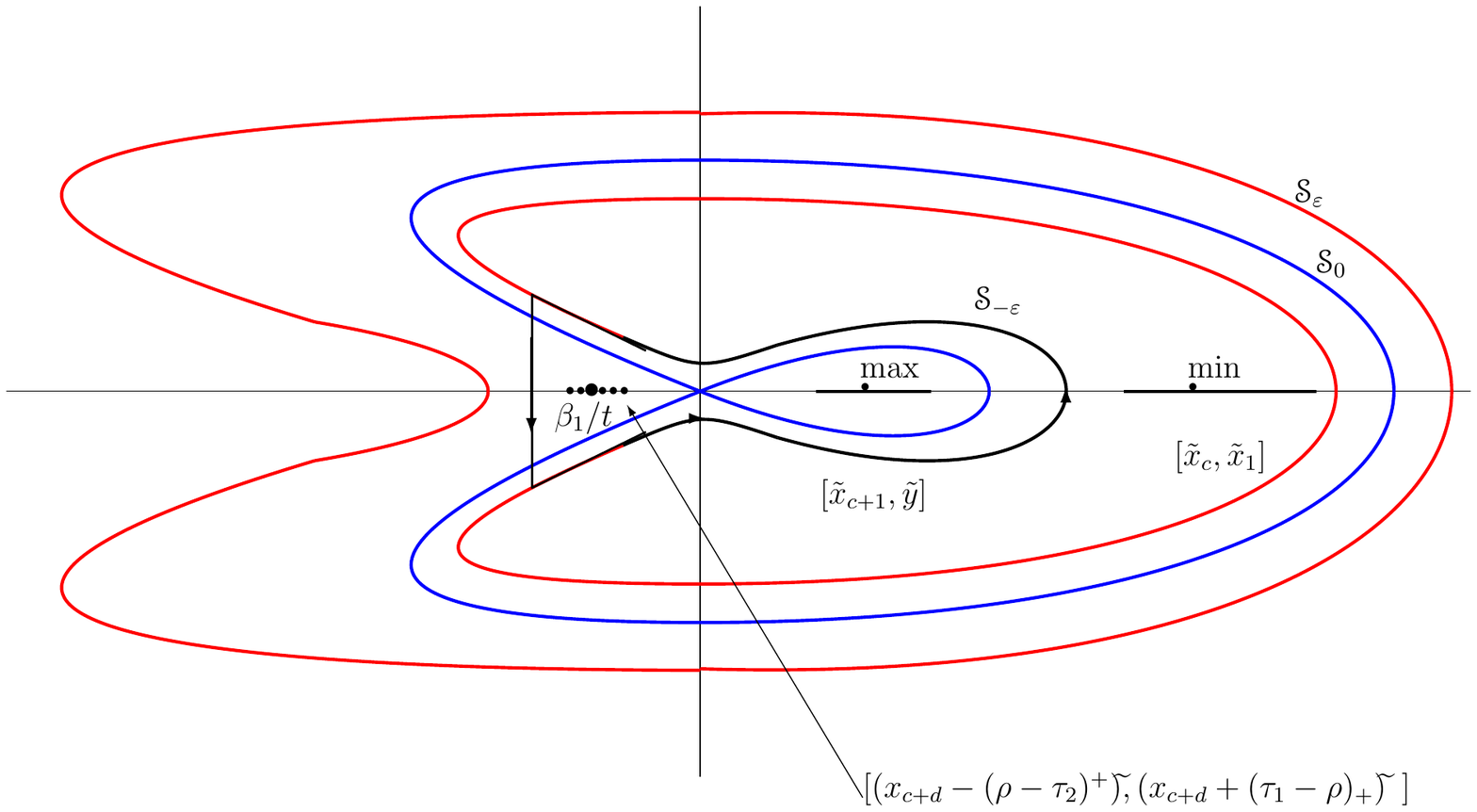} }}}
 
   \put(255,-20){\makebox(0,0) {$-\Re S_1$}}
    \put(20,-13){\makebox(0,0) {$\leftarrow~\mbox{up}$}}
  \end{picture}
 $\label{pp}$ 
  \vspace*{6cm}

  \begin{picture}(-50,0)
  
   \put(125,-70){\makebox(0,0) {\rotatebox{0}{\includegraphics[width=100 mm,height=180 mm]
 {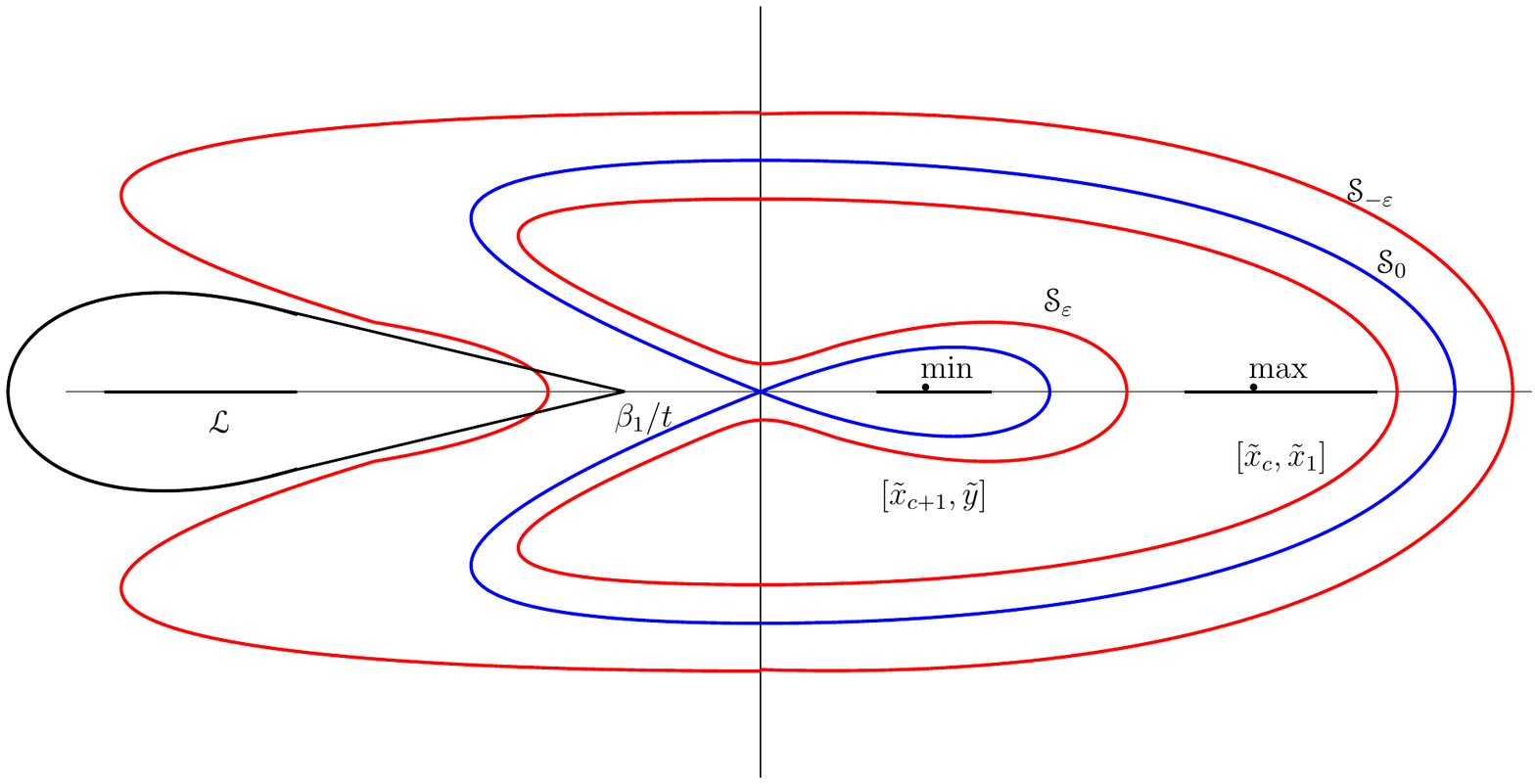} }}}
 
 \put(255,-20){\makebox(0,0) {$\Re S_1$}}
     \put(7,-14){\makebox(0,0) {$\leftarrow~\mbox{down}$}}
     
     \put(29,17){\makebox(0,0) {$\Ga_{\LR}$}}
   
 \centering
  \end{picture}
  \vspace{4cm}
  
  Fig. 16. Separatrices (blue) and level lines (red) for $-\Re S_1$ and $\Re S_1$, including the steep descent curves (black). 
   \vspace{1cm}
   
    \noindent {\bf The analysis}. We assume throughout that $\beta_1<0$.  In a sufficiently small neighborhood of the saddle point of the functions $\pm S_i$, the Taylor series (\ref{Taylor0}) of $\pm S_i$ is convergent. To be precise, setting  $F(\om):=\pm S_i(\om_0+\om)$, we have that for some compact subset $K\subset \BC$ and for $ \frac{\tom}t  \in K$, 
     \be
  t^2\left| F\Bigl(\frac{\tom}{t}\Bigr)-F(0)-\frac{F''(0)}{2}\Bigl(\frac{\tom }{ t }\Bigr)^2  \right|\leq 
  t^2\sup_{z\in K}\left|   F'''(z)\right| \Bigl(\frac{|\tom |}t\Bigr)^3\leq C\frac{|\tom|^3}{t}.
  \label{Taylor1}\ee
 Given $0<\vr''<\tfrac 13$, and for large enough $t$, a ball $|\om-\om_0|\in B(0,t^{-2/3-\vr''}) $ about the saddle point  contains the following contours involved in the integrations, expressed here in the $\om-\om_0$-scale (explaining why putting a tilde)\footnote{In (\ref{intervals1}),  $[a,b]$ , for $a,b\in \BZ$, denotes an interval of integers and $=\emptyset$ if $a>b$.} ; this uses (\ref{CoordAs}) and (\ref{poles}):
 \be\bl
\widetilde\Ga^{(v)}_{\tau_1-\rho}&\stackrel{*}{=}\Ga\left\{\frac{\beta_1}{t}+\frac{1}{t^2}\bigl[{\ga_1-r+\rho}~~~,~~~~~~ ~~~ {\ga_1-r+\max(\tau_1,\rho) } \bigr]\right\}\\
 \widetilde\Ga_{\rho-\tau_2}\!+\! \widetilde\Ga^{(v)}_{\tau_1-\rho}&\stackrel{**}{=}\Ga\left\{\frac{\beta_1}{t}+\frac1{t^2}\bigl[ {\ga_1-r+\min(\tau_2,\rho) } ~,~  {\ga_1-r+\max(\tau_1,\rho) }\bigr] \right\}\\
  \widetilde\Ga_{-\tau_2} &\stackrel{***}{=}\Ga \left\{\frac{\beta_1}{t}+\frac 1{t^2}\bigl [  {\ga_1\!-\!r\!+\!\min(\tau_2,0)} ~, ~~~  {\ga_1 -r- 1 } \bigr ]\right\}.
 \el\label{intervals1}\ee 
 These are contours about the point $ {\beta_1}/t\in   B(0,t^{-2/3-\vr''})$, all of width $1/t^2$. Then in the $\tom$-scale, $ \tom=\frac ta (\om-\om_0)\in B(0,t^{1/3-\vr''})$ and thus the Taylor expansion (\ref{Taylor1}) in $\tom/t$ converges, since the error has order
  $$C\frac{|\tom|^3}{t}\leq \frac{1}{t^{3\vr''}}.
  $$
  Moreover on the boundary of the ball $  B(0,t^{-2/3-\vr''}) $, we also have
  \be 
   t^2\left| F(\om)-F(0) \right|\simeq \tfrac{F''(0)}{2}t^{2/3-2\vr''}.
  \label{Fdiff}\ee
  
    
 Case 1: First consider the 
 $ \left\{ \begin{array}{lllllll}
 \mbox{$v$-contour in ${\mathbb L}_1,  {\mathbb L}_4$}
 \\
 \mbox{$z$-contour $ {\mathbb L}_2, {\mathbb L}_{4 }, {\mathbb L}'_{ 4} $}
 \end{array}\right\}
 $ .
When $t\to \infty$, these contours about the intervals (\ref{intervals1}) above get squeezed in the $\widetilde\om$-scale (resp.  $\widetilde\ze$-scale) to the point $\beta_1/a$ and thus the contours turn into a loop around $\beta_1/a$, which in the $V$-scale (resp. $Z$-scale) becomes a loop $\Ga_0$ about the origin.

\bigbreak

Case 2: Next consider the $z$-contours $\Ga_{\rho-\tau_2}\!+\! \Ga^{(v)}_{\tau_1-\rho}\!+\!\Ga_{y-x_{c+1}}$ in $ {\mathbb L}_1,  {\mathbb L}'_1, {\mathbb L}_3$.  In the  $\ze$-scale, this is a loop surrounding the intervals  $[\widetilde x_{c+1},~\widetilde y]$ and (\ref{intervals1}**). The function $-\Re S_1(\om_0+\om)$ is a  deformation of  $\Re (X+\I Y)^2$ about a small neighborhood of the saddle point $\om=0$, and has a maximum in the interval $[\widetilde x_{c+1},~\widetilde y]$ and a minimum in  $[\widetilde x_{c },~\widetilde x_1]$; see Fig. 12. The level curve (red in Fig. 16) 
  $$
  {\mathcal S}_{- \vr}:=\left\{\ze=X+\I Y ~ \bigr| ~-\Re S_1(\om_0+\ze)=-\Re S_1(\om_0)-\vr
  \right\}$$
for the level profile of $-\Re S_1$ is, for $\vr>0$ small enough, a loop (black in Fig. 16) inside the separatrix $
  {\mathcal S}_0$ (blue in Fig. 16), whose left-most point belongs to the region $X<\frac{\beta_1}{t}<0$, at least for $t$ large enough.
  Then, for $-\vr'<\frac{\beta_1}{t}$ close enough to the saddle point $0$ and $t$ large enough, the function $-\Re S_1(\om_0-\vr'+\I Y)$ will be decreasing in $|Y|$, even a little beyond the intersection of the line $-\vr'+\I \BR$ with the separatrix ${\mathcal S}_{0}$, as follows from the  type $ \Re (X+\I Y)^2$ of the saddle.   
  
   Consider now the curve (black in Fig. 16) formed by 
   \newline {\bf (i)} the vertical segment $L^{-\vr'}$ of $ - \vr'+\I \BR$, up to the point of intersection with a level curve $ {\mathcal S}_{- \vr}$ for small enough $0<\vr$, and
   \newline {\bf (ii)} the part $ \bar{\mathcal S}_{- \vr}$ of the level curve $ {\mathcal S}_{- \vr}$, starting form the intersection points with the vertical segments and winding around the minimum of $-\Re S_1$. 
   
  The curve $L^{-\vr'}\cup \bar{\mathcal S}_{- \vr} $ is indeed a loop about the intervals (\ref{intervals1}**) and $[\widetilde x_{c+1},~\widetilde y]$, along which the function $-\Re S_1(\om_0+\ze)$ will be decreasing, starting from the point $(X,Y)=(-\vr',0)$ and constant all along $\bar{\mathcal S}_{- \vr}$.

 For  $-\vr'<\frac{\beta_1}{t}$ and $1/t$ small enough, the segment $L^{-\vr'}$ will belong to the ball of convergence $|\ze-\om_0|\in B(0,t^{-2/3-\vr''}) $ (for $0<\vr''<\tfrac 13$) of the Taylor series (\ref{Taylor1}) about the saddle point. So, there the  intersection point  $L^{-\vr'}\cap \bar{\cal S}_{-\vr}$ is of order $t^{-2/3-\vr''}$, which in the $\tze=\tfrac ta(\ze-\om_0) $-scale has order $t^{1/3-\vr''}$. 
 Moreover the value of  $-\Re S_1(\om_0+\ze)$ along $\bar{\mathcal S}_{- \vr} $ is given by (\ref{Fdiff}), i.e.,
   \be
 t^2\left(  \Re (-S_1)(\om_0+X+\I Y)-\Re (-S_1)(\om_0)\right)\simeq -t^{2/3-2\vr''};
\label{Sdiff}\ee
so one is at the boundary of the ball $B(0,t^{-2/3-\vr''})$ in the $\ze-\om_0$-scale.     \bigbreak
  


%
 Finally, the interval (\ref{intervals1}**) 
  above gets multiplied with $\frac ta$, which contracts the interval to the point $\frac{\beta_1}a$, whereas the interval  $[\widetilde x_{c+1},~\widetilde y]$ gets send to infinity; see Fig. 14. This implies that in this limit, and since in the $\tze=\tfrac ta(\ze-\om_0) $-scale the ball of convergence has order $t^{1/3-\vr''}$, the small neighborhood along the vertical segment $L^{-\vr'}$ gets blown up to an imaginary line $\downarrow\!\! L_{\frac{\beta_1}{a}}^-$ in the $\tze$-scale and thus the line $\downarrow\!\! L_{0}^-$ in the $Z$-scale. In view of (\ref{Sdiff}), the integral in the $\ze$-scale about the curve $\bar {\cal S}_{-\vr}$ in $\ze$-scale will be in absolute value  
 $$\leq \{\mbox{rational function of $t$}\} \{\mbox{length of $\bar {\cal S}_{-\vr}$}\}e^{-t^{2/3-2\vr''}},
 $$
 and thus exponentially small.

\bigbreak

Case 3. Consider the 
 $ \left\{ \begin{array}{lllllll}
 \mbox{$v$-contour in ${\mathbb L}_2,  {\mathbb L}_3$}
 \\
 \mbox{$u$-contour in $\Om$, as in $ {\mathbb L}_1, {\mathbb L}'_{1 }, {\mathbb L}_{ 4} ,{\mathbb L}'_{4 }$}
 \end{array}\right\}
 $ . Here we are dealing with the functions $\Re (S_1+S_2)$ and $\Re S_2$, which to the left of the $\widetilde x, \widetilde y$ have a similar behavior, with a maximum along $\LR$. Therefore the contours $\Ga_\LR$ can be deformed to a loop consisting of a vertical segment passing through the saddle point, which is continued symmetrically by a level curve about the interval $\LR$. This a steep descent path and, as in Case 2, the segment can be made to belong to the neighborhood where the Taylor series argument is valid. So, the rest of the argument proceeds as in Case 2 and the segment gets blown up to the imaginary line $\uparrow\!\! L_{\frac{\beta_1}{a}}^-$ in the $\tom$-scale (resp. )and thus the line $\uparrow\!\! L_{0}^-$ in the $V$-scale. 

\bigbreak

   Case 4. For the $v$-contours $\Ga_\LR$  in ${\mathbb L}'_{1 },  {\mathbb L}'_{4 }$, we deform this contour such that the right-most point belongs to the ball $|\om-\om_0|\in B(0,t^{-2/3-\vr''}) $ of convergence of the Taylor series and such that the contour traverses the level lines as in second part of Fig. 16. This will be a curve of steep descent, such that at the right-most point we have 
    $$
 t^2\left(  \Re (S_1)(\om_0+X+\I Y)-\Re (S_1)(\om_0)\right)\simeq -t^{2/3-2\vr''}.
$$
   This implies that the integral about $\LR$ will satisfy 
   $$\leq \{\mbox{length of $\Ga_\LR$}\}e^{-t^{2/3-2\vr''}}
 $$
 which tends to $0$ for $t\to \infty$.

  









 \vspace*{.5cm}
From (\ref{Final2}) and (\ref{Final3}), and the four cases above, we thus have:
\be\label{Final}\begin{aligned}
\lim_{t\to \infty} \frac{C_t^{(2)}}{C_t^{(1)}}&{\mathbb L}(\eta_1,\xi_1;\eta_2,\xi_2) \frac12 \Dt \xi_2
\\&=-(-1)^{\tau_1-\tau_2}
{\mathbb H}^{\tau_1-\tau_2}(  \sg_2-  \sg_1)d\sg_2
\\& -\oint_{\Ga_0}\frac{dV} {(2\pi\I)^2}\oint_{\downarrow L^-_0}\frac{ dZ}{Z-V}\frac{V^{\rho-\tau_1}}{Z^{\rho-\tau_2}}
\frac{e^{-V^2 + (   \sg_1-\bar \beta_2  )V  }}
{e^{-Z^2 + ( \sg_2-\bar \beta_2  )Z  }}
      \frac{ \Xi_r(V+\tfrac{\beta_1}a,Z+\tfrac{\beta_1}a )} { \Xi_r(0,0)} d \sg_2 
\\& -\oint_{\uparrow L^-_0}
\frac{dV} {(2\pi\I)^2}\oint_{\Ga_0}\frac{ dZ}{Z-V}\frac{V^{ -\tau_1}}{Z^{ -\tau_2}}
\frac{e^{ V^2 + (   \sg_1+\bar \beta_1  )V  }}
{e^{ Z^2 + ( \sg_2+\bar \beta_1  )Z  }}
     \frac{ \Xi_r(Z+\tfrac{\beta_1}a ,V+\tfrac{\beta_1}a )} { \Xi_r(0,0)}d \sg_2 
\\& -r\oint_{\uparrow L^-_0}\frac{dV} {(2\pi\I)^2} \oint_{\downarrow L^-_0}dZ \frac{V^{ -\tau_1}}{Z^{\rho-\tau_2}}
\frac{e^{ V^2 + (   \sg_1+\bar \beta_1  )V  }}
{e^{-Z^2 + ( \sg_2-\bar \beta_2  )Z  }}
    \frac{ \Xi^+_{r-1}( V+\tfrac{\beta_1}a,Z+\tfrac{\beta_1}a )} { \Xi_r(0,0)}d \sg_2 
 \\
 & -\tfrac1{r+1}\oint_{\Ga_0}\frac{dV} {(2\pi\I)^2} \oint_{\Ga_0}dZ 
 \frac{V^{ \rho-\tau_1}}{Z^{ -\tau_2}}
\frac{e^{-V^2 + (   \sg_1-\bar \beta_2  )V  }}
{e^{ Z^2 + ( \sg_2+\bar \beta_1  )Z  }}
    \frac{ \Xi^-_{r+1}(V+\tfrac{\beta_1}a,Z+\tfrac{\beta_1}a  )} { \Xi_r(0,0)} d \sg_2 .
\end{aligned}\ee
%
Finally, multiply (\ref{Final}) by the conjugation $(-1)^{\tau_2-\tau_1}$ and set $V\to -V$ and $Z\to -Z$, and exchanging $V\leftrightarrow Z$ in the second double integral. This maps $\downarrow L^-_0 ~ \to ~ \uparrow L^+_0$ and $\Ga_0\to \Ga_0$. It also changes the signs of the double integrals, except for the last one. One also sets $\theta_i=\sg_i-\bar \beta_2
$ and remember $
\beta=-\bar\beta_1-\bar\beta_2$.
This leads to the kernel ${\mathbb L}^{\mbox{\tiny dTac}}  ( \tau_1, \theta_1 ;\tau_2, \theta_2) $, as in (\ref{Final0}), and  with $\Theta_r( V  , Z )$ and $\Theta^{\pm}_{r\mp1}(  V, Z) $, defined by
  $$   \begin{aligned}
 \Theta_r( V, Z)&:=(-1)^{\rho r}\Xi_r(-V+\tfrac {\beta_1}a,
-Z+\tfrac {\beta_1}a)
 \\
  \Theta^{\pm}_{r\mp1}(  V, Z)   &:=(-1)^{\rho(r\mp1)}\Xi^{\pm}_{r\mp1}(-V+\tfrac {\beta_1}a,
-Z+\tfrac {\beta_1}a)
 \\
\end{aligned}$$
 which gives exactly formula (\ref{Theta}). This ends the proof of the main statement, namely Theorem  \ref{Theorem2}.

  \end{document}